\documentclass[10pt]{elsarticle}

\usepackage{amsmath, amsfonts, amssymb}
\usepackage[margin=0.75in]{geometry}
\usepackage{graphicx}
\usepackage{hyperref}
\usepackage{setspace}
\usepackage{algorithm}
\usepackage{algpseudocode}
\usepackage{tabularx}
\usepackage{lscape} 
\usepackage{rotating}
\usepackage{fancyhdr}
\usepackage{hyphenat} 
\usepackage{enumitem}
\usepackage{mathrsfs}
\usepackage{bm}
\usepackage{placeins} 
\usepackage{cleveref}
\usepackage{fancyhdr}
\usepackage{color}
\usepackage{lineno}
\usepackage{subcaption}
\usepackage{multirow}
\usepackage{adjustbox} 
\usepackage{tikz} 
\usepackage{longtable} 
\usepackage{caption}
\usepackage{makecell}
\newcommand{\prompttt}{\ttfamily\spaceskip=0.5em plus 0.5em minus 0.1em\xspaceskip=\spaceskip\sloppy}

\usepackage{rotating}
\usepackage{array, booktabs, makecell} 
\usepackage{pdflscape,colortbl}            
\newcolumntype{N}{>{$}r<{$}}               
\setcitestyle{square, comma, numbers, compress, super}
\usepackage{array, booktabs, makecell} 

\begin{document}

\begin{frontmatter}

\title{Differentiable Schwarz-Coupled Solvers for Thin Boundary Layer Transport:  Inverse Modeling via Agentic AI Optimization}


\author[utah1,utah2]{Siva Viknesh}
\author[utah1,utah2,utah3]{Amirhossein Arzani\corref{cor1}}
\ead{amir.arzani@sci.utah.edu}

\address[utah1]{Department of Mechanical Engineering, University of Utah, Salt Lake City, UT, USA}
\address[utah2]{Scientific Computing and Imaging Institute, University of Utah, Salt Lake City, UT, USA}
\address[utah3]{Department of Biomedical Engineering, University of Utah, Salt Lake City, UT, USA}
\cortext[cor1]{Corresponding author}

\begin{abstract}
High-Péclet-number transport problems give rise to thin boundary layers with steep near-wall gradients whose accurate solution requires high-resolution meshes and could challenge inverse modeling. To address this challenge, we develop an overlapping Schwarz domain-decomposition framework grounded in perturbation theory, in which an inner equation governs near-wall dynamics and an outer equation governs bulk transport, and both subproblems are discretized using the finite element method and assembled into a fully differentiable multidomain physics solver. By confining the thin boundary layer to the inner subdomain, the formulation permits  coarser mesh resolution tailored to the distinct near-wall and bulk scales, thereby reducing computational cost relative to a single-domain formulation. Exploiting the differentiability of the coupled solver, we formulate inverse problems in which the governing equations and boundary conditions are imposed as hard constraints, with the Schwarz coupling enforcing interface continuity, and parameter sensitivities are obtained through a discrete-adjoint formulation.  We consider the intrinsically multidomain setting in which measurements reside in one subdomain while the unknown parameter lies in the other, thereby allowing sensitivity information to propagate across the Schwarz coupling. To accelerate the resulting adjoint-based optimization, we integrate large language models (LLMs) into the inverse workflow in two distinct configurations: a standalone optimizer that proposes successive parameter updates from the adjoint-computed sensitivities and an agentic artificial intelligence (AI) approach with access to the complete solver workflow, enabling it to perform multistage optimization procedures. The agentic AI configuration is shown to be most advantageous when the spatially varying unknown is represented in a functional basis rather than a high-dimensional nodal field representation. In this case, the AI agent grows the library of candidate basis functions during optimization under a prescribed admission policy, admitting predefined families in stages and initializing their parameters. In short, our proposed framework integrates Schwarz-coupled differentiable physics with agentic AI optimization for the inverse modeling of multiscale transport problems.
\end{abstract}



\end{frontmatter}


\section{Introduction}
\label{sec1}
Many problems in fluid mechanics involve the transport of a passive scalar, such as heat, chemical species, or biological solute, that is carried by the flow and diffuses through it without influencing the flow itself. Of particular importance are problems in which the scalar is exchanged between the fluid and a solid wall, often leading to thin boundary layer formation. A practical example is mass transport between the blood and the vessel wall in large arteries, which plays a central role in major forms of cardiovascular disease~\cite{tarbell2003mass,farghadan2019combined,mahmoudi2021story}. In general, near-wall transport is governed by the competition between two mechanisms: advection by the flow and molecular diffusion. Their relative strength is measured by the P\'eclet number, $\mathrm{Pe}=UL/D$, where $U$ and $L$ are characteristic velocity and length scales and $D$ is the molecular diffusivity. In many applications, chemical diffusivity is small, leading to $\mathrm{Pe}\gg1$ and thin boundary layers. The thickness $\delta$ of these layers follows from a balance between tangential advection and wall-normal diffusion, and it consequently depends on the velocity profile near the wall. Although the boundary layer occupies an asymptotically small fraction of the domain, it controls the wall concentration (or temperature) and the wall flux, so that any reliable prediction of these quantities must resolve the layer accurately.



Resolving boundary layers is the central difficulty in the numerical simulation of near-wall transport at high P\'eclet numbers, as discussed in~\cite{hansen2019finite}, and requires high-resolution near-wall meshing. Layer-adapted meshes tackle this challenge by concentrating resolution near the wall, but the mesh must be redesigned whenever the layer thickness changes~\cite{linss2010layer} (e.g., different diffusivity). In practice, these requirements are severe. As an example, in large arteries, where $\mathrm{Pe}\sim10^{7}$--$10^{8}$, the concentration boundary layer is of micron scale, whereas the flow must be resolved over the entire vessel, and such simulations are consequently often under-resolved~\cite{hansen2019finite}.


Singular perturbation theory provides the classical framework for exploiting this separation of scales at high P\'eclet number~\cite{vandyke1975,bender1999advanced,holmes2013perturbation}. Since $\mathrm{Pe}^{-1}$ multiplies the highest-order derivative in the dimensionless transport equation, the problem is singularly perturbed, and an approximate solution can be constructed by the method of matched asymptotic expansions. In the outer region, which comprises the bulk of the domain, diffusion is neglected at leading order and the scalar is transported by advection alone. In the inner region, adjacent to the wall, the wall-normal coordinate is stretched by $\delta$, so that the boundary layer is $\mathcal{O}(1)$ in the stretched coordinate, independently of $\mathrm{Pe}$, and wall-normal diffusion is retained to satisfy the wall condition. The inner and outer solutions are then connected through asymptotic matching conditions and combined into a composite expansion that is uniformly valid across the domain, an approach long applied to high-P\'eclet transport problems~\cite{acrivos1965asymptotic,stone1989heat,choi2005steady,lawson2021mass}. This mathematical structure has also been embedded in machine learning based computational methods for convection-dominated transport~\cite{Arzani2023BLPINN,cao2024multistep,shen2026matched,gie2026singular}. The accuracy of these approaches, however, rests on asymptotic assumptions and the nature of coupling between the inner and outer regions. This one-way structure of coupling is adequate when the boundary layer remains attached along the entire wall, but it deteriorates when complex phenomena such as flow separation occur. At separation and reattachment points, the tangential velocity at the wall vanishes, so that the boundary layer scaling breaks down locally and the interaction between the near-wall and outer regions of the flow intensifies.

Domain decomposition provides a natural framework for such coupled problems. In Schwarz coupling, which originates from the Schwarz alternating method~\cite{schwarz1870}, the domain is divided into overlapping subdomains, and the governing equation is solved in each subdomain in turn, with Dirichlet data on its artificial boundary taken from the most recent solution in the neighboring subdomain, until the solutions agree in the overlap region~\cite{lions1988schwarz,gander2008schwarz,quarteroni1999domain,toselli2005domain}. Because each subdomain problem is a separate boundary value problem, it can be discretized independently, with non-matching meshes of different resolutions~\cite{mota2017schwarz,mota2022schwarz}. For a boundary layer problem, the domain can thus be divided into an inner subdomain adjacent to the wall and an outer subdomain containing the bulk flow, and each Schwarz iteration passes information from the inner subdomain to the outer subdomain and back, providing the two-way exchange that the one-way structure of leading-order matched asymptotics lacks. Since the inner subdomain is discretized independently, it can also be posed in the stretched wall-normal coordinate of singular perturbation theory, so that, for a known boundary-layer scaling, the inner mesh need not be redesigned as $\mathrm{Pe}$ changes, unlike a layer-adapted mesh. Such combinations of domain decomposition and singular perturbation theory have been developed for advection-dominated transport since the 1980s~\cite{chin1986parallel,hedstrom1990shock,brown1991layer,scroggs1991interface,rodrigue1989domain,hedstrom1989turning,garbey1996schwarz,garbey1997,macmullen2002convergence,boglaev2003parabolic}, and Schwarz coupling has more recently been combined with reduced-order models and neural networks~\cite{moore2025opinffe,snyder2023schwarzpinn,chumbipuma2026ninn}. In these methods, however, the subdomain containing a wall layer is posed in physical coordinates~\cite{rodrigue1989domain,hedstrom1989turning} or formulated and analyzed for an advection velocity of fixed direction that is bounded away from zero~\cite{garbey1997,mathew1998uniform,macmullen2002convergence,boglaev2003parabolic}, as are the layer-adapted meshes for such layers~\cite{hegarty1995special,hegarty2022smooth}. Other approaches reduce the cost by simplifying the governing equation in one region, either near the wall, as in surface transport models~\cite{hansen2016reduced}, or in the bulk, as in heterogeneous domain decomposition~\cite{gastaldi1990coupling,gander2016factorization}. All of these methods have consequently been developed almost exclusively for flows without separation, for which their simplifications and velocity assumptions hold. At separation and reattachment points, by contrast, the near-wall tangential velocity changes sign and, as noted above, the boundary-layer approximation breaks down, so that all terms of the advection--diffusion equation must be retained, and at separation the layer leaves the wall and is carried into the bulk. To our knowledge, Schwarz coupling with the inner subdomain in the stretched wall-normal coordinate has not been extensively investigated for thin boundary layers containing critical points of the near-wall flow, such as separation and reattachment points, where the boundary layer assumptions break down and the near-wall and bulk solutions must be coupled in both directions to obtain an accurate solution. Effective inverse modeling within such formulations is also another open problem.


The present work addresses this gap with a formulation for thin boundary layers with separation and reattachment points that combines the scale separation of perturbation theory with the two-way information exchange of Schwarz coupling. The computational domain is divided into an inner subdomain adjacent to the wall and an outer subdomain containing the bulk flow, which overlap over a finite region. The inner subdomain is posed in the stretched wall-normal coordinate of perturbation theory for efficient meshing, whereas the outer subdomain is posed in physical coordinates. In contrast to classical matched asymptotics, the full advection--diffusion equation is retained in both subdomains, and the subdomains exchange Dirichlet data across their artificial boundaries at every Schwarz iteration, so that the near-wall and bulk solutions influence each other. The resulting multidomain Schwarz-coupled solver is implemented with the finite element method in FEniCS~\cite{fenics} and made differentiable with \texttt{dolfin-adjoint}~\cite{farrell2013automated,dolfin_adjoint}, which computes gradients with the discrete adjoint method~\cite{giles2000adjoint,plessix2006review}. Thus, it allows the solver to be used for inverse modeling, in which unknown parameters are recovered from available measurements by minimizing the misfit between predicted and measured quantities. Because the boundary layer is represented in stretched coordinates, the multidomain formulation is expected to reduce computational cost in two respects. In forward modeling, very thin boundary layers, at high $\mathrm{Pe}$  numbers, can be resolved on a much coarser mesh than a single-domain discretization would require, which lowers the cost of each forward solve. This reduction is compounded in inverse modeling, where each epoch involves a forward and an adjoint solve. Beyond cost, because the governing equation and boundary conditions are enforced as hard constraints at every optimization epoch, the accuracy of the inversion is limited by the available data and the discretization error of the forward solution rather than by a balance of competing loss terms. This distinguishes the adopted differentiable physics approach from physics-informed machine-learning approaches, which impose the governing physics and the data as soft constraints~\cite{raissi2019pinn,karniadakis2021piml} (see~\cite{Elhadidy_2026} for a comparison of the two approaches). Nevertheless, the efficacy of the inversion depends not only on the cost of each forward and adjoint solve but also on the optimization strategy used to translate the adjoint gradients into parameter updates, which is the second focus of this work.

Classical gradient-based optimizers (for instance, Adam and Newton methods) update the parameters through a deterministic rule that is selected a priori and remains fixed throughout the optimization. 
In contrast, large language models (LLMs) have recently been proposed as optimizers that generate the parameter update from the optimization history supplied to them as a contextual prompt, potentially allowing the optimization strategy to be chosen adaptively for the problem at hand~\cite{ma2024llm}. Depending on their degree of interaction with the solver, LLMs can serve in a passive or an active role. In a passive role, the LLM acts as a standalone optimizer in place of a classical method and proposes the next update from the history of previous iterates and their objective values~\cite{yang2024large,liu2024llambo,brahmachary2025leo,zhang2024llmpso}. In an active role, the LLM operates as an agent with access to the entire solver and can update or modify the solver on the basis of the optimization history. Such agents have automated computational fluid dynamics workflows~\cite{pandey2025openfoamgpt,chen2025optmetaopenfoam,xu2025cfdagent}, generated and debugged finite element and reduced-order solvers~\cite{deotale2026allfem,guo2026llmcae}, calibrated the parameters of flow and transport models in inverse problems~\cite{ma2026groundwater}, and discovered governing equations from data~\cite{du2024llm4ed,shojaee2025llmsr}. This recently emerging agentic AI paradigm has further been extended to multi-agent systems for scientific machine learning, in which specialized agents propose, critique, and refine the model architecture, loss formulation, and training strategy, and discover methods that outperform single-agent and human-designed baselines~\cite{jiang2026agenticsciml}. In the present work, the differentiable inverse solver is retained unchanged up to the computation of the misfit and its adjoint gradients are passed to the LLM at each optimization epoch. In the passive role, the LLM uses a limited history of the misfit, gradient, and parameters to propose the next update and can adapt its strategy rather than following a fixed classical rule. In the active role, the agent has access to the complete optimization history and the inverse-solver workflow, selects the update strategy at each epoch, and, when the unknown is represented in a functional basis, additionally decides when to admit predefined families of basis functions and initializes their parameters. In both methodological roles, the language model is wrapped around a mechanistic differentiable physics solver and seeks to accelerate its convergence for inverse problems, and its performance is compared with that of classical optimizers. Rather than claiming that LLMs and agentic AI can replace classical methods, this work investigates inverse modeling scenarios in which they may offer a viable optimization strategy~\cite{huang2024exploring}, and we also seek to understand scenarios in which they currently cannot surpass classical methods.

The main contributions of this work are as follows.
\begin{itemize}
  \item \textbf{Multidomain Schwarz-coupled solver for thin boundary layers:}
  A Schwarz-coupled solver is developed for advection--diffusion problems with thin boundary layers containing separation and reattachment points, in which a perturbation-based inner subdomain in the stretched wall-normal coordinate and an outer subdomain in physical coordinates are coupled in a two-way fashion. 
  
  
  \item \textbf{Differentiable multidomain inversion:} The Schwarz-coupled solver is made differentiable with the discrete adjoint method, with the governing equation and boundary conditions enforced as hard constraints. The adjoint sensitivities propagate across the subdomain interface, so that the measurements and the unknown can reside in different subdomains.
  
  \item \textbf{LLM-integrated and agentic optimization for inverse modeling:} LLMs are integrated into the differentiable inverse solver, which remains unchanged and supplies the misfit and adjoint gradient at each epoch. LLM reasoning is used to propose a parameter update policy during the course of optimization.

 \item \textbf{Staged growth of the basis function library with agentic AI:} For inverse problems in which the unknown varies in space, the LLM agent grows the library of basis functions during the inversion instead of using a library fixed before the optimization begins. Under a prescribed admission policy, the agent decides when to admit new pre-specified candidate functions and decides how to initialize their parameters.
    
  
  \item \textbf{Comparison with classical optimizers:} LLM-based and classical gradient-based optimizers are compared systematically for boundary layer transport problems over a range of P\'eclet numbers and unknowns of increasing spatial complexity, to identify where LLMs offer a viable alternative and where current LLM methodology does not surpass
  classical methods.
\end{itemize}

The remainder of this paper is organized as follows. In Section~\ref{sec:methods}, we present the governing equations and methods. In Section~\ref{sec3}, we first benchmark the solver in a forward setting and then provide a detailed comparison with  classical methods for inverse modeling.  In Section~\ref{sec4}, the findings, their limitations, and directions for future work are discussed, and in Section~\ref{sec5}, the paper is concluded.  The solver algorithms, as well as the prompts and human validation of the LLM-based optimizations, are provided in the Appendix.



\section{Methods}
\label{sec:methods}

\subsection{Governing equations}
\label{sec:governing}
We consider the steady transport of a passive scalar in a bounded domain $\Omega\subset\mathbb{R}^{2}$. The boundary $\partial\Omega=\Gamma_{w}\cup\Gamma_{D}$ comprises a wall $\Gamma_{w}$ subject to a prescribed flux and a remainder $\Gamma_{D}$ subject to a Dirichlet condition. The scalar concentration $C$ is governed by the advection--diffusion equation
\begin{equation}
  \mathbf{v}\cdot\nabla C \;=\; D\,\nabla^2 C
  \qquad \text{in } \Omega,
  \label{eq:adv_diff}
\end{equation}
where $\mathbf{v}$ is a prescribed incompressible velocity field satisfying $\nabla\cdot\mathbf{v}=0$, and $D$ is a constant molecular diffusivity. Here, we focus on the advection-dominated regime, characterized by a large P\'eclet number $\mathrm{Pe}=UL/D\gg1$, where $U$ is a characteristic velocity scale and $L$ is a characteristic domain length.
 
 
Near the wall, a thin boundary layer develops, across which $C$ varies rapidly in the wall-normal direction. The characteristic layer thickness follows from a balance between tangential advection and normal diffusion. For a slip wall, the tangential velocity is $\mathcal{O}(U)$ at the wall, yielding $\delta\sim L\,\mathrm{Pe}^{-1/2}$, whereas, for a no-slip wall, the tangential velocity vanishes at the wall, and the corresponding balance yields the Lévêque scaling $\delta\sim L\,\mathrm{Pe}^{-1/3}$. In either case, $\delta\ll L$ is noted for sufficiently large $\mathrm{Pe}$. Here, a Cartesian coordinate system $(x, y)$ is considered,  with $x$ parallel to the bottom wall and $y$ directed inward normal to it, such that $\Gamma_{w}$ lies at $y=0$ and the domain occupies $y>0$. The Neumann boundary condition $\partial C/\partial n=g(x)$ is imposed on $\Gamma_{w}$, where $n$ is directed into the domain. Thus, $g(x)<0$ denotes scalar injection into the domain. Here, the prescribed gradient, which is referred to as the wall flux in the manuscript, $g(x)$, may either vary spatially along $\Gamma_{w}$ or be a constant scalar value. Zero Dirichlet values are imposed on all remaining boundaries $\Gamma_{D}$. 

\subsection{Forward modeling through the differentiable physics solver}
\label{sec:forward}
\subsubsection{High-fidelity single-domain forward solver}
\label{sec:forward-baseline}

For a reference configuration, we solve the governing Eq.~\ref{eq:adv_diff} over the full domain $\Omega$ using a single high-resolution mesh that directly resolves the thin near-wall boundary layer. The corresponding weak formulation is
\begin{equation}
  \int_{\Omega}(\mathbf{v}\cdot\nabla C)\,\phi\,d\mathbf{x}
  + D\!\int_{\Omega}\nabla C\cdot\nabla \phi\,d\mathbf{x}
  - D\!\int_{\Gamma_{w}} g(x)\,\phi\,d\Gamma
  \;=\; 0,
  \label{eq:baseline_weak}
\end{equation}
where $\phi$ is a test function and the boundary integral over $\Gamma_{w}$ weakly imposes the prescribed flux $g(x)$. The problem is discretized using continuous piecewise-linear finite elements and stabilized by the streamline-upwind Petrov--Galerkin (SUPG) method. The resulting linear system is assembled and solved using a sparse direct solver.
 
The mesh is refined until the wall profile and the interior field are grid-converged. The resulting single-domain solution serves as the reference for the present work, where its concentration field provides the benchmark for assessing the accuracy of the Schwarz-coupled solver.

\subsubsection{Schwarz-coupled inner--outer perturbation theory forward solver}
\label{sec:forward-coupled}

\begin{figure}  
    \centering
    \includegraphics[width=\textwidth,height=\textheight,keepaspectratio]{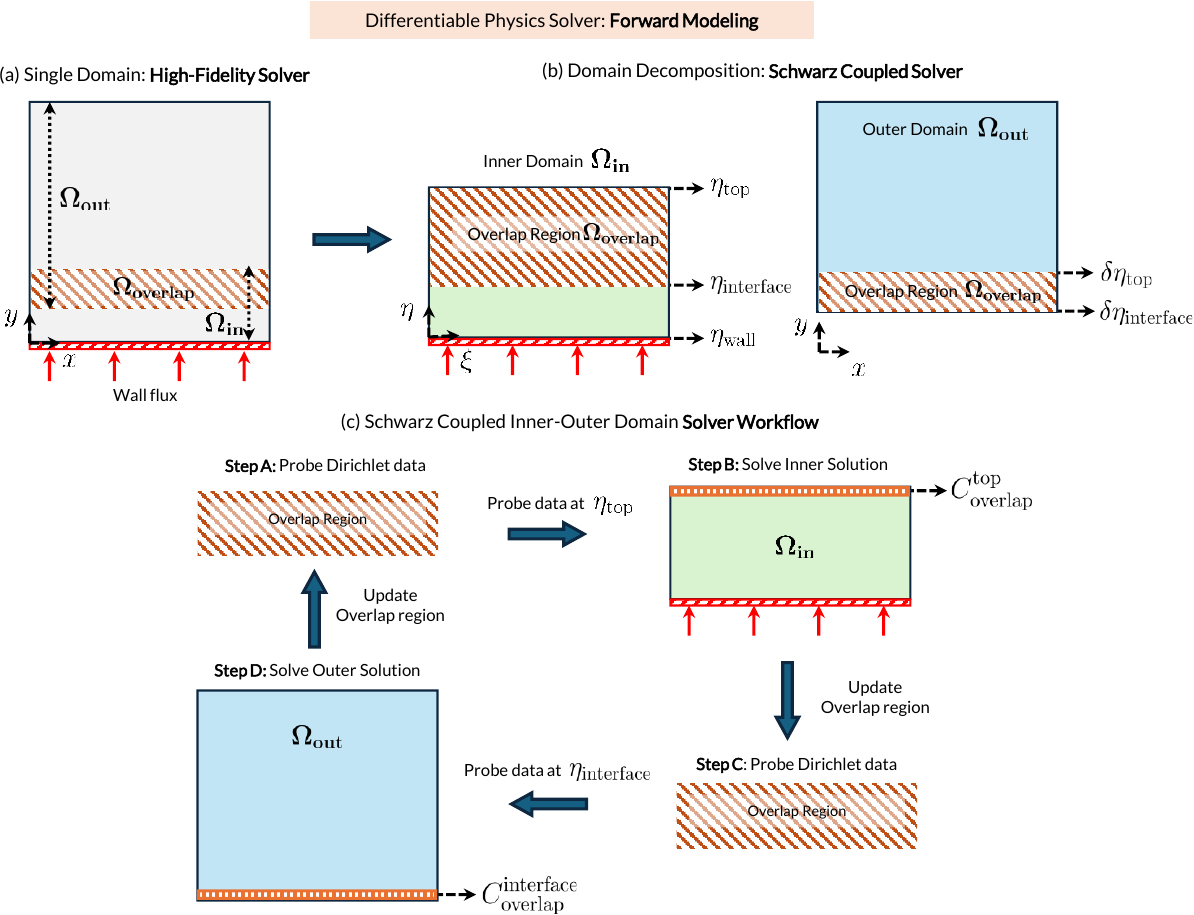}
    \caption{Schwarz-coupled inner--outer domain decomposition. 
    (a)~Single-domain configuration in Cartesian coordinates $(x,y)$, with prescribed scalar influx $g(x)$ on the lower wall. (b)~Overlapping decomposition into an inner subdomain $\Omega_{\mathrm{in}}$, formulated in the stretched coordinates $(\xi,\eta)$, and an outer subdomain $\Omega_{\mathrm{out}}$ in physical coordinates. The overlap region is bounded by $\eta_{\mathrm{interface}}$ and $\eta_{\mathrm{top}}$, which correspond to $y=\delta\eta_{\mathrm{interface}}$ and $y=\delta\eta_{\mathrm{top}}$ in the outer domain. 
    (c)~One alternating Schwarz sweep. Dirichlet data probed from the overlap at $\eta_{\mathrm{top}}$  drive the inner solve. The updated overlap is  then probed at $\eta_{\mathrm{interface}}$ to drive the outer solve. Each overlap update is under-relaxed and sweeps continue until the change falls below the tolerance $\varepsilon$.}
\label{fig:schematic}
\end{figure}

To resolve thin boundary layers efficiently, the forward problem is solved using an overlapping Schwarz domain-decomposition method~\cite{mota2017schwarz,mota2022schwarz} with singular perturbation theory~\cite{bender1999advanced,Arzani2023BLPINN}. The computational domain is divided into an inner subdomain, $\Omega_{\mathrm{in}}$, containing the near-wall boundary layer, and an outer subdomain, $\Omega_{\mathrm{out}}$, containing the bulk outer flow. The advection--diffusion equation is retained in both subdomains, and the two solutions are coupled numerically over a finite overlap domain, rather than through the more conventional asymptotic matching approach in classical perturbation theory. The overlap domain, $\Omega_{\mathrm{overlap}} = \Omega_{\mathrm{in}}\cap\Omega_{\mathrm{out}}$, has thickness $\mathcal{O}(\delta)$ and is located at a wall-normal distance $\mathcal{O}(\delta)$ from the bottom wall, $\Gamma_{w}$. Thus, the upper portion of $\Omega_{\mathrm{in}}$ and the lower portion of $\Omega_{\mathrm{out}}$ represent the same physical region. The coupled solution is obtained iteratively. At each Schwarz iteration, the concentration trace from the inner solution is imposed on the lower artificial boundary of $\Omega_{\mathrm{out}}$, while the trace from the outer solution is imposed on the upper artificial boundary of $\Omega_{\mathrm{in}}$. The iterations are continued until the exchanged traces and the solution in the overlap domain converge. An overview of the domain decomposition is shown in Fig.~\ref{fig:schematic}.
 
Within the inner subdomain $\Omega_{\mathrm{in}}$, singular perturbation theory is used and the physical Cartesian coordinates $(x, y)$ are transformed to the stretched coordinates $\xi=x/L$ and  $\eta=y/\delta$, where $\delta$ is the boundary layer thickness scale and $L$ is the characteristic domain length. Retaining the physical dimensions for the dependent variables, under this transformation, Eq.~\ref{eq:adv_diff} becomes
\begin{equation}
  \frac{\delta}{L}\,v_\xi\,\frac{\partial C}{\partial \xi}
  + v_\eta\,\frac{\partial C}{\partial \eta}
  =
  \frac{D\delta}{L^{2}}\,\frac{\partial^2 C}{\partial \xi^2}
  + \frac{D}{\delta}\,\frac{\partial^2 C}{\partial \eta^2}.
  \label{eq:inner_pde_general}
\end{equation}


For a slip wall, the boundary layer thickness scales as $\delta/L\sim\mathrm{Pe}^{-1/2}$. We set $L=1$ and $U=1$, which gives $\delta=\sqrt{D}$. Substituting this choice into Eq.~\ref{eq:inner_pde_general} and dividing by $\sqrt{D}$ gives
\begin{equation}
  v_\xi\,\frac{\partial C}{\partial \xi}
  + \frac{v_\eta}{\sqrt{D}}\,\frac{\partial C}{\partial \eta}
  =
  D\,\frac{\partial^2 C}{\partial \xi^2}
  + \frac{\partial^2 C}{\partial \eta^2},
  \label{eq:inner_pde}
\end{equation}
while for a no-slip wall, the L\'ev\^eque scaling gives $\delta/L\sim\mathrm{Pe}^{-1/3}$. Proceeding in the same way with the same choice of scales gives $\delta=D^{1/3}$. Substituting this choice into Eq.~\ref{eq:inner_pde_general} and dividing by $D^{2/3}$ gives
\begin{equation}
  \frac{v_\xi}{D^{1/3}}\,\frac{\partial C}{\partial \xi}
  + \frac{v_\eta}{D^{2/3}}\,\frac{\partial C}{\partial \eta}
  =
  D^{2/3}\,\frac{\partial^2 C}{\partial \xi^2}
  + \frac{\partial^2 C}{\partial \eta^2},
  \label{eq:inner_pde_noslip}
\end{equation}
where the second derivatives with respect to $\xi$ and $\eta$ represent wall-parallel and wall-normal diffusion, respectively. Although the wall-parallel diffusion term is asymptotically small, it is retained in the formulation, providing mild stabilization of the discrete problem numerically. The coordinate stretching renders the boundary layer $\mathcal{O}(1)$ in $\eta$, thereby relaxing the need for excessive wall-normal mesh refinement in physical coordinates. The inner domain formulation is closed using the prescribed wall flux condition from Section~\ref{sec:governing}. The flux condition in stretched coordinates is imposed as $\partial C/\partial\eta=\delta\,g(L\xi)$ at $\eta=0$. Homogeneous zero Dirichlet conditions are applied on the lateral boundaries, and the concentration trace supplied by the outer solution is imposed as a Dirichlet condition at the upper boundary of the inner subdomain,  $C_{\mathrm{overlap}}^{\mathrm{top}}$ probed at $\eta=\eta_{\mathrm{top}}$, during the Schwarz iteration. The outer subdomain retains the original, unscaled form of Eq.~\ref{eq:adv_diff} in the physical Cartesian coordinates $(x,y)$. The outer problem is stabilized using the SUPG formulation. It is closed by the prescribed lateral boundary conditions together with a Dirichlet concentration trace supplied by the inner solution at the lower artificial boundary of $\Omega_{out}$. 

At each Schwarz iteration, the concentration trace obtained from one subdomain is prescribed as a Dirichlet boundary condition on the artificial boundary of the other subdomain and the overlap concentration field \(C_{\mathrm{overlap}}\) is under-relaxed according to
 
\begin{equation}
  C^{(k)}_{\mathrm{overlap}} \;=\; \omega\,\widehat{C}^{(k)}_{\mathrm{overlap}}
  + (1-\omega)\,C^{(k-1)}_{\mathrm{overlap}},
  \qquad \omega\in(0,1],
  \label{eq:relax}
\end{equation}
where $\widehat{C}^{(k)}_{\mathrm{overlap}}$ denotes the newly computed concentration solution at a Schwarz iteration $k$, $C^{(k-1)}_{\mathrm{overlap}}$ is the relaxed concentration solution from the preceding iteration, and $\omega$ is the relaxation parameter. Convergence is assessed from the change in the relaxed overlap concentration field between two successive Schwarz iterations, and the iteration is terminated when $\big\|C_{\mathrm{overlap}}^{(k)}-C_{\mathrm{overlap}}^{(k-1)}\big\|_{\Omega_{\mathrm{overlap}}}$ falls below the prescribed tolerance $\varepsilon$. Alternatively, the convergence criterion may be
evaluated only on either artificial boundary, \(\eta=\eta_{\mathrm{top}}\) or \(\eta=\eta_{\mathrm{interface}}\). Based on our experiments, this reduced criterion produces a coupled solution that is indistinguishable, within numerical accuracy, from that obtained by evaluating the mismatch throughout the overlap region.

\subsubsection{Forward solution procedure}
Rather than resolving the entire domain on a single, globally refined mesh, the Schwarz-coupled solver divides the problem into two subproblems, each discretized to suit the physics it resolves. The reference configuration, shown in Fig.~\ref{fig:schematic}a, is posed over the full domain $\Omega$ in Cartesian coordinates $(x,y)$ and solved with the high-fidelity solver of Section~\ref{sec:forward-baseline}. 
The corresponding decomposed configuration is presented in Fig.~\ref{fig:schematic}b. The inner subdomain $\Omega_{\mathrm{in}}$ extends from the wall, $\eta=\eta_{\mathrm{wall}}=0$, to its upper artificial boundary,
$\eta=\eta_{\mathrm{top}}$, and is formulated in the stretched coordinates $(\xi,\eta)$. The outer subdomain $\Omega_{\mathrm{out}}$ extends from its lower artificial boundary, $y=\delta\eta_{\mathrm{interface}}$, to the far field and is formulated in the physical coordinates $(x,y)$. Since $\eta_{\mathrm{interface}}<\eta_{\mathrm{top}}$, the two subdomains share the overlap region $\Omega_{\mathrm{overlap}}$: $\eta_{\mathrm{interface}}\le\eta\le\eta_{\mathrm{top}}$, equivalently
$\delta\eta_{\mathrm{interface}}\le y\le\delta\eta_{\mathrm{top}}$. Each artificial boundary therefore lies in the interior of the other subdomain, where that subdomain's solution provides the required Dirichlet data. The two
subdomains are meshed separately with the same spatial resolution. Data are transferred between them by evaluating the donor finite-element solution at the receiver's nodes, which are located through the mapping $x=L\xi$, $y=\delta\eta$.

The coupled problem is solved by the alternating Schwarz sweep illustrated in Fig.~\ref{fig:schematic}c. At sweep $k$, the following four steps are performed.
 
\begin{itemize}
  \item \textbf{Step A: Probe Dirichlet data for the inner problem.} Evaluate $C_{\mathrm{overlap}}$ at $\eta=\eta_{\mathrm{top}}$ to obtain the trace $C_{\mathrm{overlap}}^{\mathrm{top}}$. For $k>1$, this trace originates from the outer solution of the preceding sweep.
  
  \item \textbf{Step B: Inner solve.} Solve the inner equation, Eq.~\ref{eq:inner_pde} for a slip wall or Eq.~\ref{eq:inner_pde_noslip} for a no-slip wall, on $\Omega_{\mathrm{in}}$. The wall flux $\partial C/\partial\eta=\delta\,g(x)$  is imposed at $\eta=\eta_{\mathrm{wall}}$, and the Dirichlet condition $C_{\mathrm{overlap}}^{\mathrm{top}}$ is imposed at $\eta=\eta_{\mathrm{top}}$. The inner solution restricted to the overlap then serves as $\widehat{C}_{\mathrm{overlap}}$ in Eq.~\ref{eq:relax}, which updates $C_{\mathrm{overlap}}$.
  \item \textbf{Step C: Probe Dirichlet data for the outer problem.} Evaluate the updated $C_{\mathrm{overlap}}$ at $\eta=\eta_{\mathrm{interface}}$ to obtain the trace $C_{\mathrm{overlap}}^{\mathrm{interface}}$. This trace originates from the inner solution of the current sweep.
  \item \textbf{Step D: Outer solve.} Solve the SUPG-stabilized form of Eq.~\ref{eq:adv_diff} on $\Omega_{\mathrm{out}}$. The prescribed far-field  and lateral conditions are imposed, together with the Dirichlet condition  $C=C_{\mathrm{overlap}}^{\mathrm{interface}}$ at $y=\delta\eta_{\mathrm{interface}}$. The outer  solution restricted to the overlap then serves as $\widehat{C}_{\mathrm{overlap}}$ in  Eq.~\ref{eq:relax}, which updates $C_{\mathrm{overlap}}$ to $C_{\mathrm{overlap}}^{(k)}$.
\end{itemize}
Each sweep thus performs one inner-to-outer and one outer-to-inner exchange,
and each subdomain always receives the most recent information from the other.
The sweep is repeated until the change of $C_{\mathrm{overlap}}$ between two successive sweeps falls below the tolerance $\varepsilon$. Upon convergence, the inner and
outer solutions agree throughout the overlap to within $\varepsilon$ and
together constitute the coupled solution, which is compared with the
single-domain reference.
 
\textit{One-way method:} To assess the role of the two-way coupling, we also consider a one-way variant
in which the procedure advances from Step A to Step D only once and never completes the Step D to Step A loop.  The one-way approach is what is used in classical perturbation theory~\cite{Arzani2023BLPINN}, which limits accuracy when complex phenomena such as flow separation are present. Comparing the one-way and fully coupled solutions, therefore, isolates the contribution of the
outer-to-inner feedback. It demonstrates that the accuracy of the method stems from cycling the data exchange between the subdomains, not merely from performing subdomain solves.
 

The complete forward procedure is summarized in Algorithm~\ref{alg:forward} of~\ref{app:algorithms}. Within this algorithm, the one-way variant corresponds to a single pass through the Schwarz loop, obtained by setting the iteration limit to $K=1$.

\subsection{Inverse modeling through the differentiable physics solver}
\label{sec:inverse}
The forward solvers developed above serve as the basis for an inverse problem, in which an unknown parameter $\theta$ of the transport model is recovered from measurements available only on a subregion $\Omega_{\mathrm{obs}}$ of the domain. The parameter may enter the governing operator, as in the diffusivity $D$, or the boundary condition, as in the wall flux $g(x)$. The recovery is posed as minimization of the data misfit
\begin{equation}
  J(\theta) \;=\; \frac{1}{2}\int_{\Omega_{\mathrm{obs}}}
  \bigl(C(\theta)-C_{\mathrm{obs}}\bigr)^{2}\,d\mathbf{x},
  \label{eq:misfit}
\end{equation}
where $C(\theta)$ denotes the forward prediction and $C_{\mathrm{obs}}$ the partial available data. In contrast to soft-constrained formulations, in which the governing equations are appended to the objective as penalty terms, the differentiable solver minimizes the misfit $J(\theta)$ subject to the governing equations and boundary conditions being satisfied exactly, to the accuracy of the discretization chosen, and the Schwarz coupling to the prescribed iterative tolerance. Minimization requires the gradient $dJ/d\theta$, whose computation differs between the single-domain and coupled treatments developed below: the single-domain case is presented first as a reference scenario, followed by the coupled solver methodology, in which the sensitivity is propagated across the subdomain interface through a differentiable strategy.

\subsubsection{High-fidelity single-domain inverse solver formulation}
 
\label{sec:inverse-baseline}
In the single-domain solver, the forward problem is solved exactly at each parameter $\theta$ and the state is thereby determined by the parameter, $\mathbf{u}=\mathbf{u}(\theta)$. Writing the discretized forward problem as $\mathbf{R}(\mathbf{u},\theta)=\mathbf{0}$, the discrete adjoint yields the required derivative by introducing an adjoint state
$\boldsymbol{\mu}$, which is first obtained from a transpose solve and then contracted to give the reduced gradient,
\begin{equation}
  \Big(\frac{\partial \mathbf{R}}{\partial \mathbf{u}}\Big)^{\!\top}
  \boldsymbol{\mu}
  \;=\;\Big(\frac{\partial J}{\partial \mathbf{u}}\Big)^{\!\top},
  \qquad
  \frac{dJ}{d\theta}
  \;=\;\frac{\partial J}{\partial \theta}
  -\boldsymbol{\mu}^{\!\top}\frac{\partial \mathbf{R}}{\partial \theta}.
  \label{eq:adjoint}
\end{equation}
The adjoint solve on the left is driven by the misfit sensitivity $\partial J/\partial\mathbf{u}$, which is nonzero only on $\Omega_{\mathrm{obs}}$. Since the forward operator is already factorized, this solve reuses the same factorization and is correspondingly inexpensive. Given $\boldsymbol{\mu}$, the gradient on the right follows from a single contraction against the parametric sensitivity $\partial\mathbf{R}/\partial\theta$ and is exact for the discrete problem. In practice, neither relation is assembled by hand and the forward solve is recorded and reverse-differentiated by algorithmic automatic differentiation in FEniCS, which returns this gradient directly. 

\subsubsection{Schwarz-coupled inverse solver formulation}
\label{sec:inverse-coupled}
In the coupled solver, the forward evaluation is recorded as a directed sequence of subdomain solves connected by the interface transfer, and the corresponding adjoint is obtained by applying reverse-mode differentiation to this sequence. Since the two subdomains are coupled only through their interface traces, the transfer provides the sole pathway for sensitivity propagation between them. Each transfer samples one subdomain at the interface and injects the value as boundary data for the other. Denoting the inner-to-outer transfer by $\mathbf{P}$, the receiving near-wall datum is $C_{\mathrm{interface}}=\mathbf{P}\,\mathbf{C}_{\mathrm{in}}$. As $\mathbf{P}$ is a fixed linear map, consisting of an interface probe composed with interpolation onto the receiving boundary, its adjoint is exactly the transpose $\mathbf{P}^{\!\top}$.
 
At a given $\theta$, the two subdomains are advanced alternately until the Schwarz iteration converges, after which the interface trace feeding back into the subdomain containing $\theta$ is held fixed. Importantly, this frozen trace is now the converged interface trace, which evolved from the initialized interface trace over the Schwarz iteration. Thus, the recorded evaluation must be taken at the converged state for the gradient to correspond to the coupled problem. With the feedback frozen, the recorded model constitutes a single directed evaluation and $J$ reduces to a function of $\theta$ alone, so that its gradient follows from the same discrete adjoint Eq.~\ref{eq:adjoint}, as in the single-domain case. The misfit sensitivity $\partial J/\partial\mathbf{u}$, which remains nonzero only on $\Omega_{\mathrm{obs}}$, seeds the adjoint in the subdomain containing the data, and the subsequent path of the adjoint is governed by the placement of $\theta$ relative to the data region, leading to two regimes:
\begin{itemize}[noitemsep]
  \item \emph{Common subdomain:} When $\theta$ resides in the same subdomain as the data, the adjoint is seeded and closed within it, $\mathbf{P}^{\!\top}$ does not enter, and the recovery reduces to the single-domain adjoint on a single scale. Nevertheless, the coupling iteration remains necessary in the forward prediction, as the interface trace closing that subdomain acts as its boundary condition and determines the accuracy of the recovered field. 
  \item \emph{Distinct subdomains:} When $\theta$ lies in the opposing subdomain, the adjoint must cross, and $\mathbf{P}^{\!\top}$ conveys the sensitivity from the data side across the overlap, with the adjoint of the remaining domain completing the path to $\theta$. This transposed transfer operator is the coupling mechanism in the reverse pass, without which the two subdomain adjoints decouple, and the cross-domain gradient vanishes identically. 
\end{itemize}
 
 
 
\subsubsection{Parametrization of the unknown variable}
\label{sec:param}
The construction is independent of how the unknown variable $\theta$ is represented. In each case, the chosen quantities are declared as controls to the adjoint framework, and since the forward solve and the adjoint computations are unchanged, the representation affects only the form of the returned gradient. Throughout, $\theta$ denotes the set of controls, and the cases differ in their definitions as follows.
 
\begin{itemize}
\item\emph{Scalar control:} When the unknown variable is a single scalar, such as the diffusivity $D$ or a spatially uniform boundary flux, $\theta$ contains only that value.
 
\item\emph{Discrete field:} When the unknown variable is a spatially varying field, such as a non-uniform wall flux  $g(x)$, it may be recovered pointwise, with $\theta=\{g_i\}$ the nodal values on the mesh and the adjoint returning the sensitivity at each node. This representation imposes no functional form on $g$, at the cost of one control variable per node. As a consequence, the dimension of $\theta$ scales proportionally with the chosen mesh resolution.
 
\item\emph{Functional basis:} The same field may instead be expanded in a set of basis functions, $g(x)=\sum_j c_j\, \phi_j(x;\boldsymbol{\beta}_j)$, which represent it with substantially fewer controls than a nodal discretization. The controls $\theta=\{c_j,\boldsymbol{\beta}_j\}$ comprise the linear coefficients $c_j$ that scale the basis functions and the internal parameters $\boldsymbol{\beta}_j$, such as the frequency of a sinusoidal function, which enter nonlinearly. The adjoint returns the gradient with respect to both through the same recorded differentiation. The basis set may be fixed and selected from a predefined library or grown during the inversion.

\end{itemize}

\subsubsection{Inverse solution procedure}
\label{sec:inverse-procedure}
 
\begin{figure}
    \centering
    \includegraphics[width=0.8\textwidth,height=\textheight,keepaspectratio]{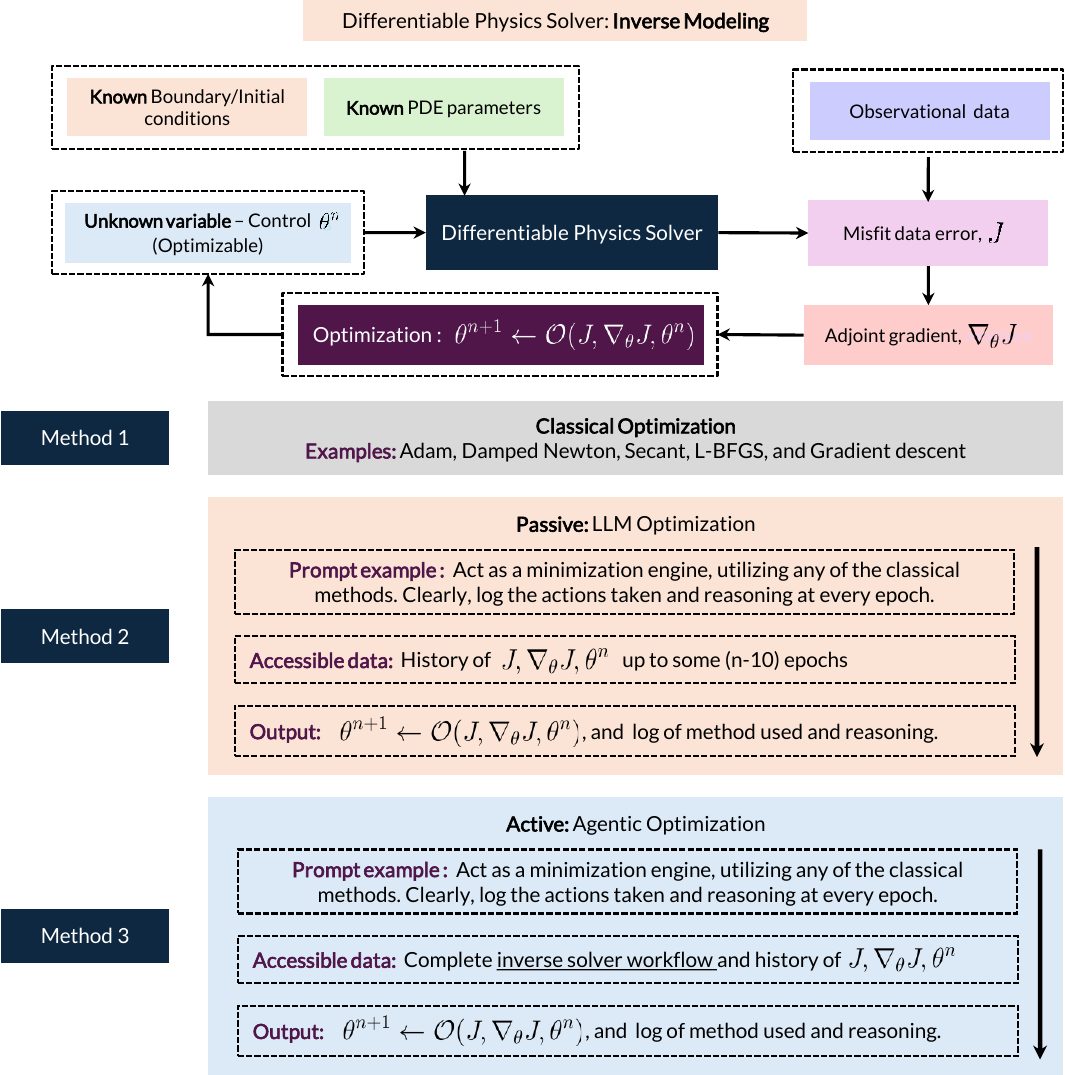}
    \caption{Inverse modeling with the differentiable physics solver. At each
  optimization epoch $n$, the solver maps the control $\theta^{n}$, together
  with the known boundary conditions and PDE parameters, to a prediction that
  is compared with the observational data to form the misfit $J$. The adjoint
  gradient $\nabla_\theta J$ is then passed to an update operator
  $\theta^{n+1}\gets\mathcal{O}(J,\nabla_\theta J,\theta^{n})$. Three
  realizations of $\mathcal{O}$ are considered: Method~1, classical
  gradient-based optimizers; Method~2, a passive LLM optimizer with access to
  a limited history of $J$, $\nabla_\theta J$, and $\theta$; and Method~3, an
  active agentic optimizer with access to the complete history and the full
  inverse-solver workflow.}
    \label{fig:inverse}
\end{figure}

With the gradient computation and the parametrizations established, the complete inverse problem is solved by an iterative loop that separates two roles. The differentiable physics solver evaluates the objective
and its gradient, and an optimizer uses this information to update the unknown
parameter. This workflow is shown in Fig.~\ref{fig:inverse}. Starting from an initial guess $\theta^{0}$, each optimization epoch $n$ consists of the following steps.
\begin{itemize}
  \item \textbf{Step 1: Forward evaluation.} Solve the forward problem at
  $\theta^{n}$. For the single-domain solver, this is a direct solve on the
  full mesh. For the Schwarz-coupled solver, the Schwarz iteration of
  Section~\ref{sec:forward-coupled} is first advanced to convergence without
  being recorded for differentiation. The interface trace feeding the
  subdomain containing $\theta$ is then frozen at its converged value, and a
  single directed evaluation of the coupled model is recorded with this trace
  held fixed.
  \item \textbf{Step 2: Misfit evaluation.} Compare the forward prediction with
  the observations on $\Omega_{\mathrm{obs}}$ to obtain $J(\theta^{n})$ from
  Eq.~\eqref{eq:misfit}.
  \item \textbf{Step 3: Adjoint gradient.} Apply reverse-mode algorithmic
  differentiation to the recorded evaluation to obtain
  $\nabla_\theta J(\theta^{n})$, following Eq.~\eqref{eq:adjoint}. Because
  only the final evaluation is recorded, the cost and memory of the reverse
  pass are independent of the number of Schwarz sweeps. When $\theta$ and the
  data lie in distinct subdomains, the sensitivity crosses the overlap through
  $\mathbf{P}^{\!\top}$, as described in Section~\ref{sec:inverse-coupled}.
  \item \textbf{Step 4: Parameter update.} Pass the objective, its gradient,
  and the current iterate to the update operator $\mathcal{O}$, which returns
  $\theta^{n+1}=\mathcal{O}\big(J,\nabla_\theta J,\theta^{n}\big)$.
\end{itemize}
 
The loop is repeated until the misfit or the parameter update falls below a
prescribed tolerance, or until a maximum number of epochs is reached. The
optimization epoch $n$ is distinct from the Schwarz sweep index $k$, and every
epoch contains a complete, converged Schwarz iteration within Step 1.

Steps  1--3 are identical in all cases. The optimization methods differ only in how Step 4 is realized. Three methods are considered, shown as Methods 1--3 in Fig.~\ref{fig:inverse}. The classical method establishes the
reference case, while the two large language model (LLM) configurations differ in the extent of their access to the inverse solver workflow:
\begin{itemize}
  \item \textit{Method 1 -- Classical optimization:} In this method, the update
  operator $\mathcal{O}$ is a conventional gradient-based optimizer, such as
  gradient descent, Adam, L-BFGS, a secant method, or a damped Newton method.
  The update rule is fixed in advance and applied identically at every epoch.
  Among these, the first-order Adam scheme serves as the reference update rule
  throughout, and the second-order damped Newton scheme is applied where the
  control dimension keeps the curvature information tractable. 
  
  \item \textit{Method 2 -- Passive LLM optimization:} In this method, the
  update operator is realized by an LLM, the open-source
  Qwen3-8B~\cite{qwen}, which acts as a standalone optimizer in place of the
  classical update rule. The model is prompted to act as a minimization engine.
  It may employ any of the classical methods, and it must record the action
  taken and its reasoning at every epoch. Its access is limited to the recent
  optimization history, namely a text-formatted record of $J$,
  $\nabla_\theta J$, and $\theta$ over the ten most recent epochs. It returns the selected method and its controls, from which the controller computes
  $\theta^{n+1}$, together with a log of the reasoning behind it. The model is passive in the sense that it acts only on this
  history and has no access to the solver or the wider workflow. This
  configuration assesses whether a language model can generate effective
  updates from the optimization history alone in a very controlled fashion.

  \item \textit{Method 3 -- Active agentic optimization:} In this method, the update operator is realized by an autonomous LLM agent, GPT~5.6 Terra, built on Codex~\cite{codex}, which receives a similar prompt and produces the same type of output as in Method~2. However, the agent is given access to the complete optimization history and to the full inverse solver workflow. It has read access to the entire code but write access only to the definition of $\theta$, so that the forward and adjoint computations remain untouched and the agent operates solely on their output. At each epoch, after Steps~1--3 are complete, the agent examines the misfit and adjoint sensitivities and returns the update method and its controls, from which the controller computes $\theta^{n+1}$. In addition, the write access allows the agent to change the representation of $\theta$ during the inversion, which is exploited only for the functional basis parametrization (Section~\ref{sec:llm-basis}). This cycle continues for a prescribed number of epochs.
\end{itemize}

The distinction between the three methods is most pronounced for the
functional basis parametrization. Method 1 can adjust only the coefficients of a fixed library. In contrast, the agent of Method 3 can additionally grow the library during the inversion, deciding under a prescribed admission policy when to admit new candidate functions and initializing their parameters. The idea is motivated by the recent use of LLMs for equation discovery and symbolic regression~\cite{pang2026deliberate,islam2026discovery}, although here the search is restricted to the staged admission of prescribed candidate functions. The complete inverse procedure is summarized in Algorithm~\ref{alg:inverse} of~\ref{app:algorithms}.



\subsection{Differentiable physics solver implementation}
\label{sec:implementation}
 
The forward and inverse formulations are implemented in a differentiable
programming workflow built on the FEniCS finite element
framework~\cite{fenics}, with the variational forms expressed in the Unified Form Language (UFL)~\cite{ufl}. Differentiation is performed with
\texttt{dolfin-adjoint}~\cite{dolfin_adjoint,farrell2013automated}, whose
\texttt{pyadjoint} backend records the forward operations on a computational
tape. The unknown $\theta$ is registered as a control and the misfit as a
reduced functional, so that the state remains implicitly constrained by the
discrete forward problem. In the reverse pass, the UFL residual is
differentiated symbolically to assemble the transpose adjoint system and
accumulate the reduced gradient. Because differentiation acts on the discrete
variational form, the adjoint is consistent with the assembled forward
problem, and the gradients are exact for the discrete problem up to solver
tolerance, without manual derivation of the adjoint equations. This differentiable physics framework for transport equations was developed and validated in our recent study~\cite{Elhadidy_2026}.

\subsection{Benchmark Problem}
\label{sec:benchmark}
\textit{2D Steady Double Gyre Flow:}
A passive scalar is transported on the
unit square $\Omega=[0,1]^2$ by a steady double-gyre flow, following the
boundary layer transport benchmark of~\cite{Arzani2023BLPINN}. The double gyre flow was chosen as it induces flow separation at the bottom wall. The incompressible velocity field is generated from the streamfunction $\psi(x,y)=0.1\,\sin(2\pi x)\sin(\pi y)$ and is given by
\begin{equation}
  \mathbf{v}(x,y) =
  \big[\,0.1\pi\sin(2\pi x)\cos(\pi y),\;
        -0.2\pi\cos(2\pi x)\sin(\pi y)\,\big],
  \label{eq:double_gyre}
\end{equation}
so that $\nabla\cdot\mathbf{v}=0$. Three diffusivities are considered,
$D\in\{10^{-4},10^{-5},10^{-6}\}$, with domain scale $L=1$ and characteristic
speed $U=\max|\mathbf{v}|=0.2\pi$. The corresponding P\'eclet numbers,
$\mathrm{Pe}=UL/D\approx6.3\times10^{3}$, $6.3\times10^{4}$, and
$6.3\times10^{5}$, place the transport in the advection-dominated
regime. A thin diffusive boundary layer of thickness $\mathcal{O}(\delta)$
forms adjacent to the lower wall $y=0$, while the bulk remains advection
dominated. Since the near-wall tangential velocity
$v_x(x,0)=0.1\pi\sin(2\pi x)$ is $\mathcal{O}(U)$, the layer follows the
slip-wall scaling $\delta\sim\mathrm{Pe}^{-1/2}$. Accordingly, $\delta$
ranges from $10^{-2}$ to $10^{-3}$ across the three cases. A
nonhomogeneous Neumann condition, $\partial C/\partial y|_{y=0}=g(x)$, is
imposed on the lower wall, and homogeneous Dirichlet conditions are imposed on
the remaining three boundaries. 


The wall flux is written $g(x)=-A\,K(x)$, where $A>0$ sets the peak influx magnitude and $K(x)$ its spatial shape. To test the solver under forcing of increasing spatial complexity, four peak-normalized shapes are used:
\begin{equation}
  \mathrm{K}(x)=
  \begin{cases}
    \hspace{2mm} 1, & \text{Constant},\\[2pt]
    \hspace{2mm} 4\;x\,(1-x), & \text{Polynomial envelope},\\[2pt]
    \hspace{2mm} 0.5\;\big[1-\cos(2k\pi x)\big], & \text{Fourier envelope},\\[2pt]
    \hspace{2mm} 2\,x(1-x)\,\big[1-\cos(2k\pi x)\big], & \text{Hybrid Poly-Fourier envelope},
  \end{cases}
  \qquad \text{with } k=3.
  \label{eq:flux_shapes}
\end{equation}
Each shape is nonnegative with unit peak, $0\leq K(x)\leq1$. With $A=10$,
\begin{equation}
  g_{\text{const}}=-10,\quad
  g_{\text{poly}}=-40\,x(1-x),\quad
  g_{\text{Fourier}}=-5\big[1-\cos(6\pi x)\big],\quad
  g_{\text{hybrid}}=-20\,x(1-x)\big[1-\cos(6\pi x)\big].
  \label{eq:flux_explicit}
\end{equation}
The four profiles represent uniform, single-lobed, oscillatory, and combined lobed--oscillatory forcing. They benchmark the Schwarz-coupled inverse solver under bottom-wall forcing of increasing spatial complexity.

\textit{Geometric definition:} The single-domain solver discretizes the full unit square $\Omega=[0,1]^2$ in physical coordinates. For the Schwarz-coupled solver, the domain is divided along the wall-normal direction into two overlapping subdomains, $\Omega_{\mathrm{in}}$ and $\Omega_{\mathrm{out}}$, both spanning the full wall length $0\le x\le1$. The inner subdomain extends from the wall, $\eta=0$, to its upper artificial boundary, $\eta_{\mathrm{top}}=7$, and is discretized in the stretched coordinate $\eta$. The outer subdomain extends from its lower artificial boundary, $\eta_{\mathrm{interface}}=2$, to the far-field boundary, $y=1$, and is discretized in the physical coordinate $y$. In physical units, the artificial boundaries lie at $y=2\delta$ and $y=7\delta$, so that the two subdomains overlap over $2\delta\le y\le7\delta$, a thickness of $5\delta$. Because the inner subdomain is defined in the stretched coordinate, its extent in $\eta$ and hence the number of grid divisions across the boundary layer are independent of $D$ and only its physical height, $7\delta$, contracts as $D$ decreases. The outer subdomain, in contrast, resolves the bulk flow, where the solution varies on the scale of the domain. Six mesh resolutions are considered for the grid-refinement study, labeled M0 to M5 and summarized in Table~\ref{tab:mesh_counts}, with the nominal resolution doubling from $m=128$ to $m=4096$ between successive levels. At each level, the single-domain solver employs an $m\times m$ grid with $(m+1)^2$ degrees of freedom (DOFs). The Schwarz-coupled solver employs an $m\times(m/8)$ inner grid, with $m$ divisions along the wall and $m/8$ divisions across $\eta\in[0,\eta_{\mathrm{top}}]$, and an $m\times(15m/16)$ outer grid over $y\in[\delta\eta_{\mathrm{interface}},1]$, giving $6.3$--$7.0\%$ more DOFs than the single-domain solver at the same $m$. The Schwarz iteration employs an under-relaxation factor $\omega=0.5$ and a convergence tolerance $\varepsilon=10^{-10}$, with the overlap field initialized to the far-field concentration, $C_{\mathrm{overlap}}^{(0)}=0$. Accuracy is assessed against the high-fidelity reference (the single-domain solution on M4). All computations were performed on a single workstation with an Intel Xeon W-1370P processor (3.6~GHz, 8 cores, 16 threads) and 32~GB of memory.



\begin{table}[h!]
\centering
\small
\setlength{\tabcolsep}{6pt}
\renewcommand{\arraystretch}{1.5}
\begin{tabular}{|c|c|c|c|c|c|c|}
\hline
\multirow{2}{*}{\textbf{Mesh tag}} & \multirow{2}{*}{\textbf{Mesh resolution}} & \multicolumn{2}{c|}{\textbf{Single-domain solver}} & \multicolumn{3}{c|}{\textbf{Schwarz-coupled solver}} \\
\cline{3-7}
 & & \textbf{Grid} & \textbf{DOFs} & \textbf{Inner grid} & \textbf{Outer grid} & \textbf{Total DOFs} \\
\hline
M0 & Very coarse & $128\times128$   & $16{,}641$       & $128\times16$   & $128\times120$   & $17{,}802$       \\
\hline
M1 & Coarse      & $256\times256$   & $66{,}049$       & $256\times32$   & $256\times240$   & $70{,}418$       \\
\hline
M2 & Medium      & $512\times512$   & $263{,}169$      & $512\times64$   & $512\times480$   & $280{,}098$      \\
\hline
M3 & Fine        & $1024\times1024$ & $1{,}050{,}625$  & $1024\times128$ & $1024\times960$  & $1{,}117{,}250$  \\
\hline
M4 & Very fine   & $2048\times2048$ & $4{,}198{,}401$  & $2048\times256$ & $2048\times1920$ & $4{,}462{,}722$  \\
\hline
M5 & Super fine  & $4096\times4096$ & $16{,}785{,}409$ & $4096\times512$ & $4096\times3840$ & $17{,}838{,}338$ \\
\hline
\end{tabular}
\caption{Grid dimensions, given as $m_x\times m_y$ divisions, and degrees of freedom (DOFs) for each mesh level. The piecewise-linear discretization places one DOF at each of the $(m_x+1)(m_y+1)$ grid vertices. The single-domain solver uses a single $m\times m$ grid over $\bm{\Omega}$. The Schwarz-coupled solver uses an $m\times(m/8)$ inner grid over $\eta\in[0,\eta_{\mathrm{top}}]$ in stretched coordinates and an $m\times(15m/16)$ outer grid over $y\in[\delta\eta_{\mathrm{interface}},1]$ in physical coordinates. The coupled solver therefore contains between $6.3\%$ (M5) and $7.0\%$ (M0) more DOFs than the single-domain solver at the same $m$.}
\label{tab:mesh_counts}
\end{table}

\section{Results}
\label{sec3}

\subsection{Differentiable Schwarz Coupled Solver: Forward and Inverse Modeling}
\label{sec:results-benchmark}
 

We first benchmark the proposed Schwarz-coupled solver as a forward map against a high-fidelity single-domain solver. The single-domain solver resolves the thin boundary layer directly on a globally refined mesh. In addition, a feedback-free \textit{one-way} method, in which interface information is transferred without alternating outer--inner updates, is included as an additional case to isolate the contribution of ``bidirectional'' Schwarz coupling. The forward comparison is performed for $D\in\{10^{-4},10^{-5},10^{-6}\}$ under the constant Neumann bottom wall condition $g(x)=-10$. Next, the Schwarz-coupled solver is employed as the differentiable forward model in adjoint-based inverse recovery experiments. Two inverse problems are considered: identification of the unknown diffusion coefficient $D$ and reconstruction of the unknown Neumann boundary condition $g$ at the bottom wall. 

\subsubsection{Forward Modeling}
\label{sec:results-forward}
\begin{figure}
    \centering
    \includegraphics[width=0.95\textwidth,height=\textheight,keepaspectratio]{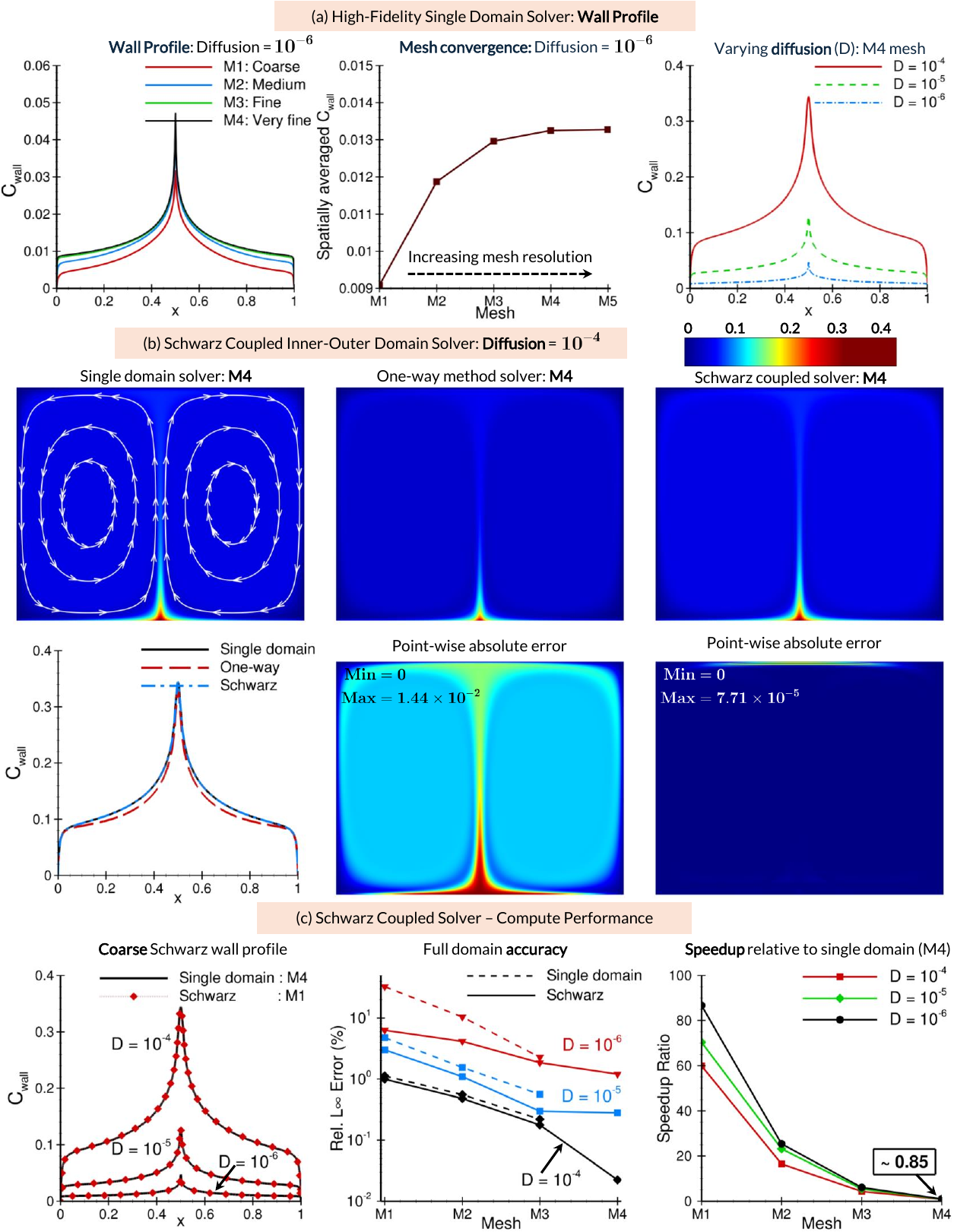}
    \caption{Forward solver verification of the Schwarz-coupled solver against
  the high-fidelity single-domain reference on the double-gyre benchmark.
  (a)~High-fidelity solver: the wall profile $C_{\mathrm{wall}}(x)$ for meshes
  M1--M4 at $D=10^{-6}$ (left); the spatially averaged wall concentration
  $\langle C_{\mathrm{wall}}\rangle$ against mesh refinement (center); and
  $C_{\mathrm{wall}}(x)$ on M4 for $D\in\{10^{-4},10^{-5},10^{-6}\}$ (right).
  (b)~At $D=10^{-4}$ on M4, the concentration fields for the high-fidelity
  (with superimposed streamlines), one-way, and Schwarz solvers, together with
  the corresponding $C_{\mathrm{wall}}(x)$ traces and the pointwise error maps
  of the one-way and Schwarz solutions, sharing the same colormap.
  (c)~Computational performance: the Schwarz $C_{\mathrm{wall}}(x)$ on the coarse M1 mesh
  overlaid on the M4 reference (left); the relative full-domain $L_\infty$ error of
  the high-fidelity (dashed, M1--M3) and Schwarz (solid, M1--M4) solutions with
  respect to the high-fidelity M4 reference (center); and the Schwarz speedup relative to the high-fidelity M4
  solve (right).}
    \label{fig:forward}
\end{figure}

As a prerequisite for the benchmarking, a grid-convergence study is first performed to establish a sufficiently resolved high-fidelity reference solution. This is conducted at the smallest diffusivity considered ($D=10^{-6}$) where the concentration boundary layer is thinnest (numerically, the most challenging case). For this purpose, the single-domain mesh is successively refined from M1 to M5, as listed in Table~\ref{tab:mesh_counts}, and the resulting wall concentration profiles on M1--M4 are shown in the left panel of Fig.~\ref{fig:forward}a. The profiles converge systematically with refinement, and the M3 and M4 solutions closely agree over the entire wall. This convergence is quantified in the center panel through the spatially averaged wall concentration, $\langle C_{\mathrm{wall}}\rangle=\int_0^1 C(x,0)\,dx$, which approaches an asymptotic value as the mesh is refined, the change between M4 and M5 being negligible compared with that between the coarser levels.  The relative wall profile error with respect to M4 decreases from $29.81\%$ on M1 to $10.16\%$ on M2 and $2.17\%$ on M3 (Table~\ref{tab:forward-solver-sweep}). The single-domain solution on M4 is therefore adopted as the high-fidelity reference for all diffusivities, and the corresponding wall profiles are compared in the right panel of Fig.~\ref{fig:forward}a. 

\begin{table}[h!]
\centering
\small
\setlength{\tabcolsep}{10pt}
\renewcommand{\arraystretch}{1.3}
\begin{tabular}{|c|c|c|c|c|c|c|}
\hline
\multicolumn{7}{|c|}{\textbf{Benchmarking Schwarz-Coupled Forward-Solver}} \\
\hline
Method & Mesh & \makecell{Schwarz\\iterations} & \makecell{Rel. $L_\infty$ error\\Full domain} & \makecell{Rel. $L_2$ error\\Full domain} & \makecell{Rel. $L_2$ error\\Wall profile} & Runtime (s) \\
\hline
\multicolumn{7}{|c|}{$\boldsymbol{D=10^{-4}}$} \\
\hline
\multirow{4}{*}{\textbf{Single domain}} & M1 & --- & 1.1297\% & 0.2058\% & 0.7040\% &   1.52 \\ \cline{2-7}
                                        & M2 & --- & 0.5571\% & 0.0589\% & 0.2260\% &   6.36 \\ \cline{2-7}
                                        & M3 & --- & 0.2205\% & 0.0162\% & 0.0628\% &  34.67 \\ \cline{2-7}
                                        & M4 & --- & \textit{Reference} & \textit{Reference} & \textit{Reference} & 203.51 \\
\hline
\textbf{One-way} & M4 & 1 & 4.1991\% & 10.8442\% & 7.3910\% & 195.36 \\
\hline
\multirow{4}{*}{\textbf{Schwarz}} & M1 & 48 & 0.9981\% & 1.2050\% & 0.7691\% &   3.39 \\ \cline{2-7}
                                  & M2 & 47 & 0.4811\% & 0.1592\% & 0.2283\% &  12.32 \\ \cline{2-7}
                                  & M3 & 47 & 0.1775\% & 0.0378\% & 0.0570\% &  47.31 \\ \cline{2-7}
                                  & M4 & 47 & 0.0224\% & 0.0076\% & 0.0080\% & 240.11 \\
\hline
\multicolumn{7}{|c|}{$\boldsymbol{D=10^{-5}}$} \\
\hline
\multirow{4}{*}{\textbf{Single domain}} & M1 & --- & 4.7966\% & 1.5127\% & 4.8779\% &   1.54 \\ \cline{2-7}
                                        & M2 & --- & 1.5512\% & 0.3328\% & 1.1779\% &   7.01 \\ \cline{2-7}
                                        & M3 & --- & 0.5620\% & 0.0632\% & 0.2549\% &  35.45 \\ \cline{2-7}
                                        & M4 & --- & \textit{Reference} & \textit{Reference} & \textit{Reference} & 254.92 \\
\hline
\textbf{One-way} & M4 & 1 & 3.7295\% & 10.9711\% & 7.3690\% & 251.42 \\
\hline
\multirow{4}{*}{\textbf{Schwarz}} & M1 & 45 & 3.0210\% & 4.2335\% & 1.2549\% &   3.62 \\ \cline{2-7}
                                  & M2 & 45 & 1.0953\% & 0.8767\% & 0.4388\% &  11.01 \\ \cline{2-7}
                                  & M3 & 45 & 0.2981\% & 0.1866\% & 0.1316\% &  46.99 \\ \cline{2-7}
                                  & M4 & 45 & 0.2800\% & 0.0368\% & 0.0682\% & 277.71 \\
\hline
\multicolumn{7}{|c|}{$\boldsymbol{D=10^{-6}}$} \\
\hline
\multirow{4}{*}{\textbf{Single domain}} & M1 & --- & 32.7257\% & 13.7340\% & 29.8072\% &   1.48 \\ \cline{2-7}
                                        & M2 & --- & 10.3612\% & 2.3819\% & 10.1571\% &   6.45 \\ \cline{2-7}
                                        & M3 & --- & 2.2785\% & 0.4591\% & 2.1689\% &  31.68 \\ \cline{2-7}
                                        & M4 & --- & \textit{Reference} & \textit{Reference} & \textit{Reference} & 254.83 \\
\hline
\textbf{One-way} & M4 & 1 & 4.3020\% & 10.4922\% & 6.9450\% & 243.64 \\
\hline
\multirow{4}{*}{\textbf{Schwarz}} & M1 & 39 & 6.3448\% & 13.1975\% & 3.6150\% &   2.94 \\ \cline{2-7}
                                  & M2 & 43 & 4.1676\% & 8.9514\% & 1.8091\% &  10.05 \\ \cline{2-7}
                                  & M3 & 43 & 1.8609\% & 2.2243\% & 0.7292\% &  41.57 \\ \cline{2-7}
                                  & M4 & 43 & 1.2081\% & 0.5807\% & 0.6780\% & 265.18 \\
\hline
\end{tabular}%

\caption{Forward solver accuracy and computational cost on the double-gyre
  benchmark, grouped by diffusivity $D$. The single-domain solution on M4 serves
  as the high-fidelity reference. The single-domain solver is evaluated on
  M1--M4, the one-way method on M4, and the alternating Schwarz method on
  M1--M4, using relaxation factor $\omega=0.5$ and convergence tolerance
  $\varepsilon=10^{-10}$. Relative $L_\infty$ errors are normalized by
  $\max|C_{\mathrm{ref}}|$, and relative $L_2$ errors by the $L_2$ norm of the
  reference over the same region (full domain or wall). All errors are
  reported as percentages. For all methods, runtimes are end-to-end wall-clock
  times for one complete forward solve, including operator
  assembly and factorization, and, for the decomposed methods, all subdomain
  solves and interface transfers.}
\label{tab:forward-solver-sweep}
\end{table}

Before comparing the Schwarz and single-domain solvers, we examine whether bidirectional coupling is necessary for resolving the thin boundary layer accurately by comparing the one-way and Schwarz-coupled solutions at $D=10^{-4}$ on the M4 mesh and with identical boundary conditions. The concentration fields and error distributions are presented in Fig.~\ref{fig:forward}b. The double-gyre streamlines, superimposed on the high-fidelity reference, show that the flow lifts the concentration away from the wall near $x=0.5$, carrying it from the boundary layer through the bulk toward the upper wall. The one-way method reproduces the overall structure of the field and qualitatively captures the boundary layer, but it quantitatively deviates from the reference, whereas the Schwarz profile is nearly indistinguishable from it. The one-way error, with a maximum of $1.44\times10^{-2}$, extends well beyond the near-wall region and persists throughout the bulk, showing that an accurate local representation of the boundary layer alone is insufficient to recover the global field due to flow separation. The maximum Schwarz error, $7.71\times10^{-5}$, is more than two orders of magnitude smaller, corresponding to $0.02\%$ of the peak concentration, and is confined to the region near the upper wall where the plume impinges and generates local gradients. Across all three diffusivities, the one-way method yields relative full-domain $L_2$ errors of $10.5$--$11.0\%$ and wall-profile errors of approximately $7\%$, despite being evaluated on the finest mesh. 

With the coupling established, we turn to the central comparison between the Schwarz-coupled and single-domain solvers in resolving the thin boundary layer as both are coarsened from M4 to M1, as summarized in Fig.~\ref{fig:forward}c. The left panel shows that, on the coarsest mesh, M1, the Schwarz wall profiles closely follow the M4 reference for all three diffusivities. In contrast, the single-domain wall profiles at $D=10^{-6}$, shown for the different mesh resolutions in the left panel of Fig.~\ref{fig:forward}a, deviate markedly from the reference on the coarse meshes M1 and M2, indicating that the single-domain solver under-resolves the boundary layer at these resolutions. As the boundary layer thins, the Schwarz solver leveraging perturbation theory continues to resolve the near-wall field accurately on a coarse mesh and offers a clear advantage in accuracy over the single-domain solver at the coarser mesh resolution. This qualitative observation is quantified in the center panel of Fig.~\ref{fig:forward}c and in Table~\ref{tab:forward-solver-sweep}. The center panel compares the relative full-domain $L_\infty$ error of both solvers on each mesh with respect to the single-domain reference solution on M4. Since the solution attains its largest values and steepest gradients in the thin boundary layer, this error is sensitive to how well each solver resolves the layer. For every diffusivity, the Schwarz error decreases monotonically with mesh refinement, while the number of Schwarz iterations varies only between $39$ and $48$, indicating that the convergence of the coupling procedure depends only weakly on the mesh resolution. On every mesh from M1 to M3 and for all diffusivities, the Schwarz $L_\infty$ error is lower than that of the single-domain solver, and the difference widens progressively as $D$ decreases. At $D=10^{-4}$, the boundary layer is thick enough to be captured on the coarse single-domain mesh, and the two solvers attain comparable errors. The benefit of the Schwarz solver becomes evident as the diffusivity is reduced. At $D=10^{-6}$, the single-domain error becomes considerably larger and remains so on all coarser meshes, reaching $32.73\%$, $10.36\%$, and $2.28\%$ on M1, M2, and M3, compared with $6.34\%$, $4.17\%$, and $1.86\%$ for the Schwarz solver, a fivefold reduction on the coarsest mesh. The relative $L_2$ errors of the wall profile in Table~\ref{tab:forward-solver-sweep} confirm this advantage at the wall itself. On M1, the Schwarz wall error remains between $0.77\%$ and $3.62\%$ for all three diffusivities, and at $D=10^{-6}$ the Schwarz wall errors of $3.62\%$, $1.81\%$, and $0.73\%$ on M1, M2, and M3 are three to eight times lower than the corresponding single-domain errors of $29.81\%$, $10.16\%$, and $2.17\%$. The full-domain $L_2$ error, in contrast, is higher for the Schwarz solver in most cases. This metric is dominated by low-level errors over the bulk of the domain, where the concentration itself is close to zero, whereas the non-zero concentration is noticeable only in the thin region adjacent to the wall. The full-domain $L_2$ error is therefore less sensitive to the resolution of the thin boundary layer and is not an appropriate measure of accuracy for thin boundary layer problems, which are the scope of the present work; however, it is reported for completeness.

The computational cost of the Schwarz-coupled solver is presented in the right panel of Fig.~\ref{fig:forward}c as its speedup relative to the single-domain solver on M4. On M4 itself, the Schwarz-coupled solver is slower than the single-domain solver, with a speedup of $0.85$--$0.96$, owing to the coupling iterations between the inner and outer subdomain solves. In contrast, on the coarser meshes, speedups of $60$--$87$ on M1, $16.5$--$25.4$ on M2, and $4.3$--$6.1$ on M3 are achieved. Consequently, the Schwarz-coupled solver resolves the near-wall field accurately at a substantially lower cost than the fine-mesh single-domain solution. Nevertheless, on any given mesh, the single-domain solver is faster than the Schwarz-coupled solver, although it does not resolve the wall profile with the same accuracy. For a fair assessment, the two solvers should be compared at equal wall accuracy, by pairing the Schwarz-coupled solver on M1 with the single-domain solver on the coarsest mesh whose wall error does not exceed that of the Schwarz-coupled solver. At $D=10^{-4}$, this required mesh is M1 itself, since the boundary layer is thick enough for the single-domain solver to resolve it, and the Schwarz-coupled solver is approximately two times slower. As $D$ decreases, the single-domain solver requires finer meshes to match the wall accuracy, namely M2 at $D=10^{-5}$ and M3 at $D=10^{-6}$, and the Schwarz-coupled solver becomes approximately two and eleven times faster, respectively. Thus, the advantage of the Schwarz-coupled solver grows with the P\'eclet number.

\subsubsection{Inverse Modeling}
\label{sec:results-inverse}

\begin{figure}
    \centering
    \includegraphics[width=\textwidth,height=\textheight,keepaspectratio]{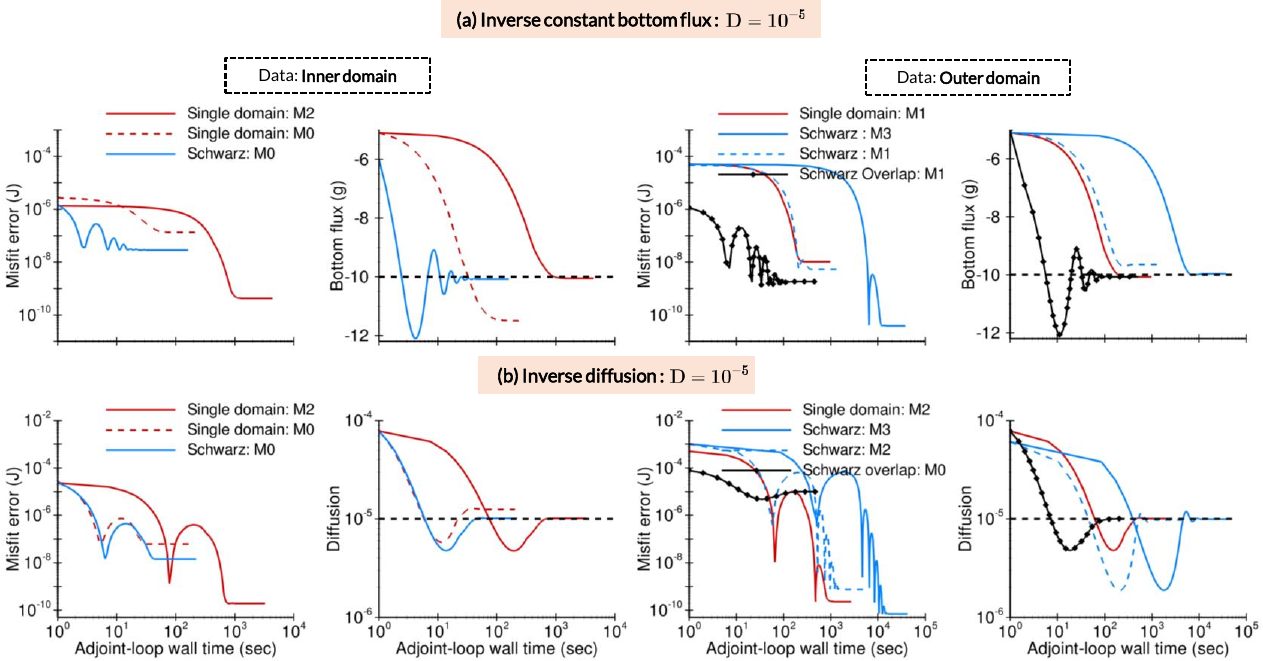}
    \caption{Convergence of the inverse recovery at $D=10^{-5}$ for the single-domain and Schwarz-coupled solvers. (a)~Recovery of the constant wall flux, with true value $g=-10$; (b)~recovery of the diffusivity, with true value $D=10^{-5}$. In each row, the left pair of panels corresponds to observations in the inner domain and the right pair to observations in the outer domain, along with the overlap domain using only the Schwarz-coupled solver. Within each pair, the misfit $J$ (left) and the recovered parameter (right) are plotted against the wall-clock time of the adjoint optimization loop only, excluding the solver setup, with each run offset to start at $t=1$~s on the logarithmic axis. The wall times in Tables~\ref{tab:inv-grad} and~\ref{tab:inv-diff}, in contrast, include both the setup and the adjoint optimization. Solid curves: each method on its coarsest mesh attaining the $1\%$ target. Dashed curves: the competing method on the same mesh, failing to attain the target. Black markers: Schwarz-coupled solver with overlap-zone observations. Dashed horizontal line: true value.}
\label{fig:inverse_convergence}
\end{figure}

Following the verification of the forward solvers, each solver is now employed as the differentiable forward model in an inverse setting. For comparison, the single-domain solver is applied to the same inverse problems, together with the one-way variant, which is included to demonstrate the importance of bidirectional coupling for inverse simulation of a multi-domain problem. Two unknowns of distinct character are considered: diffusivity $D$ and constant wall flux $g$. More complex spatially varying fluxes are considered in the next section.

The observations are taken from three zones, each spanning the full wall length ($0\le x\le1$): the inner zone, $0.1\delta\le y\le1.95\delta$ ($0.1\le\eta\le1.95$), between the wall and the overlap; the outer zone, $2\delta\le y\le1$, extending from the overlap into the bulk flow; and the overlap zone, $2\delta\le y\le7\delta$ ($2\le\eta\le7$), used in the overlap strategy introduced below. Observations on the wall and immediately above it, $0\le y<0.1\delta$, are excluded. Data in this region would allow flux to be estimated directly, for instance by finite differences, without using the governing equation. The noise-free synthetic data are generated by the single-domain solver on M4 with the true parameter values and interpolated onto the nodes of each solver's mesh within the respective zone, and the misfit of Eq.~\eqref{eq:misfit} is evaluated on that mesh. The observation zones are therefore identical for the single-domain, Schwarz-coupled, and one-way solvers, whereas the number of observation nodes depends on the solver and the mesh. In particular, the stretched inner mesh of the Schwarz-coupled solver contains more nodes within the inner zone than the single-domain mesh of the same level. For the single-domain solver, the placement of the data affects only their information content, whereas for the Schwarz-coupled solver it also determines which operations are recorded for the adjoint computation.  In all inverse simulations of this section, the parameter is updated by the Adam optimizer, a first-order gradient-based method, applied here with the exact adjoint gradient and a step-decay learning-rate schedule, with an initial learning rate of $10^{-1}$ decayed by a factor of $0.75$ every $50$ epochs. The epoch budget is increased as $D$ decreases, since the thinning boundary layer makes the recovery more demanding, and is set to $250$, $500$, and $750$ epochs for $D=10^{-4}$, $10^{-5}$, and $10^{-6}$, respectively. Each forward evaluation of the Schwarz-coupled solver uses a relaxation factor $\omega=0.5$, a tolerance $\varepsilon=10^{-10}$, and at most $200$ Schwarz iterations. All optimization settings are likewise identical for the three solvers.



The accuracy of the recovered parameter is affected by the discretization error of the forward model, which is determined by the mesh resolution. For each method and data placement, the mesh is therefore refined progressively, starting from the coarsest level until the recovered parameter meets the target relative error, and the coarsest mesh that meets this target is reported. Consistent with the forward solver comparison of Section~\ref{sec:results-forward}, the methods are thus compared at equal accuracy rather than on the same mesh, and the procedure identifies the coarsest resolution, and hence the lowest computational cost, at which each method achieves the required accuracy. Furthermore, since the recovered parameter evolves over the epochs and may overshoot or oscillate before settling, meeting the target must be defined precisely, both to decide whether a recovery is successful and to measure its cost. The definition adopted here reflects how an inversion is assessed in practice. With real measurements, the true parameter is unknown, and the only available indication of convergence beyond the data loss is that the estimate ceases to change over the epochs. After an initial transient, the optimizing parameter settles onto a plateau and remains there over many epochs, and this stabilized value is adopted as the final estimate. Since the observations here are synthetic, the accuracy of this plateau can be assessed against the true value. Formally, let $\theta^{n}$ denote the recovered parameter at epoch $n$, $e^{n}=|\theta^{n}-\theta|/|\theta|$ its relative error with respect to the true value $\theta$, and $N$ the final epoch. A recovery is successful if $e^{N}\le1\%$. For a successful recovery, the epochs to target, $n_{\mathrm{t}}$, are determined by scanning the history backward from $N$ until the error first exceeds $1\%$, and $n_{\mathrm{t}}$ is taken as the epoch immediately following. By construction, the condition $e^{n}\le1\%$ holds at $n_{\mathrm{t}}$ and at every subsequent epoch up to $N$, thus $n_{\mathrm{t}}$ marks the onset of the final uninterrupted plateau within the target band. A forward criterion, which records the first epoch at which $e^{n}\le1\%$, would not be reliable, since a trajectory may pass through the true value during the transient and subsequently settle elsewhere (as seen in our results), which can happen due to the inaccurate estimation of data by the solver.  For each method and data placement, these values are reported on the coarsest successful mesh.


\begin{table}[h!]
\centering
\footnotesize
\setlength{\tabcolsep}{10pt}
\renewcommand{\arraystretch}{1.3}
\begin{tabular}{|c|c|c|c|c|c|}
\hline
\multicolumn{6}{|c|}{\textbf{Inverse Wall Flux Recovery}} \\
\hline
Method & Data zone & Coarsest mesh & Epochs to target & Wall time (s) & Converged value \\
\hline
\multicolumn{6}{|c|}{$\boldsymbol{D=10^{-4}}$} \\
\hline
\multirow{3}{*}{\textbf{Single domain}} & Inner   & M1 (Coarse)      & $91$  & $236.72$          & $-10.0174$ \\ \cline{2-6}
                                        & Overlap & M1 (Coarse)      & $86$  & $204.79$          & $-10.0675$ \\ \cline{2-6}
                                        & Outer   & M1 (Coarse)      & $87$  & $\mathbf{206.81}$ & $-10.0493$ \\
\hline
\multirow{3}{*}{\textbf{Schwarz}}       & Inner   & M1 (Coarse)      & $65$  & $\mathbf{113.37}$ & $-9.9907$  \\ \cline{2-6}
                                        & Overlap & M1 (Coarse)      & $91$  & $\mathbf{158.37}$ & $-10.0784$ \\ \cline{2-6}
                                        & Outer   & M1 (Coarse)      & $82$  & $285.56$          & $-9.9222$  \\
\hline
\textbf{One-way} & Inner & \emph{Not achieved} & --- & --- & $-11.1426$ \\
\hline
\multicolumn{6}{|c|}{$\boldsymbol{D=10^{-5}}$} \\
\hline
\multirow{3}{*}{\textbf{Single domain}} & Inner   & M2 (Medium)      & $91$  & $788.93$          & $-10.0524$ \\ \cline{2-6}
                                        & Overlap & M1 (Coarse)      & $89$  & $174.17$          & $-10.0938$ \\ \cline{2-6}
                                        & Outer   & M1 (Coarse)      & $86$  & $\mathbf{167.95}$ & $-10.0700$ \\
\hline
\multirow{3}{*}{\textbf{Schwarz}}       & Inner   & M0 (Very coarse) & $93$  & $\mathbf{29.70}$  & $-10.0771$ \\ \cline{2-6}
                                        & Overlap & M1 (Coarse)      & $93$  & $\mathbf{94.44}$  & $-10.0690$ \\ \cline{2-6}
                                        & Outer   & M3 (Fine)        & $79$  & $5914.58$         & $-9.9685$  \\
\hline
\textbf{One-way} & Inner & \emph{Not achieved} & --- & --- & $-11.1304$ \\
\hline
\multicolumn{6}{|c|}{$\boldsymbol{D=10^{-6}}$} \\
\hline
\multirow{3}{*}{\textbf{Single domain}} & Inner   & M3 (Fine)        & $615$ & $29066.16$         & $-9.9081$  \\ \cline{2-6}
                                        & Overlap & M2 (Medium)      & $409$ & $3445.69$          & $-9.9732$  \\ \cline{2-6}
                                        & Outer   & M3 (Fine)        & $93$  & $\mathbf{4396.51}$ & $-10.0000$ \\
\hline
\multirow{3}{*}{\textbf{Schwarz}}       & Inner   & M1 (Coarse)      & $68$  & $\mathbf{56.69}$  & $-10.0139$ \\ \cline{2-6}
                                        & Overlap & M2 (Medium)      & $58$  & $\mathbf{156.83}$ & $-10.0643$ \\ \cline{2-6}
                                        & Outer   & M3 (Fine)        & $91$  & $16318.39$        & $-9.9661$  \\
\hline
\textbf{One-way} & Inner & \emph{Not achieved} & --- & --- & $-11.7466$ \\
\hline
\end{tabular}
\caption{Efficiency and accuracy of the inverse wall flux recovery under constant wall flux, grouped by the diffusivity $D$. The data zone indicates the region containing the observations. For each method and data zone, the mesh is refined from M0 until the target is met,  and the coarsest successful mesh is reported. On that mesh, the epoch count is the first epoch of the final uninterrupted interval in which the error remains below $1\%$, and the wall time is the cumulative end-to-end wall-clock time up to that
epoch, including, for all methods, operator assembly and factorization, and every forward and adjoint solve, together with, for the
decomposed methods, the Schwarz iterations, and all interface transfers. Bold marks the shorter wall time between the single domain and Schwarz methods for the same data zone. The converged value is the final recovered wall flux. A method is marked \emph{Not achieved} when no tested mesh meets the target, in which case the value from the mesh with the smallest final error is reported.}
\label{tab:inv-grad}
\end{table}

The convergence behavior of the inverse parameter recovery is illustrated at the representative intermediate diffusivity $D=10^{-5}$, for both the constant wall flux and the diffusivity, as shown in Fig.~\ref{fig:inverse_convergence}a and Fig.~\ref{fig:inverse_convergence}b, respectively. In each case, the left pair of panels corresponds to inner data and the right pair to outer data, and the misfit and the recovered parameter are plotted against the wall-clock time of the adjoint optimization loop alone. The solver setup, comprising operator assembly and factorization, depends on the mesh and the solver and is therefore excluded, so that the curves compare the optimization cost of all recoveries from a common time origin value. The wall times in Tables~\ref{tab:inv-diff} and~\ref{tab:inv-grad}, in contrast, include both the solver setup and the adjoint optimization loop. The recovery of the constant wall flux is examined first, with the initial estimate set to $g=-5$ in all experiments. With inner data, the Schwarz-coupled solver meets the target on the coarsest mesh, M0. Its recovered wall flux overshoots to approximately $-12$ during an initial transient before settling onto the true value, while the misfit decreases, with corresponding oscillations. The single-domain solver, in contrast, fails on the same coarse mesh, M0 (dashed curve). Its recovered wall flux crosses the true value and settles at approximately $-11.5$. The single-domain solver meets the target only on the considerably finer mesh, M2, on which the recovered wall flux decreases monotonically and settles onto the true value after a longer runtime. Both successful recoveries require nearly the same number of epochs, $93$ for the Schwarz-coupled solver on M0 and $91$ for the single-domain solver on M2 (Table~\ref{tab:inv-grad}), so that the difference in wall time arises from the cost of each epoch, which is determined by the mesh. Consistent with the forward results, the ability of the Schwarz-coupled solver to resolve the boundary layer on a coarse mesh thus translates directly into a lower cost of the inversion.

For the outer data, the Schwarz-coupled solver requires a finer mesh than the single-domain solver at $D=10^{-5}$. The single-domain solver meets the target on M1. On the same mesh, the Schwarz-coupled solver fails (dashed curve), with its recovered wall flux settling at approximately $-9.6$, and it meets the target only on M3. The need for this finer mesh is attributed to the interface transfer operator $\mathbf{P}$. With outer data, the misfit is evaluated on the outer subdomain solution, which receives the influence of the wall flux only through the interface values transferred from the inner solution by $\mathbf{P}$. Since $\mathbf{P}$ interpolates the inner solution \textit{linearly} onto the interface nodes of the outer mesh, spatial variations of the inner solution between these nodes may be lost or misrepresented on a coarse outer mesh, so that $\mathbf{P}$ effectively acts as a filter on the information passed from the inner to the outer subdomain.  In addition to requiring a finer mesh, the Schwarz-coupled solver incurs a higher cost per epoch with outer data. Since the wall flux enters only the inner subdomain while the misfit is evaluated on the outer subdomain solution, this placement realizes the distinct subdomain regime of Section~\ref{sec:inverse-coupled}, in which each forward evaluation recorded for the adjoint computation includes the inner solve, the interface transfer, and the outer solve, all of which the adjoint computation traverses in reverse. Both successful recoveries require a similar number of epochs, $79$ for the Schwarz-coupled solver on M3 and $86$ for the single-domain solver on M1 (Table~\ref{tab:inv-grad}), and the higher wall time of the Schwarz-coupled solver arises from  the finer mesh and the larger recorded evaluation per epoch. In principle, this computational cost could be reduced by employing a \textit{differentiable} higher-order interface transfer operator that preserves variations on a coarser mesh, although the development of such an operator lies beyond the scope of the present work.  Instead, a different strategy, in which outer observations in the overlap region are considered but assigned to the inner subdomain, is proposed and pursued below.

Within the present Schwarz-coupled setting, the recovery from outer data can nevertheless be performed efficiently by exploiting the overlap region shared by the two subdomains. Specifically, the overlap region occupies the lower part of the outer zone, and in this strategy, only the outer data located within the overlap region are considered for the inverse simulation. Interestingly, since the inner subdomain also extends over the overlap region, the outer observations within the overlap region can also be assigned to the inner subdomain rather than to the outer subdomain, and the precise domain of the overlap region is also flexible. With the misfit evaluated on the inner subdomain solution, the influence of the wall flux reaches the data without passing through the interface transfer operator $\mathbf{P}$. Accordingly, with the overlap strategy, shown by the black markers in the right pair of panels of Fig.~\ref{fig:inverse_convergence}a, the Schwarz-coupled solver meets the target on M1 after approximately $10^{2}$~s despite fewer data observations, instead of on M3 after approximately $5\times10^{3}$~s with the full outer data observation. The Schwarz-coupled solver with overlap data is thus competitive with the single-domain solver with the full outer data, which meets the target after approximately $10^{2}$~s. For a fair comparison and completeness, the single-domain solver, which involves no interface transfer, is also applied with the same overlap data. The single-domain solver meets the target on the same mesh, M1, with both the overlap data and the full outer data, and at a comparable wall time (Table~\ref{tab:inv-grad}), which confirms that the improvement of the Schwarz-coupled solver arises from avoiding the interface transfer rather than from the restriction of the observations. Including the solver setup, the Schwarz-coupled solver with overlap data requires $94$~s, against $5915$~s with the full outer data and $174$~s for the single-domain solver with the same data (Table~\ref{tab:inv-grad}). 

The complete set of wall flux recoveries, covering all three diffusivities and both data placements together with the overlap strategy, is summarized in Table~\ref{tab:inv-grad}, and the trends observed at $D=10^{-5}$ persist throughout. The one-way variant, in contrast, fails at every diffusivity, converging to wall fluxes biased by $11$--$18\%$ and to diffusivities biased by $21$--$32\%$ on every tested mesh (Tables~\ref{tab:inv-grad} and~\ref{tab:inv-diff}), which confirms that bidirectional coupling is required for an unbiased recovery.

\begin{table}[h!]
\centering
\footnotesize
\setlength{\tabcolsep}{10pt}
\renewcommand{\arraystretch}{1.3}
\begin{tabular}{|c|c|c|c|c|c|}
\hline
\multicolumn{6}{|c|}{\textbf{Inverse Diffusivity Recovery}} \\
\hline
Method & Data zone & Coarsest mesh & Epochs to target & Wall time (s) & Converged value \\
\hline
\multicolumn{6}{|c|}{$\boldsymbol{D=10^{-4}}$} \\
\hline
\multirow{3}{*}{\textbf{Single domain}} & Inner   & M1 (Coarse)      & $100$ & $\mathbf{274.48}$  & $1.0035\times 10^{-4}$ \\ \cline{2-6}
                                        & Overlap & M1 (Coarse)      & $136$ & $299.61$           & $1.0006\times 10^{-4}$ \\ \cline{2-6}
                                        & Outer   & M1 (Coarse)      & $131$ & $\mathbf{334.28}$  & $1.0025\times 10^{-4}$ \\
\hline
\multirow{3}{*}{\textbf{Schwarz}}       & Inner   & M1 (Coarse)      & $106$ & $337.28$           & $9.9498\times 10^{-5}$ \\ \cline{2-6}
                                        & Overlap & M0 (Very coarse) & $165$ & $\mathbf{61.99}$   & $1.0034\times 10^{-4}$ \\ \cline{2-6}
                                        & Outer   & M2 (Medium)      & $144$ & $2535.02$          & $9.9294\times 10^{-5}$ \\
\hline
\textbf{One-way} & Inner & \emph{Not achieved} & --- & --- & $1.3157\times 10^{-4}$ \\
\hline
\multicolumn{6}{|c|}{$\boldsymbol{D=10^{-5}}$} \\
\hline
\multirow{3}{*}{\textbf{Single domain}} & Inner   & M2 (Medium)      & $108$ & $1031.47$           & $9.9203\times 10^{-6}$ \\ \cline{2-6}
                                        & Overlap & M2 (Medium)      & $156$ & $1618.20$          & $9.9986\times 10^{-6}$ \\ \cline{2-6}
                                        & Outer   & M2 (Medium)      & $131$ & $\mathbf{1120.86}$ & $1.0027\times 10^{-5}$ \\
\hline
\multirow{3}{*}{\textbf{Schwarz}}       & Inner   & M0 (Very coarse) & $172$ & $\mathbf{76.03}$   & $1.0084\times 10^{-5}$ \\ \cline{2-6}
                                        & Overlap & M0 (Very coarse) & $134$ & $\mathbf{88.89}$   & $1.0052\times 10^{-5}$ \\ \cline{2-6}
                                        & Outer   & M3 (Fine)        & $139$ & $11696.10$         & $9.9377\times 10^{-6}$ \\
\hline
\textbf{One-way} & Inner & \emph{Not achieved} & --- & --- & $1.2103\times 10^{-5}$ \\
\hline
\multicolumn{6}{|c|}{$\boldsymbol{D=10^{-6}}$} \\
\hline
\multirow{3}{*}{\textbf{Single domain}} & Inner   & M3 (Fine)        & $127$ & $6767.13$          & $9.9932\times 10^{-7}$ \\ \cline{2-6}
                                        & Overlap & M2 (Medium)      & $199$ & $2083.82$          & $9.9075\times 10^{-7}$ \\ \cline{2-6}
                                        & Outer   & M3 (Fine)        & $125$ & $\mathbf{6071.57}$ & $1.0003\times 10^{-6}$ \\
\hline
\multirow{3}{*}{\textbf{Schwarz}}       & Inner   & M1 (Coarse)      & $108$ & $\mathbf{127.12}$  & $9.9800\times 10^{-7}$ \\ \cline{2-6}
                                        & Overlap & M1 (Coarse)      & $149$ & $\mathbf{339.60}$  & $1.0003\times 10^{-6}$ \\ \cline{2-6}
                                        & Outer   & M4 (Very fine)   & $107$ & $16947.50$         & $9.9809\times 10^{-7}$ \\
\hline
\textbf{One-way} & Inner & \emph{Not achieved} & --- & --- & $1.2272\times 10^{-6}$ \\
\hline
\end{tabular}
\caption{Efficiency and accuracy of the inverse diffusivity recovery, grouped
  by the true diffusivity $D$. The data zone indicates the region containing
  the observations. The target is a relative error of the recovered
  diffusivity below $1\%$. For each method and data zone, the mesh is refined
  from M0 until the target is met, and the coarsest successful mesh is
  reported. On that mesh, the epoch count is the first epoch of the final
  uninterrupted interval in which the error remains below $1\%$, obtained by
  scanning the history backward from the final epoch, and the wall time is the
  cumulative end-to-end wall clock time up to that epoch, including, for all
  methods, mesh generation, operator assembly and factorization, and every
  forward and adjoint solve, together with, for the decomposed methods, the
  Schwarz iterations and all interface transfers. Bold marks the shorter wall
  time between the single-domain and Schwarz methods for the same data zone.
  The converged value is the final recovered diffusivity. A method is marked
  \emph{Not achieved} when no tested mesh meets the target, in which case the
  value from the mesh with the smallest final error is reported.}
\label{tab:inv-diff}
\end{table}

In the second inverse experiment, the diffusivity is recovered. The boundary layer scale is held fixed at $\delta=\sqrt{D}$ of the true diffusivity, so that the domain decomposition, the inner mesh, and the observation zones remain unchanged over the epochs, and only $D$ in Eq.~\eqref{eq:inner_pde} is updated. The convergence of the recovered diffusivity at $D=10^{-5}$ is shown in Fig.~\ref{fig:inverse_convergence}b, where, as in all diffusivity experiments, the initial estimate is set one order of magnitude above the true value, here $D_{0}=10^{-4}$. With inner data, the Schwarz-coupled solver meets the target on the coarsest mesh, M0. Its recovered diffusivity first undershoots to approximately $5\times10^{-6}$ and then settles onto the true value within approximately $10^{2}$~s. On the same mesh, however, the single-domain solver under-resolves the boundary layer and converges to approximately $1.2\times10^{-5}$, and it meets the target only once the mesh is refined to M2, after approximately $10^{3}$~s. As expected from the wall flux recovery, the outcome is reversed with outer data, for which the Schwarz-coupled solver requires a finer mesh than the single-domain solver. The single-domain solver meets the target on M2 within approximately $5\times10^{2}$~s, whereas the Schwarz-coupled solver fails on M2 (dashed curve) and succeeds only on M3, after approximately $10^{4}$~s. The overlap strategy restores the efficiency of the Schwarz-coupled solver for the diffusivity as well. When the outer observations within the overlap region are assigned to the inner subdomain, as shown by the black markers, the Schwarz-coupled solver meets the target on M0 within approximately $10^{2}$~s despite the reduced number of observations. The single-domain solver, in contrast, requires M2 with both the overlap data and the full outer data (Table~\ref{tab:inv-diff}). Table~\ref{tab:inv-diff} extends the observations made at $D=10^{-5}$ to all three diffusivities. 

\subsection{Language Model Integrated Differentiable Physics Solver for Inverse Modeling}
\label{sec:llm}
The benchmark inverse problems above compared the forward models with the optimizer held fixed, whereas the efficiency of an inverse recovery can depend  on the optimizer. Classical optimizers such as Adam, in particular, follow a fixed update rule, whose hyperparameters, such as the learning rate and its decay schedule, are problem-dependent, selected empirically, and fixed before the optimization begins. Recent advances in large language models (LLMs) and agentic AI have opened a different route, in which the language model acts as the optimizer~\cite{yang2024large, pang2026deliberate,islam2026discovery}. At each epoch, the language model selects the optimization method \textit{adaptively} as the recovery progresses. In this section, we investigate whether this route can accelerate the inverse recovery using two strategies: \textit{passive LLM optimization} and \textit{active agentic optimization}. The recovery of the wall flux is considered here with increasing spatial complexity.


\subsubsection{Constant flux: a scalar reconstruction}
The constant wall flux is recovered with the true value $g=-10$ and the initial estimate $g_{0}=-5$, as in Section~\ref{sec:results-inverse}. For all methods, the Schwarz-coupled solver is used on M1 at $D=10^{-4}$ and $10^{-5}$ and on M2 at $D=10^{-6}$, so that the inner and overlap data are treated on a common mesh for each diffusivity. Every method is given the same budget of $150$ epochs, which is long enough to show whether the recovered wall flux settles after its initial transient. Success is judged by the criterion of Section~\ref{sec:results-inverse}, which requires the recovered wall flux to remain within the $1\%$ band from the onset of its final plateau until the last epoch.

Since the unknown is a single scalar, the recovery constitutes a one-dimensional optimization problem, and three groups of optimizers are compared. The first group consists of classical methods that use increasing amounts of derivative information, namely Adam and the secant method, which use the adjoint gradient, and the Newton method, which additionally uses the adjoint Hessian. The second and third groups are the passive LLM optimization, based on the language model Qwen3-8B~\cite{qwen} served locally through the Ollama API, and the active agentic AI optimization, based on the agent GPT~5.6 Terra built on Codex. The first LLM input is a prompt that instructs the language model to act as a minimization engine. The second input is the optimization policy, which consists only of gradient-based methods, namely the normalized gradient (G), Adam (A), momentum (M), damped Newton (N), and secant (S) methods, linear regression of the gradient (Rg), a quadratic fit of the objective (Rj), bracketed root finding of the gradient (Bg), and bracketed refinement of the objective (Bj). The third input is the epoch history, which comprises the parameter, the misfit, and the adjoint gradient of the ten most recent epochs for the passive LLM optimization and of all preceding epochs, together with the previously selected methods, for the agent. Based on the prompt and the epoch history, the language model selects one method of the optimization policy, specifies any damping and step controls for the selected method, and logs the reasoning behind the selection. The controller then executes the selected method with the specified controls and returns the updated parameter to the solver for the next epoch. Apart from the underlying model and the length of the epoch history, the active optimization differs from the passive optimization in that the agent is additionally given the inverse code and thereby has access to the problem being solved, which may enable the agent to select a more suitable method at each epoch. Finally, Brent's method, a derivative-free method that combines golden-section search with parabolic interpolation, is included as a competitive baseline, since it is particularly well suited for the present \textit{one-dimensional} inverse problem.

The classical methods are deterministic, so that a single run suffices for each case. The optimizers based on language models, in contrast, are probabilistic, since the same history of states can yield different updates across runs, and their robustness must therefore be assessed over repeated runs. The passive LLM optimization could, in principle, be made deterministic, either by setting the sampling temperature of the Ollama API to zero or by fixing its random seed, which controls the randomness of the generated response. To expose its variability, it is instead run $30$ times independently with different seeds for each diffusivity and data placement, at a moderate sampling temperature of $0.5$, which shows a degree of variability across the runs while keeping the responses focused. The agentic AI approach, by comparison, offers no seed, and its runs differ through its own nondeterminism. Because the proprietary agent is considerably more expensive to run than the open-source model, it is limited to five independent runs for each case, each started in a fresh project. For both optimizers, the results are reported as the median and the interquartile range (IQR) over the runs, together with the number of runs that meet the target.

\begin{figure}
    \centering
    \includegraphics[width=0.8\textwidth,height=\textheight,keepaspectratio]{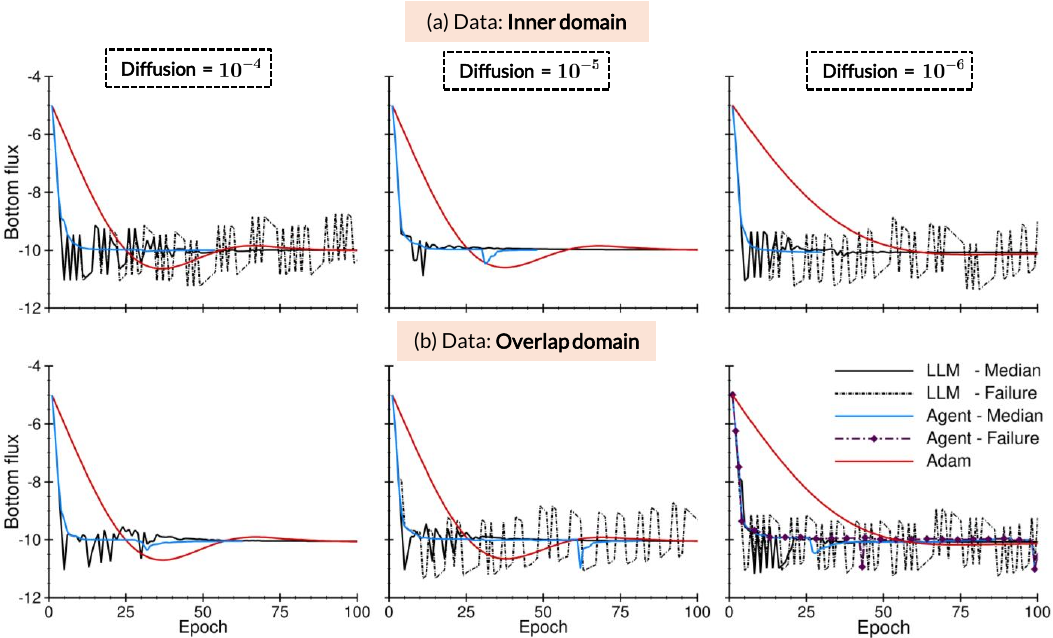}
    \caption{Convergence of the constant wall flux recovery with the Schwarz-coupled solver for (a)~inner data and (b)~outer data within the overlap region under the overlap strategy, at $D=10^{-4}$, $10^{-5}$, and $10^{-6}$ (left to right), on the meshes listed in Table~\ref{tab:qwen-classical-codex-work-comparison}. The recovered wall flux is plotted against the epoch, starting from the initial estimate $g_{0}=-5$, with the true value $g=-10$; only the first $100$ of the $150$ epochs are shown. Solid black and blue curves show the median runs of the passive LLM optimization and of the active agentic AI optimization, respectively, and the red curve shows the Adam optimizer. Dashed black curves show a failed run of the passive LLM optimization, and the dash-dotted curve with markers shows the single failed run of the active agentic AI optimization, which occurs only for overlap data at $D=10^{-6}$. At $D=10^{-6}$, Adam enters its final interval within the $1\%$ band only at epochs $122$ and $130$ for inner and overlap data, respectively, beyond the range shown.}
  \label{fig:inverse_recovery}
\end{figure}

In Table~\ref{tab:qwen-classical-codex-work-comparison}, we report the results of all optimizers. Overall, both optimizations based on language models prove robust and accurate. Newton, secant, and Brent converge to the solution with the fewest epochs. However, this advantage, as discussed next, does not hold when the unknown flux is spatially varying and an Adam optimization is necessary. With inner data, the passive LLM optimization meets the success criterion in $88$ of its $90$ trials and the agent in all of its $15$ trials, and their median converged values deviate from those of the Newton and secant methods by at most $0.001\%$ and $0.11\%$, respectively. In terms of wall time, Brent's method is the fastest in all cases with the exception of overlap data at $D=10^{-6}$, where the damped Newton method is faster. The LLM-based optimizers are generally slower than the classical methods for this one-dimensional problem, although they remain faster than the tuned Adam. The only exception is the agent with inner data at $D=10^{-6}$, which, at $50.8$~s, outperforms both the damped Newton and the secant methods. With overlap data, the passive LLM optimization remains equally robust and accurate, whereas the agent is less consistent. The passive LLM optimization meets the success criterion in $87$ of its $90$ trials and again deviates from those of the Newton and secant methods by at most $0.001\%$. The agent meets the success criterion in $14$ of its $15$ trials, but its median converged value deviates by up to $0.6\%$ for $D=10^{-5}$, and for $D=10^{-4}$ and $10^{-5}$ its trials show a considerably larger spread than with inner data. For both inner and overlap data, the difference between the two optimizations based on language models is reflected in the selection mix, that is, the percentage of queries in which each method of the optimization policy is selected. In the passive LLM optimization, the language model selects predominantly the Newton (N) and normalized gradient (G) methods, whose alternation is consistent with the oscillations observed in Fig.~\ref{fig:inverse_recovery}. The agent, in contrast, relies mainly on the secant method (S) and on the regression, quadratic-fit, and bracketing methods (Rg, Rj, Bg, Bj), which draw on several entries of the epoch history and are consistent with its smaller oscillations.


\begin{table*}
\centering
\footnotesize
\setlength{\tabcolsep}{4pt}
\renewcommand{\arraystretch}{1.3}

\begin{tabular}{|c|c|c|c|c|c|c|c|}
\hline
\multicolumn{8}{|c|}{\textbf{Inverse Constant Wall Flux Recovery: Language-Model and Classical Optimizers}} \\
\hline
Data & Metric & Passive LLM & Active agentic & Adam & Newton & Secant & Brent \\
\hline
\multicolumn{8}{|c|}{$\boldsymbol{D=10^{-4}}$ (M1 Mesh)} \\
\hline
\multirow{5}{*}{\textbf{Inner}}
  & Successful trials   & 29/30             & 5/5               & 1/1      & 1/1      & 1/1      & 1/1      \\ \cline{2-8}
  & Converged value     & -9.9908 [0.0006]  & -9.9798 [0.0036]  & -9.9892  & -9.9907  & -9.9907  & -9.9758  \\ \cline{2-8}
  & Epochs to target    & 49 [65]           & 11 [4]            & 74       & 4        & 7        & 3        \\ \cline{2-8}
  & Wall time (s)       & 177.7 [229.8]     & 23.4 [5.75]      & 202.1    & 17.1     & 22.7     & 14.0     \\ \cline{2-8}
  & Selection mix (\%)  & {\scriptsize N65.4/G34.0/S0.4/Bj0.1} & \makecell{{\scriptsize G0.6/N6.1/S29.9/Rg16.5}\\{\scriptsize Rj7.9/Bg11.0/Bj28.0}} & -- & -- & -- & -- \\
\hline
\multirow{5}{*}{\textbf{Overlap}}
  & Successful trials     & 30/30             & 5/5               & 1/1      & 1/1      & 1/1      & 1/1      \\ \cline{2-8}
  & Converged value     & -10.0560 [0.0004] & -10.0562 [0.0427] & -10.0544 & -10.0561 & -10.0561 & -10.0069 \\ \cline{2-8}
  & Epochs to target    & 41 [25]           & 39 [17]           & 55       & 4        & 7        & 3        \\ \cline{2-8}
  & Wall time (s)       & 153.2 [95.3]      & 61.2 [31.40]     & 154.1    & 14.8     & 23.2     & 13.1     \\ \cline{2-8}
  & Selection mix (\%)  & {\scriptsize N77.7/G21.9/S0.3/Bj0.1} & \makecell{{\scriptsize G0.5/A0.5/N2.9/S30.1}\\{\scriptsize Rg23.9/Rj3.3/Bg15.3/Bj23.4}} & -- & -- & -- & -- \\
\hline
\multicolumn{8}{|c|}{$\boldsymbol{D=10^{-5}}$ (M1 Mesh)} \\
\hline
\multirow{5}{*}{\textbf{Inner}}
  & Successful trials     & 30/30             & 5/5               & 1/1      & 1/1      & 1/1      & 1/1      \\ \cline{2-8}
  & Converged value     & -9.9832 [0.0002]  & -9.9723 [0.0006]  & -9.9814  & -9.9832  & -9.9832  & -9.9692  \\ \cline{2-8}
  & Epochs to target    & 33 [24]           & 35 [4]            & 77       & 3        & 7        & 3        \\ \cline{2-8}
  & Wall time (s)       & 112.9 [81.3]      & 46.8 [4.53]      & 189.8    & 12.8     & 21.2     & 12.8     \\ \cline{2-8}
  & Selection mix (\%)  & {\scriptsize N90.0/G9.7/S0.3/Bj0.1} & \makecell{{\scriptsize N1.3/S42.0/Rg20.4}\\{\scriptsize Rj5.7/Bg7.0/Bj23.6}} & -- & -- & -- & -- \\
\hline
\multirow{5}{*}{\textbf{Overlap}}
  & Successful trials     & 28/30             & 5/5               & 1/1      & 1/1      & 1/1      & 1/1      \\ \cline{2-8}
  & Converged value     & -10.0507 [0.0002] & -9.9897 [0.0661]  & -10.0489 & -10.0508 & -10.0508 & -9.9752  \\ \cline{2-8}
  & Epochs to target    & 31 [24]           & 68 [18]           & 56       & 4        & 7        & 3        \\ \cline{2-8}
  & Wall time (s)       & 109.7 [79.3]      & 79.4 [29.91]     & 143.8    & 16.2     & 20.8     & 12.8     \\ \cline{2-8}
  & Selection mix (\%)  & {\scriptsize N78.6/G20.8/S0.4/Bj0.2} & \makecell{{\scriptsize N0.9/S29.7/Rg13.8}\\{\scriptsize Rj22.0/Bg12.5/Bj21.1}} & -- & -- & -- & -- \\
\hline
\multicolumn{8}{|c|}{$\boldsymbol{D=10^{-6}}$ (M2 Mesh)} \\
\hline
\multirow{5}{*}{\textbf{Inner}}
  & Successful trials     & 29/30             & 5/5               & 1/1      & 1/1      & 1/1      & 1/1      \\ \cline{2-8}
  & Converged value     & -10.0725 [0.0005] & -10.0640 [0.0018] & -10.0823 & -10.0725 & -10.0725 & -10.0606 \\ \cline{2-8}
  & Epochs to target    & 55 [49]           & 9 [1]             & 122      & 4        & 6        & 3        \\ \cline{2-8}
  & Wall time (s)       & 475.9 [434.7]     & 50.8 [6.29]      & 739.8    & 52.8     & 65.6     & 49.7     \\ \cline{2-8}
  & Selection mix (\%)  & {\scriptsize N77.5/G22.4/Bj0.1} & \makecell{{\scriptsize N2.1/S37.1/Rg29.3}\\{\scriptsize Bg7.1/Bj24.3}} & -- & -- & -- & -- \\
\hline
\multirow{5}{*}{\textbf{Overlap}}
  & Successful trials     & 29/30             & 4/5               & 1/1      & 1/1      & 1/1      & 1/1      \\ \cline{2-8}
  & Converged value     & -10.0806 [0.0003] & -10.0811 [0.0006] & -10.0889 & -10.0807 & -10.0807 & -9.9162  \\ \cline{2-8}
  & Epochs to target    & 45 [46]           & 43 [8]            & 130      & 4        & 6        & 7        \\ \cline{2-8}
  & Wall time (s)       & 338.3 [330.2]     & 180.2 [43.26]    & 769.3    & 54.1     & 65.4     & 78.7     \\ \cline{2-8}
  & Selection mix (\%)  & {\scriptsize N76.0/G23.3/S0.4/Bj0.3} & \makecell{{\scriptsize G1.2/N0.6/S20.0/Rg22.3}\\{\scriptsize Rj18.8/Bg17.7/Bj19.4}} & -- & -- & -- & -- \\
\hline
\end{tabular}

\caption{Recovery of the constant wall flux ($g=-10$, $g_{0}=-5$, $150$ epochs) for inner data and for outer data under the overlap strategy. The passive LLM optimization ($30$ seeded trials) and the active agentic AI optimization ($5$ trials, no seed) are reported as median [interquartile range], and the classical methods and Brent's method, which is not part of the optimization policy, are run once. A trial is successful if the relative error of the recovered wall flux remains below $1\%$ until the final epoch. The converged value is the wall flux at the final epoch, the epochs to target is the first epoch of the final interval within the $1\%$ band, and the wall time is measured up to that epoch. The selection mix is the percentage of optimization policy per trial selecting G (normalized gradient), A (Adam), N (damped Newton), S (secant), Rg (linear regression of the gradient), Rj (quadratic fit of the objective), Bg (bracketed root finding of the gradient), or Bj (bracketed refinement of the objective).}
\label{tab:qwen-classical-codex-work-comparison}
\end{table*}

The convergence of the recovered wall flux is presented in Fig.~\ref{fig:inverse_recovery} for one optimizer of each group, namely Adam from the classical methods, the passive LLM optimization, and the active agentic AI optimization, for inner data and for outer data under the overlap strategy, at the three diffusivities. Despite the clear benefit of other classical optimization methods in this test case, we focus on comparing Adam with the LLM models as Adam becomes necessary for more complex spatially varying test cases presented next. For the passive and the active optimization, the figure shows the median trial and, wherever a failure occurred, one of the failed trials. Adam serves as the reference in the figure, and its hyperparameters were tuned to meet the success criterion in as few epochs as possible, unlike in the benchmarking of Section~\ref{sec:results-inverse}.  A single trial of the agent consumes on average $0.55$ million input tokens and $16{,}000$ reasoning tokens, whereas a single trial of the passive LLM optimization, in which the language model is queried once per epoch, consumes up to approximately $72{,}000$ input tokens and $10{,}400$ output tokens, that is, approximately one eighth of the input tokens of the agent. With five trials per case for the agent and $30$ trials per case for the passive LLM optimization, the six cases comprise $30$ and $180$ trials, respectively. In total, the agent consumed $16.54$ million input tokens and $0.47$ million reasoning tokens, and the passive LLM optimization up to approximately $13.0$ million input tokens and $1.9$ million output tokens.

To validate the LLMs, one successful trial of the passive LLM optimization and one of the agent are selected at random, and at each epoch the method selected by the language model is implemented manually, together with its logged step controls. For both trials, the manually reproduced trajectory of the wall flux is closely aligned with the trajectory of the trial, although the two are not exactly identical. The prompt and the optimization policy, together with the reasoning, the log, and the actions of the language model at each epoch of the selected trials and the corresponding manual validation, are detailed in~\ref{app:llm-const}.

\subsubsection{Spatially varying wall flux: discrete profile reconstruction}
\label{sec:llm-discrete}

We now turn to the recovery of a spatially varying wall flux $g(x)$, in which the entire profile along the bottom wall is reconstructed at each spatial nodal point. As a result, the number of unknowns grows with the mesh resolution, namely $257$ nodal values on M1, used at $D=10^{-4}$ and $10^{-5}$, and $513$ on M2, used at $D=10^{-6}$. For each profile, the synthetic data are generated with the single-domain solver on M4.  Since every nodal value of the wall flux is recovered as a separate unknown, the result may depend on the initial estimate, and the recovery is therefore started from three initial estimates, namely a flat profile with the constant value $-5$ along the entire wall (Flat), which differs entirely from the true profiles, and the true profile with random nodal perturbations of  $10\%$ and $20\%$ of its peak magnitude (P10 and P20). Owing to the high dimensionality of this inverse problem, several optimizers of the preceding experiment are not considered here, even though the Newton, secant, and Brent methods outperformed Adam and the LLM optimizers in the constant-flux recovery. The secant and Brent methods are more suited for univariate problems. The Newton method requires the Hessian with respect to every nodal value, which is memory-intensive, and in our testing converged to an incorrect profile. The passive LLM optimization is likewise excluded as it did not perform well in the multi-dimensional setting. This observation is consistent with Yang et al.~\cite{yang2024large}. The active agentic optimization is not subject to this limitation, since the language model operates at the workflow level rather than at the parameter level. It receives the current nodal profile of the wall flux and its adjoint gradient, together with the history of the misfit, and selects the update method and its controls, whereas the update of the nodal values is computed by the controller, so that the agent never proposes nodal values directly. Adam and the agentic AI optimization are therefore employed in this experiment.

The discrete reconstruction is challenging, since a large number of nodal values must be recovered from data that are most informative close to the wall (due to the thin boundary layer). To improve the profile reconstruction, the misfit is augmented with a Tikhonov regularization term that penalizes the curvature of $g(x)$, weighted by the regularization parameter $\alpha$. Since the appropriate value of $\alpha$ is not known in advance, it is determined as a hyperparameter by a sweep over the nine values $\alpha\in\{10^{-9},\, 5\times10^{-9},\, 10^{-8},\, 5\times10^{-8},\, 10^{-7},\, 5\times10^{-7},\, 10^{-6},\, 5\times10^{-6},\, 10^{-5}\}$. For every case, each run of the sweep employs Adam over $2000$ epochs with an initial learning rate of $0.1$, reduced by a factor of $0.75$ every $200$ epochs, and the optimal value $\alpha^{*}$ is selected as the value that minimizes the relative $L_\infty$ error of the recovered profile with respect to the true profile. Since this selection uses the true profile, which is unavailable in practice, $\alpha^{*}$ represents the most favorable regularization for Adam.


For the active agentic optimization, the controller of the preceding experiment is retained, whereas its optimization policy is replaced by methods that update the entire profile at once, namely Adam, normalized gradient descent, momentum gradient descent, the Barzilai--Borwein method, the conjugate gradient method, and projected gradient descent. At each epoch, the controller passes to the agent the prompt, the available methods of the optimization policy, the current nodal profile of the wall flux and its adjoint gradient, and the history of the misfit and of the previously selected methods. The optimal value of the regularization parameter $\alpha^{*}$, corresponding to the Schwarz-coupled solver, is used as the reference value $\alpha_{\mathrm{ref}}$ for the agent, which is run over $300$ epochs in each of five trials per case. Since the reconstruction error depends on both the value of $\alpha$ and the optimizer, fixing $\alpha$ ensures that the agent and the Schwarz-coupled solver with Adam solve the same regularized problem.

\begin{figure}
    \centering
    \includegraphics[width=\textwidth,height=\textheight,keepaspectratio]{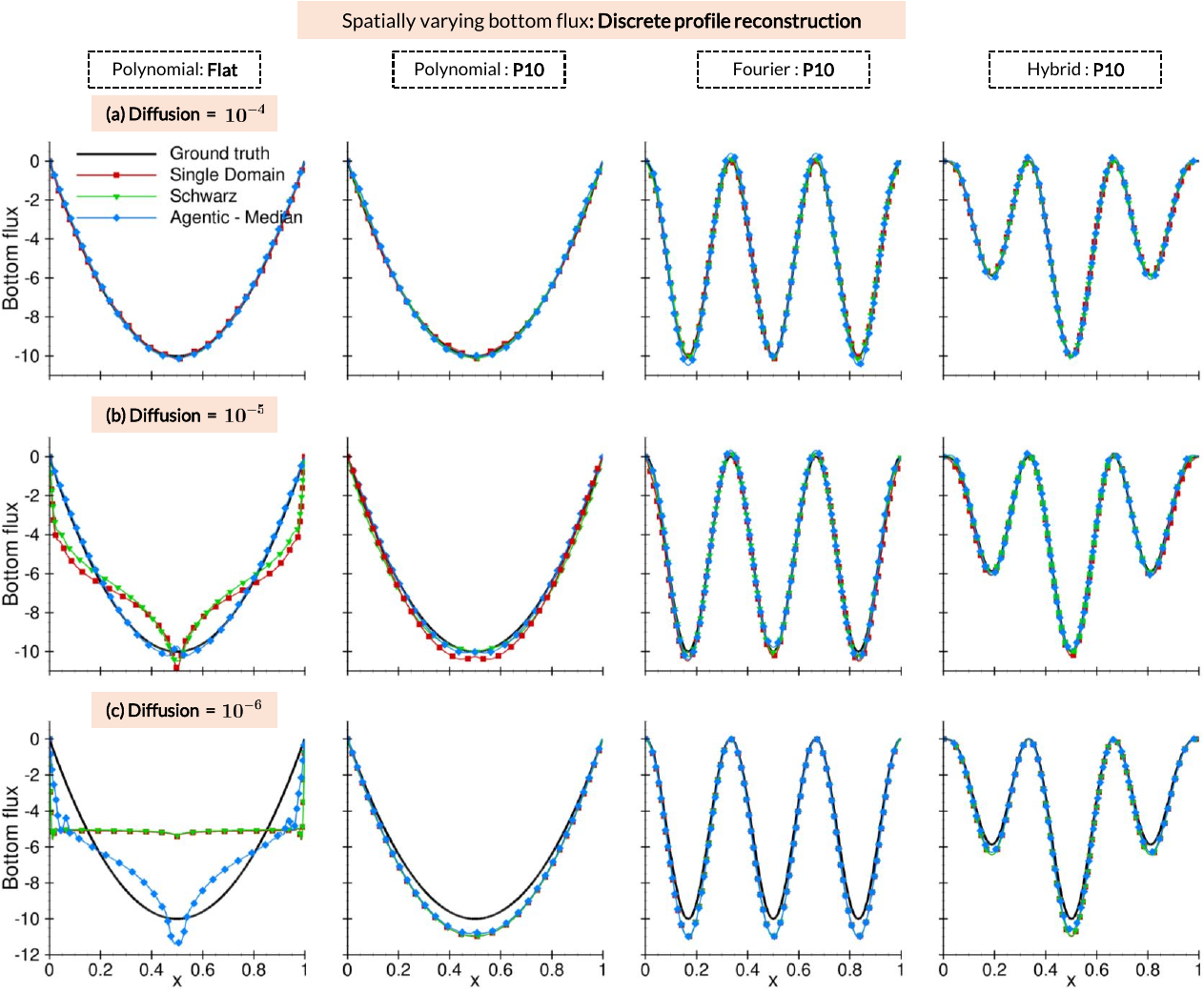}
    \caption{Discrete reconstruction of the spatially varying wall flux with inner data at (a) $D=10^{-4}$, (b) $D=10^{-5}$, and (c) $D=10^{-6}$. The first column shows the polynomial profile initialized from Flat, and the remaining columns show the polynomial, Fourier, and hybrid profiles initialized from P10. The true profile (black) is compared with the profiles recovered by the single-domain solver with Adam (red) and the Schwarz-coupled solver with Adam (green), both run over $2000$ epochs, and by the median trial of the active agentic optimization with the Schwarz-coupled solver (blue), run over $300$ epochs.}
  \label{fig:discrete_recovery}
\end{figure}

\begin{sidewaystable*}[p]
\centering
\small
\setlength{\tabcolsep}{3pt}
\renewcommand{\arraystretch}{1.2}

\begin{tabular}{|*{11}{c|}}
\hline
\multicolumn{11}{|c|}{\textbf{Spatially varying wall flux: Discrete profile reconstruction -- Inner data}}\\
\hline
\multirow{2}{*}{\textbf{Profile}} &
\multirow{2}{*}{\textbf{Init.}} &
\multicolumn{3}{c|}{\textbf{Single-domain -- Adam}} &
\multicolumn{3}{c|}{\textbf{Schwarz -- Adam}} &
\multicolumn{3}{c|}{\textbf{Schwarz -- Agentic}}\\
\cline{3-11}
& &
$\alpha^*$ & Rel.\ $L_\infty$ (\%) & Wall time (min) &
$\alpha^*$ & Rel.\ $L_\infty$ (\%) & Wall time (min) &
$\alpha_{\rm ref}$ & Rel.\ $L_\infty$ (\%) & Wall time (min)\\
\hline

\multicolumn{11}{|c|}{$\boldsymbol{D=10^{-4}}$}\\
\hline
\multirow{3}{*}{Polynomial}
& Flat & 5e-7 & 1.615 & 65.319 & 1e-8 & 1.389 & 23.820
& 1e-8 & 1.454 [0.002] & 59.195 [4.382] \\
\cline{2-11}
& P10 & 1e-6 & 1.412 & 66.285 & 5e-7 & 1.668 & 22.222
& 5e-7 & 3.883 [0.001] & 52.874 [7.483] \\
\cline{2-11}
& P20 & 1e-6 & 1.309 & 63.382 & 1e-7 & 1.509 & 19.831
& 1e-7 & 2.075 [0.001] & 51.684 [5.917] \\
\hline
\multirow{3}{*}{Fourier}
& Flat & 1e-7 & 3.561 & 49.689 & 5e-8 & 4.339 & 23.286
& 5e-8 & 5.909 [0.375] & 60.826 [6.264]  \\
\cline{2-11}
& P10 & 5e-7 & 2.113 & 89.058 & 1e-6 & 2.595 & 25.803
& 1e-6 & 4.836 [0.311] & 58.655 [4.758] \\
\cline{2-11}
& P20 & 1e-7 & 2.797 & 60.786 & 1e-7 & 3.249 & 38.793
& 1e-7 & 4.054 [0.285] & 51.850 [7.031] \\
\hline
\multirow{3}{*}{Hybrid}
& Flat & 1e-7 & 3.890 & 43.344 & 1e-7 & 4.376 & 28.796
& 1e-7 & 8.023 [1.208] & 61.855 [5.342] \\
\cline{2-11}
& P10 & 1e-6 & 0.888 & 43.418 & 1e-6 & 0.441 & 23.465
& 1e-6 & 2.294 [0.018] & 59.474 [6.889] \\
\cline{2-11}
& P20 & 5e-7 & 1.531 & 45.064 & 5e-7 & 1.317 & 27.011
& 5e-7 & 4.031 [0.061] & 52.001 [4.519] \\
\hline

\multicolumn{11}{|c|}{$\boldsymbol{D=10^{-5}}$}\\
\hline
\multirow{3}{*}{Polynomial}
& Flat & 1e-9 & 31.151 & 51.466 & 1e-9 & 26.425 & 33.500
& 1e-9 & 2.747 [0.018] & 57.625 [7.762] \\
\cline{2-11}
& P10 & 1e-7 & 4.294 & 51.677 & 1e-7 & 2.628 & 29.448
& 1e-7 & 3.655 [0.122] & 53.195 [5.684] \\
\cline{2-11}
& P20 & 1e-7 & 5.256 & 51.835 & 1e-7 & 4.377 & 30.902
& 1e-7 & 8.082 [2.083] & 56.052 [6.107] \\
\hline
\multirow{3}{*}{Fourier}
& Flat & 1e-9 & 43.005 & 51.402 & 1e-9 & 45.585 & 48.678
& 1e-9 & 5.065 [0.055] & 54.028 [4.293]  \\
\cline{2-11}
& P10 & 1e-7 & 6.075 & 60.240 & 1e-7 & 2.739 & 27.293
& 1e-7 & 4.173 [0.119] & 51.562 [7.218] \\
\cline{2-11}
& P20 & 1e-7 & 9.869 & 60.300 & 1e-7 & 7.694 & 53.053
& 1e-7 & 8.473 [3.917] & 54.872 [5.856] \\
\hline
\multirow{3}{*}{Hybrid}
& Flat & 1e-9 & 32.575 & 65.567 & 1e-9 & 32.933 & 55.250
& 1e-9 & 4.203 [1.279] & 59.708 [6.473] \\
\cline{2-11}
& P10 & 1e-7 & 3.616 & 43.146 & 1e-7 & 1.312 & 48.384
& 1e-7 & 2.157 [0.117] & 57.882 [4.941] \\
\cline{2-11}
& P20 & 1e-7 & 5.359 & 64.597 & 1e-7 & 3.672 & 52.067
& 1e-7 & 4.609 [1.881]& 52.464 [7.594] \\
\hline

\multicolumn{11}{|c|}{$\boldsymbol{D=10^{-6}}$}\\
\hline
\multirow{3}{*}{Polynomial}
& Flat & 1e-9 & 51.246 & 383.615 & 1e-9 & 50.856 & 177.371
& 1e-9 & 33.830 [5.749] & 72.224 [5.028]  \\
\cline{2-11}
& P10 & 5e-8 & 9.661 & 219.526 & 5e-8 & 9.580 & 157.625
& 5e-8 & 8.562 [1.082]& 67.623 [6.738] \\
\cline{2-11}
& P20 & 5e-8 & 19.051 & 219.246 & 5e-8 & 19.048 & 168.240
& 5e-8 & 18.209 [3.038] & 41.827 [4.667] \\
\hline
\multirow{3}{*}{Fourier}
& Flat & 1e-9 & 54.133 & 223.591 & 1e-9 & 53.865 & 97.528
& 1e-9 & 45.655 [0.220] & 66.977 [7.146]  \\
\cline{2-11}
& P10 & 1e-7 & 9.814 & 217.491 & 1e-7 & 9.671 & 90.835
& 1e-7 & 9.933 [2.989]& 47.737 [5.491] \\
\cline{2-11}
& P20 & 1e-7 & 19.567 & 218.140 & 1e-7 & 19.467 & 82.610
& 1e-7 & 19.908 [3.913] & 39.947 [6.925] \\
\hline
\multirow{3}{*}{Hybrid}
& Flat & 1e-9 & 54.283 & 224.103 & 1e-9 & 54.245 & 89.424
& 1e-9 & 30.104 [0.684] & 71.931 [4.375]  \\
\cline{2-11}
& P10 & 1e-7 & 9.766 & 219.066 & 1e-7 & 9.479 & 93.005
& 1e-7 & 7.073 [1.082] & 67.970 [7.839] \\
\cline{2-11}
& P20 & 1e-7 & 19.267 & 219.252 & 1e-7 & 19.023 & 97.672
& 1e-7 & 16.617 [4.074] & 45.478 [5.763]\\
\hline
\end{tabular}%

\caption{Discrete reconstruction of the spatially varying wall flux with inner data for the polynomial, Fourier, and hybrid profiles, started from the Flat, P10, and P20 initial estimates. For the single-domain and the Schwarz-coupled solver with Adam ($2000$ epochs), $\alpha^*$ is the optimal regularization parameter of the sweep. The active agentic AI optimization ($300$ epochs, five trials) uses $\alpha_{\rm ref}=\alpha^*$ of the Schwarz-coupled solver with Adam and is reported as median [interquartile range]. The relative $L_\infty$ error is the peak error relative to the maximum magnitude of the true profile, and the wall time covers all epochs, including the inference of the language model for the agent.}
\label{tab:inner_inverse_agentic}
\end{sidewaystable*}

Figure~\ref{fig:discrete_recovery} shows the profiles recovered with inner data for the polynomial profile from the Flat initialization and for all three profiles from the P10 initialization. The single-domain solver with Adam serves as a reference. At $D=10^{-4}$, all three methods recover the true profile closely for both initializations and all three profiles. At $D=10^{-5}$, the methods differ for the Flat initialization. Starting from this estimate, Adam fails for both solvers. Since the single-domain and the Schwarz-coupled solver converge to nearly the same incorrect profile, the failure lies with the optimizer rather than with the forward solver. The agent, in contrast, recovers the true profile along the entire wall from the same initialization. This recovery is reproducible across the five trials, with an interquartile range of the relative error of $0.018\%$ (Table~\ref{tab:inner_inverse_agentic}). For the P10 initialization, all three methods recover the three profiles, with only small deviations at the minima, so that the advantage of the agent at this diffusivity is confined to cases in which the initial estimate is far from the true profile. At $D=10^{-6}$, the recovery deteriorates for all methods. For the Flat initialization, the Schwarz-coupled solver with Adam remains close to the initial value of $-5$, as does the single-domain solver with Adam. The agent moves toward the true profile and locates its central minimum, although it overestimates its depth and deviates from the true profile near the ends of the wall. For the P10 initialization, all three methods produce nearly identical profiles that remain close to the initial estimate, with minima approximately $10\%$ more negative than those of the true profile. 

Table~\ref{tab:inner_inverse_agentic} extends the observations from the figure to all profiles and initializations. The advantage of the agent for the Flat initialization at $D=10^{-5}$ and $10^{-6}$ is not limited to the polynomial profile shown in the figure but holds for all three profiles. At $D=10^{-5}$, the relative $L_\infty$ error of the Schwarz-coupled solver with Adam lies between $26$ and $46\%$, whereas that of the agent lies between $2.7$ and $5.1\%$, and at $D=10^{-6}$, the agent reduces the error to between $30$ and $46\%$, compared with $51$ to $54\%$ for the Schwarz-coupled solver with Adam. In the remaining cases, by contrast, the agent offers no such advantage. At $D=10^{-4}$ for all initializations, and for the P10 and P20 initializations at $D=10^{-5}$, its errors of $1.5$ to $8.5\%$ are larger than those of the Schwarz-coupled solver with Adam in every case, although of the same order. At $D=10^{-6}$, the errors of all methods remain close to the initial perturbations, between $7.1$ and $9.9\%$ for P10 and between $16.6$ and $19.9\%$ for P20. The single-domain solver with Adam fails in the same cases as the Schwarz-coupled solver with Adam, with nearly identical errors at $D=10^{-6}$, which confirms that the difficulty of the recovery is inherent to the discrete reconstruction and is not introduced by the Schwarz coupling. 

In terms of wall time, the Schwarz-coupled solver with Adam is faster than the single-domain solver with Adam in all but one of the $27$ cases, by a factor of up to $3.5$. Despite its considerably smaller budget of $300$ epochs, compared with the $2000$ epochs of Adam, the agent requires more wall time than the Schwarz-coupled solver with Adam at $D=10^{-4}$ and $10^{-5}$, namely $52$ to $62$~min against $20$ to $55$~min, since each of its epochs includes the inference of the language model, which influences its wall time. Only at $D=10^{-6}$, where each epoch of the solver is more expensive, does the agent require less wall time, namely $40$ to $72$~min against $83$ to $177$~min. In addition, a single trial of the agent consumes approximately $0.85$ million input tokens and $0.43$ million output tokens, including the reasoning tokens. With five trials for each of the $27$ cases, the $135$ trials with inner data amount to approximately $115$ million input tokens and $58$ million output tokens.

The reconstruction was repeated with overlap data, and the results are reported in Table~\ref{tab:overlap_inverse_agentic} of~\ref{app:supplement_data}. As with inner data, the largest improvement of the agent over Adam occurs for the Flat initialization at $D=10^{-5}$, although the reduction in error is smaller, and in several remaining cases the two optimizers yield comparable errors.



\subsubsection{Spatially varying wall flux: functional-basis identification}
\label{sec:llm-basis}

The preceding discrete reconstruction experiment imposes no functional form on the wall flux, but it requires one control per wall node. The dimension of the inverse problem therefore grows with the mesh resolution, and the reconstruction must be stabilized by Tikhonov regularization, whose optimal hyperparameter $\alpha^{*}$ is not known a priori. As a result, the discrete reconstruction is expensive and can fail in some cases. A spatially varying wall flux can, however, often be described by a compact analytical expression. We thus consider the functional-basis parametrization of Section~\ref{sec:param}, in which $g(x)$ is expanded as a linear combination of a small number of basis functions, where the controls are the linear coefficients and the nonlinear parameters, $\theta=\{c_j,\boldsymbol{\beta}_j\}_{j=1}^{K}$. The number of unknowns is now determined by the library rather than by the mesh used for the inverse problem, and it is typically considerably smaller than the number of wall nodes. Both the linear coefficients and the nonlinear parameters are optimized jointly with the adjoint gradients of the differentiable Schwarz-coupled solver.

In the present libraries, the nonlinear parameters $\boldsymbol{\beta}_j$ reduce to a single frequency $\beta_j$ of each oscillatory basis function, which is optimized jointly with the linear coefficients, similar to the ADAM-SINDy (sparse identification of nonlinear dynamical systems) approach~\cite{viknesh2026adam}. The identification is examined through four experiments, which differ in the library dictionary $\mathcal{D}$ of basis functions.

\begin{itemize}
    \item \textbf{Experiment 1 -- Ideal library: Adam optimization.} For each envelope, the library dictionary contains exactly the basis functions of the true envelope in Eq.~\eqref{eq:flux_explicit}, together with the corresponding controls,
    \begin{equation}
  \mathcal{D}_{\mathrm{ideal}}=
  \left\{
  \begin{array}{lll}
    \{x,\;x^2\}, & \theta=\{c_1,c_2\}, & \text{polynomial envelope},\\[4pt]
    \{1,\;\cos(\pi\beta_1 x)\}, & \theta=\{c_1,c_2,\beta_1\}, & \text{Fourier envelope},\\[4pt]
    \{x,\;x^2,\;x\cos(\pi\beta_1 x),\;x^2\cos(\pi\beta_2 x)\}, & \theta=\{c_1,\ldots,c_4,\beta_1,\beta_2\}, & \text{hybrid polynomial--Fourier envelope}.
  \end{array}
  \right.
  \label{eq:ideal_library}
    \end{equation}
    Since the structure of $g$ is known, only its linear coefficients and frequencies are recovered. All linear coefficients are initialized to $c_j=10$ and all frequencies to $\beta_j=2$, independently of the true envelope. The controls are then updated by Adam over $1200$ epochs, with a constant learning rate of $0.1$ for both the linear coefficients and the frequencies.

    \item \textbf{Experiment 2 -- Common library: Adam optimization.} This experiment represents the realistic setting in which the functional form of the wall flux is unknown, and it serves as the baseline for the agentic experiments. A single library dictionary is shared by all three envelopes, comprising constant, polynomial, cosine, and sine basis functions and their products,
    \begin{equation}
    \begin{aligned}
    \mathcal{D}_{\mathrm{com}}&=\{1,\;x,\;x^2,\;\cos(\pi\beta_1 x),\;\sin(\pi\beta_2 x),\;x\cos(\pi\beta_3 x),\;x\sin(\pi\beta_4 x),\;x^2\cos(\pi\beta_5 x),\;x^2\sin(\pi\beta_6 x)\},\\
    \theta&=\{c_1,\ldots,c_9,\;\beta_1,\ldots,\beta_6\},
    \end{aligned}
    \label{eq:common_library}
    \end{equation}
    with nine linear coefficients and six frequencies, that is, fifteen controls. The library dictionary contains the ideal library of every envelope as a subset but carries no information on which of its members are active. The controls are initialized as in Experiment~1 and updated by Adam with the same settings. No sparsity-promoting regularization is applied as in SINDy~\cite{brunton2016discovering}, since it did not reduce the error of the recovered profile.

    \item \textbf{Experiment 3 -- Common library: Agentic AI optimization.} This experiment retains the library dictionary  from Experiment~2, and it changes only the update operator, replacing Adam with an AI agent. The optimization policy of the agent comprises Adam, normalized gradient descent, momentum gradient descent, the Barzilai--Borwein method, the conjugate gradient method, Levenberg–Marquardt, and projected gradient descent. At every epoch, the agent receives the current linear coefficients and nonlinear parameters together with their complete history, as well as the misfit and its adjoint gradient, and selects  the method expected to provide the best update of $\theta$. The agent has no access to the true envelope and cannot modify the differentiable inverse workflow. It is run over $500$ epochs in each of five trials, without early termination.

    \item \textbf{Experiment 4 -- Staged library: Agentic AI optimization.} This experiment removes the fixed library dictionary, so that the agent grows the library under a prescribed admission policy in addition to updating its controls. The agent starts from a polynomial library of second order, $\mathcal{D}^{(0)}=\{1,\;x,\;x^2\}$, and assigns the initial values of its controls. The library is revised at most once every $\Delta n$ epochs, so that the controls of the present library can adjust before the next revision. At each revision, the agent admits the next family of basis functions if the misfit has reached a plateau with the present library, indicating that the present library cannot reduce the misfit further. No numerical threshold is prescribed for the plateau, and it is identified by the agent from the misfit history. The families are admitted in a fixed order: first the polynomials of third to fifth order, then the Fourier functions, and finally the products of Fourier functions and polynomials. At each admission, the initial values of the controls of the newly admitted basis functions are decided by the agent, in contrast to Experiments~1--3, in which all coefficients and frequencies are initialized to $c_j=10$ and $\beta_j=2$. The admitted basis functions and their initial values are written by the agent into the definition of $\theta$, which parametrizes only the wall flux boundary condition, so that the forward and adjoint computations remain unchanged. Once admitted, a basis function is retained in the library until the end of the optimization, although its linear coefficient may converge to zero. The library dictionary is limited to at most six polynomial, two Fourier, and twelve polynomial--Fourier basis functions. Between revisions, the agent updates the controls at every epoch with the optimization policy of Experiment~3, based on their current values and complete history, the misfit, and its adjoint gradient. In contrast to Experiments~2 and~3, in which the library is fixed, both the members of the library and the dimension of $\theta$ grow during the optimization. The procedure is thus loosely related to symbolic regression, in which the structure of an expression is identified together with its parameters, and to recent language-model-driven approaches to equation discovery~\cite{pang2026deliberate}, although here the structure is restricted to prescribed candidate families admitted in a fixed order. It further differs from these approaches in two respects. First, the unknown function is not fitted to observations of itself but enters the problem as a boundary condition, and it is identified only through the PDE constraint and the adjoint optimization. Second, the objective is the recovery of the wall flux profile rather than of its exact analytical expression. In equation discovery for dynamical systems, identifying the true governing terms is the foremost goal, since they carry the physical interpretation of the model. Here, by contrast, any functional representation that best approximates the profile is the primary focus. The agent is run over $75$ epochs in each of five trials, without early termination. This budget is smaller than that of Experiment~3, since the admission policy confines the search to a fixed sequence of families, and the admission of families ends once all families in the sequence have been admitted. The revision intervals $\Delta n=5$, $10$, and $15$ were explored, all of which led to a successful recovery, and the results are shown for $\Delta n=10$.

\end{itemize}

\begin{figure}
    \centering
    \includegraphics[width=\textwidth,height=\textheight,keepaspectratio]{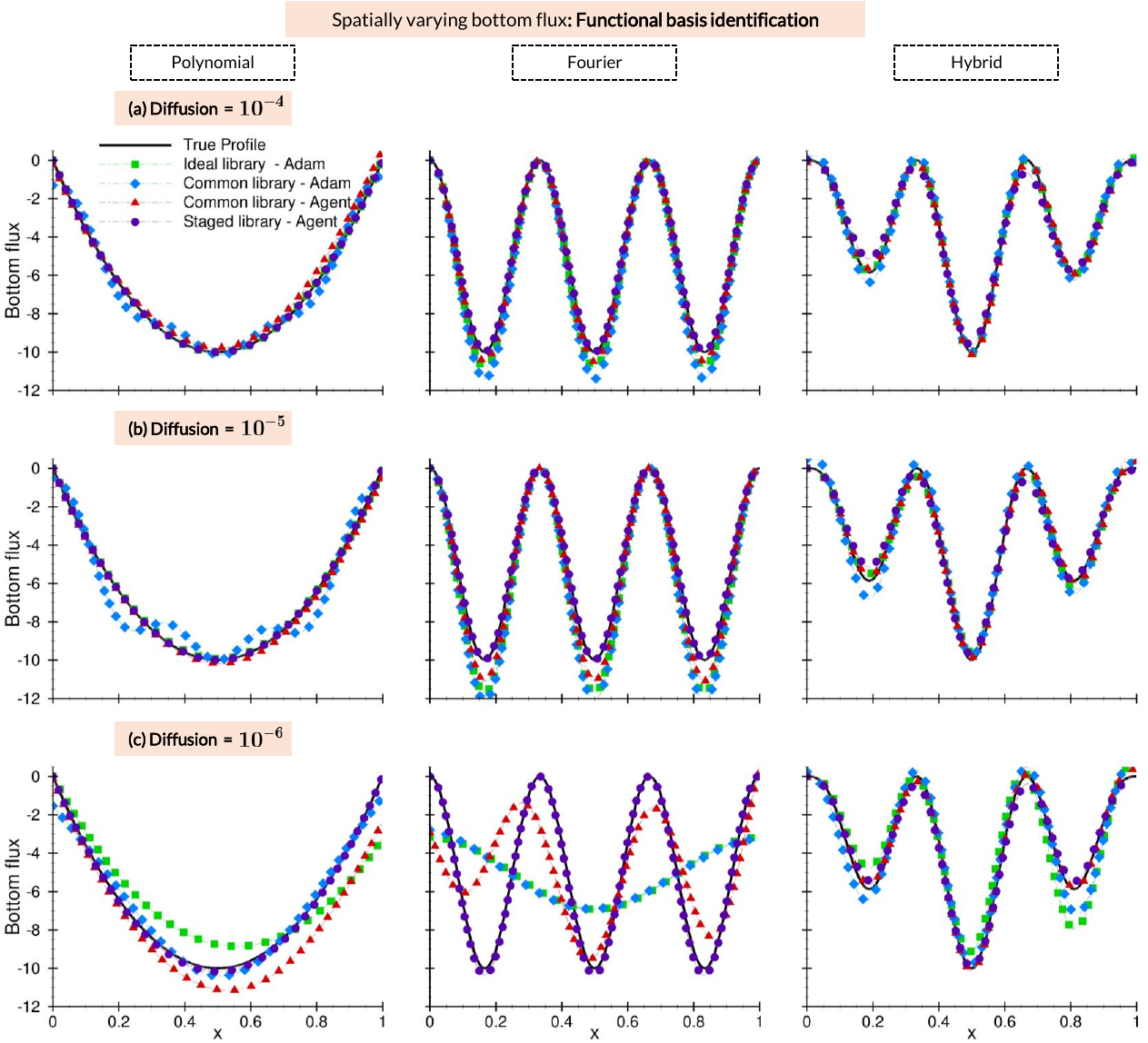}
    \caption{Functional basis identification of the spatially varying wall flux with inner data at (a) $D=10^{-4}$, (b) $D=10^{-5}$, and (c) $D=10^{-6}$, for the polynomial (left), Fourier (center), and hybrid (right) profiles. The true profile (black) is compared with the profiles recovered with the ideal library and Adam (Experiment~1, green) and with the common library and Adam (Experiment~2, blue), both run over $1200$ epochs. It is further compared with the profiles of the median trials of the agentic optimization, with the common library (Experiment~3, red), run over $500$ epochs, and with the library grown by the agent (Experiment~4, purple), run over $75$ epochs with $\Delta n=10$. The errors with outer data under the overlap strategy are reported in Table~\ref{tab:basis_recovery}.}

  \label{fig:basis_recovery}
\end{figure}

\begin{sidewaystable*}[p]
\centering
\footnotesize
\setlength{\tabcolsep}{3pt}
\renewcommand{\arraystretch}{1.2}

\begin{tabular}{|*{11}{c|}}
\hline
\multicolumn{11}{|c|}{\textbf{Spatially varying wall flux: Functional-basis identification}}\\
\hline
\multicolumn{11}{|c|}{\textbf{True envelopes:}\quad Polynomial: $-40x + 40x^{2}$; \quad Fourier: $-5 + 5\cos(6\pi x)$; \quad Hybrid: $-20x + 20x^{2} + 20x\cos(6\pi x) - 20x^{2}\cos(6\pi x)$} \\
\hline
\multirow{2}{*}{\textbf{Profile}} &
\multirow{2}{*}{\textbf{Data}} &
\multicolumn{2}{c|}{\textbf{Exp.~1: Ideal library -- Adam}} &
\multicolumn{2}{c|}{\textbf{Exp.~2: Common library -- Adam}} &
\multicolumn{2}{c|}{\textbf{Exp.~3: Common library -- Agentic}} &
\multicolumn{3}{c|}{\textbf{Exp.~4: Staged library -- Agentic}}\\
\cline{3-11}
& &
Rel.\ $L_\infty$ (\%) & Wall time (min) &
Rel.\ $L_\infty$ (\%) & Wall time (min) &
Rel.\ $L_\infty$ (\%) & Wall time (min) &
Terms & Rel.\ $L_\infty$ (\%) & Wall time (min)\\
\hline

\multicolumn{11}{|c|}{$\boldsymbol{D=10^{-4}}$}\\
\hline
\multirow{2}{*}{Polynomial}
& Inner   & 0.714  & 7.5 & 13.020 & 12.9 & 7.093 [2.118]  & 358.7 [28.08] & 2 [0] & 0.152 [0.080] & 19.8 [1.72]\\
\cline{2-11}
& Overlap & 0.447  & 7.2 & 16.792 & 12.4 & 3.372 [0.404] & 340.6 [11.92] & 2 [0]& 0.371 [0.117] & 35.7 [4.29] \\
\hline
\multirow{2}{*}{Fourier}
& Inner   & 7.139 & 7.9  & 12.221 & 11.1 & 5.576 [1.733] & 204.4 [23.87] & 3 [0] & 0.919 [0.224] & 21.3 [3.67] \\
\cline{2-11}
& Overlap & 8.413 & 7.8  & 14.095 & 11.4 & 5.797 [1.401]& 204.3 [21.03]& 3 [0] & 0.515 [0.310]& 23.2 [6.72]\\
\hline
\multirow{2}{*}{Hybrid}
& Inner   & 1.182 & 13.2 & 5.075 & 12.1 & 3.684 [0.834] & 327.0 [11.51] & 7 [1] & 7.193 [2.159] & 15.7 [1.47] \\
\cline{2-11}
& Overlap & 4.145 & 10.3 & 8.813 & 11.2 & 6.744 [3.173] & 338.4 [18.08] & 8 [1] & 6.962 [2.498] & 47.2 [4.90] \\
\hline

\multicolumn{11}{|c|}{$\boldsymbol{D=10^{-5}}$}\\
\hline
\multirow{2}{*}{Polynomial}
& Inner   & 1.053  & 13.6 & 7.422 & 16.2 & 4.242 [1.931]  & 336.8 [5.92] & 4 [0] & 0.810 [0.049] & 69.4 [5.07]\\
\cline{2-11}
& Overlap & 2.271  & 13.2 & 4.920  & 14.6 & 4.062 [2.008]  & 344.6 [8.09] & 3 [0] &  0.440 [0.021] & 65.4 [7.73] \\
\hline
\multirow{2}{*}{Fourier}
& Inner   & 16.918 & 7.7  & 14.895 & 15.0 & 10.869 [1.969] & 331.6 [1.39] & 5 [1] & 1.038 [0.510]& 77.9 [3.19] \\
\cline{2-11}
& Overlap & 17.567 & 7.7  & 6.799 & 14.7 & 5.244 [1.610] & 304.1 [9.22] & 5 [1] & 1.247 [0.381] & 74.2 [8.90]\\
\hline
\multirow{2}{*}{Hybrid}
& Inner   & 9.119  & 12.4 & 9.293  & 15.0 & 5.250 [2.771] & 223.4 [6.33] & 7 [0] & 7.135 [2.638] & 94.1 [11.04]\\
\cline{2-11}
& Overlap & 16.178 & 12.6 & 11.166 & 15.0 & 13.729 [4.219] & 326.7 [16.47] & 7 [1] & 7.084 [1.949] & 114.2 [7.88]\\
\hline

\multicolumn{11}{|c|}{$\boldsymbol{D=10^{-6}}$}\\
\hline
\multirow{2}{*}{Polynomial}
& Inner   & 32.232 & 52.0 & 15.425 & 73.3 & 23.481 [3.880] & 398.5 [22.09] & 4 [1] & 1.680 [0.494] & 71.9 [4.89] \\
\cline{2-11}
& Overlap & 34.696 & 50.8 & 17.028 & 71.3 & 13.432 [1.791] & 439.1 [5.16] & 4 [1] & 1.332 [0.822] & 71.6 [5.12] \\
\hline
\multirow{2}{*}{Fourier}
& Inner   & 60.330 & 46.5 & 61.107 & 68.4 & 56.876 [4.157] & 386.8 [9.51] & 6 [1] & 4.531 [1.201] & 59.0 [2.49]  \\
\cline{2-11}
& Overlap & 61.428 & 46.1 & 62.685 & 99.3 & 54.127 [6.388] & 331.5 [5.80]& 8 [1]& 3.687 [0.662] & 49.0 [2.10] \\
\hline
\multirow{2}{*}{Hybrid}
& Inner   & 21.126 & 51.9 & 15.075 & 71.2  & 3.739 [2.706] & 415.3 [31.45] & 11 [1] & 4.987 [1.032] & 52.0 [6.38] \\
\cline{2-11}
& Overlap & 24.842 & 51.4 & 18.110 & 106.9 & 15.784 [1.462] & 407.5 [39.27] & 12 [0] & 5.603 [2.951] & 81.1 [13.63] \\
\hline
\end{tabular}%

\caption{Functional basis identification of the spatially varying wall flux for the polynomial, Fourier, and hybrid profiles, with inner data and outer data under the overlap strategy (Overlap). Experiments~1 and~2 employ Adam over $1200$ epochs with the ideal and the common library, respectively, with all linear coefficients initialized to $c_j=10$ and all frequencies to $\beta_j=2$. Experiment~3 employs the agentic optimization with the common library and the same initial values over $500$ epochs. Experiment~4 employs the agentic optimization with the library grown by the agent over $75$ epochs, with the revision interval $\Delta n=10$ and with the initial values of the admitted controls decided by the agent. Both agentic experiments are run in five trials per case, and their results are reported as the median over the trials, with the interquartile range (IQR) given in brackets, for the number of terms, the relative $L_\infty$ error, and the wall time. \emph{Terms} denotes the number of basis functions that remain in the final expression of Experiment~4, that is, those with a linear coefficient of magnitude $|c_j|\geq10^{-4}$ and, for the sine-type functions, a frequency of magnitude $|\beta_j|\geq10^{-4}$. The relative $L_\infty$ error is the peak error relative to the maximum magnitude of the true profile, and the wall time covers all epochs, including the inference of the language model for the agent.}
\label{tab:basis_recovery}
\end{sidewaystable*}

Figure~\ref{fig:basis_recovery} compares the recovered profiles with the true profile for inner data, and Table~\ref{tab:basis_recovery} reports the corresponding relative $L_\infty$ errors and wall times for both data placements. With the ideal library, Experiment~1 recovers all three profiles at $D=10^{-4}$, with errors of $0.4$--$8.4\%$. Its accuracy declines as the diffusivity decreases. Since the ideal library contains only the true basis functions, it offers no additional degrees of freedom to reduce the misfit. The common library of Experiment~2 introduces such degrees of freedom, but their effect depends on the diffusivity. At $D=10^{-4}$, the additional candidates degrade the recovered profile, raising the error. At $D=10^{-6}$, they improve the reconstruction of the polynomial and hybrid profiles. This compensation is not general, since neither experiment recovers the Fourier profile at $D=10^{-6}$.

Replacing Adam with the agent while retaining the common library improves the convergence within that library. Experiment~3 is more accurate than Experiment~2 in sixteen of the eighteen cases, by up to a factor of five, and at $D=10^{-6}$ it reaches $3.7\%$ for the hybrid profile with inner data. However, the Fourier profile at $D=10^{-6}$ is still not recovered, where its oscillation is present but displaced, with an error of $54$--$57\%$. The improvements in Experiment 3 are costly. Since the language model is queried at every epoch, Experiment~3 requires substantially more wall time compared to Experiment~2. It also consumes about $78$ million tokens per inversion, since the complete history of the fifteen controls and their gradients is supplied to the agent at every epoch, so that the token consumption grows approximately quadratically with the number of epochs.

Experiment~4 addresses these limitations by growing the library in stages. Since a new family of basis functions is admitted only when the misfit stagnates with the present library, each family is admitted only after the controls of the present library have adjusted. Experiment~4 attains the smallest error in every polynomial and Fourier case and is the only experiment that recovers the Fourier profile at $D=10^{-6}$, with errors of $3.7$--$4.5\%$. The number of non-negligible terms in the recovered expression increases as the diffusivity decreases. For the polynomial profile at $D=10^{-4}$, the recovered expression contains exactly the two terms of the true envelope, and at lower diffusivities, one or two additional terms. For the Fourier profile, the number of active terms grows from three to as many as eight. For the hybrid profile, the recovered expression contains higher-order polynomials and Fourier functions beyond those of the true envelope at every diffusivity, with seven to twelve active terms compared with the four terms of the true envelope. Consequently, Experiment~4 is less accurate than Experiment~1 for the hybrid profile at $D=10^{-4}$, where the ideal library suffices. It should be noted that the objective is the recovery of the flux profile rather than of its exact expression, so that additional terms are acceptable. The token consumption is also reduced to about $3.3$ million per inversion, roughly $24$ times less than in Experiment~3, owing mainly to its smaller epoch budget of $75$ instead of $500$ epochs.

Beyond its accuracy, Experiment~4 is also robust across trials and data placements under the present admission policy. Across the five trials, its interquartile ranges remain below $0.9\%$, $1.3\%$, and $3.0\%$ for the polynomial, Fourier, and hybrid profiles, whereas those of Experiment~3 reach $6.4\%$ and $4.2\%$ for the Fourier and hybrid profiles. Its errors with inner and outer data differ by less than one percentage point in all cases. For Experiments~1--3, by contrast, the data placement can change the error considerably and without a consistent trend. In terms of wall time, Experiment~4 requires $16$--$114$\,min, a factor of $2$ to $21$ less than Experiment~3. Relative to Experiments~1 and~2, it is up to ten times slower at $D\geq10^{-5}$ but comparable at $D=10^{-6}$, with $49$--$81$\,min against $46$--$107$\,min. Finally, a randomly selected successful trial of each of Experiments~3 and~4 was replayed from its logged methods, controls, and library revisions without querying the agent for validation purposes, as detailed in~\ref{app:llm-basis}.

\section{Discussion}
\label{sec4}
In this work, we addressed two aspects of inverse modeling for transport problems with thin boundary layers: the cost of resolving the boundary layer in a forward solve, and the choice of the optimization method for inverse modeling with differentiable physics. With regard to the first aspect, the results show that, at high P\'eclet numbers, the cost associated with thin boundary layers can be reduced through the formulation of the problem rather than through near-wall mesh refinement. Posing the inner subdomain in the stretched coordinate of perturbation theory together with a two-way Schwarz coupling can relax the need for a high-resolution mesh and retains the interaction between the near-wall and bulk solutions. With regard to the second aspect, the results show that the benefit of language-model-based optimization depends on the complexity of the inverse problem. For the one-dimensional problem, the classical methods remained more efficient, although language-model-based configurations required less wall time than a tuned Adam. For the high-dimensional problems, the active agentic AI optimization recovered the wall flux in cases in which Adam alone did not. 


Our results suggest that the convergence of an inverse simulation can be accelerated when the optimization method is selected according to the progress in the problem solution rather than fixed in advance. This observation is relevant to a practical difficulty in inverse modeling. No single optimization method is suitable for all problems, and the performance of any method further depends on hyperparameters that are themselves problem-specific. In PDE-constrained inversion, every epoch can involve a forward and an adjoint solve, so that identifying a suitable method by comparing candidates over repeated inversions is computationally expensive. The method is therefore commonly chosen a priori, on the basis of experience or convention, and retained throughout the optimization, while it might not be suitable for the problem or, more crucially, for a given stage of the convergence. LLM-based optimization provides an alternative to this practice. In this approach, the choice of the optimization method (and possibly its hyperparameters) is revisited at every epoch on the basis of the history of the misfit and the adjoint gradient, and each choice can be accompanied by a logged justification that makes the decision traceable. For the staged library of the functional-basis identification, however, the improvement is more likely associated with the staged admission of basis functions and their initialization choice than with the selection of the optimization method during the inversion.

The assessment of the active agentic optimization is subject to three limitations. First, the agent relies on a proprietary language model accessed under academic pricing with a per-token charge, which restricted the study to five trials per case, compared with thirty trials for the locally served passive optimization. Second, the agentic AI results are not strictly reproducible. The passive optimization employed a fixed seed per trial, whereas the agent was executed without a seed; although some proprietary interfaces now accept a seed parameter, they guarantee deterministic outputs only on a best-effort basis, and the underlying model may change between versions~\cite{openai2023seed}. Third, in the functional basis identification, the candidate families and their order of admission were prescribed, and the agent was just in charge of initialization. A more ambitious approach where the agent itself decides how to grow the basis function remains to be investigated. These limitations suggest two directions for future research. The first concerns the choice of the language model. Since a single model was examined in each configuration, the dependence of the present findings on this choice remains to be established, and a systematic evaluation of several language models, including current state-of-the-art models, within the differentiable physics solver would quantify their ability to accelerate convergence of the inversion. The second concerns access to the agent. Whereas the passive optimization already operates with an open-source model served locally, the active optimization relied on a proprietary agent. The broader adoption of agentic optimization therefore depends on openly available agents that can be served locally, which would eliminate the per-token cost, permit a larger number of trials, and provide full control over the seed and the model version.

In addition, several extensions of the framework itself remain to be explored. Since the agent acts only on the optimization step, the agentic optimization is not expected to be inherently limited to two-dimensional problems or to a small number of trials, although its extension to three-dimensional problems has not been demonstrated in the present work. The present benchmark is two-dimensional and steady, and it involves only the slip-wall scaling of the boundary layer. The application of the Schwarz-coupled solver to no-slip walls with the L\'ev\^eque scaling formulated in Section~\ref{sec:forward-coupled}, to curved walls, and to three-dimensional and unsteady flows remains to be investigated. In the inverse setting, all recoveries in the present study were obtained from noise-free synthetic observations. Although the robustness of the underlying single-domain framework to measurement noise has been examined previously~\cite{Elhadidy_2026}, noisy and sparse observations are expected to pose a greater challenge for thin boundary layers. 

Finally, the basis admission policy, which was prescribed in the present work, could itself be proposed and revised by the agent, which would extend the agentic construction of the basis library toward data-driven model discovery. In SINDy, the unknown is represented as a combination of candidate functions drawn from a library fixed in advance~\cite{brunton2016discovering}, including variants in which the nonlinear parameters of the candidates are optimized jointly with their coefficients~\cite{viknesh2026adam}, so that a function absent from the library cannot be recovered. Symbolic regression (SR) removes this restriction by searching the space of expressions directly, and language models have recently been employed to guide this search~\cite{pang2026deliberate,islam2026discovery}, at the cost of a considerably larger search space. These considerations suggest an intermediate between the two approaches, an agentic SINDy, in which the agent decides when and which candidate functions to admit during the identification. It would retain the library-based structure and interpretability of SINDy while removing the bottleneck of a fixed library, without requiring the full search of SR. The present work takes only a first modest step in this direction, since the candidate functions and their order of admission were prescribed.

\section{Conclusion}
\label{sec5}
This study presented a differentiable multidomain Schwarz-coupled solver for the forward and inverse modeling of transport problems with thin boundary layers, in which an inner subdomain posed in stretched coordinates is coupled bidirectionally with an outer subdomain in physical coordinates. At high P\'eclet numbers, the solver resolves thin boundary layers on relatively coarse meshes. Through the discrete adjoint method, with the physics imposed as hard constraints, it enables efficient inversion from near-wall data and, with the overlap strategy, from data away from the wall. Language models were further integrated into the optimization step and compared to other optimization methods, and their advantage was demonstrated for recovering spatially varying parameters and for growing a basis-function library during the inversion under a prescribed admission policy.  These findings suggest that combining physics-based scale separation with language-model-based optimization offers a promising route toward accelerating the inverse modeling of multiscale transport problems.

\section*{Conflict of Interest}
The authors declare no conflict of interest.

\section*{Acknowledgments}
This work was supported by NSF Award No.~2247173.

\section*{Data Availability}
The Schwarz-coupled differentiable FEniCS solver will be made publicly available at
\url{https://github.com/siva-viknesh/Differentiable-FEM-Solver} following peer review.

\appendix
\section{Solver algorithms}
\label{app:algorithms}

This appendix summarizes the Schwarz-coupled forward solver and the adjoint-based inverse-recovery procedure developed in Sections~\ref{sec:forward} and~\ref{sec:inverse}, respectively. 

\subsection{Forward solver}
Algorithm~\ref{alg:forward} summarizes the construction, initialization, and alternating Schwarz iteration used to obtain the coupled inner--outer solution. 

\begin{algorithm}[H]
\caption{\textbf{Schwarz-Coupled Inner--Outer Forward Solver}}
\label{alg:forward}
\begin{algorithmic}
    \State \textbf{Input:}
    \Statex \quad Characteristic domain length: $L$, \; boundary layer thickness scale: $\delta$
    \Statex \quad Physical Cartesian coordinates $(x,y)$; \; stretched coordinates $\xi=x/L$, \; $\eta=y/\delta$
    \Statex \quad Wall flux profile: $g(x)$
    \Statex \quad Artificial boundaries: $\eta_{\mathrm{interface}}$ (outer bottom), \; $\eta_{\mathrm{top}}$ (inner top), \; with $\eta_{\mathrm{interface}}<\eta_{\mathrm{top}}$
    \Statex \quad Relaxation factor: $\omega \in (0,1]$
    \Statex \quad Convergence tolerance: $\varepsilon$, \; iteration limit: $K$
    \State \textbf{Domain Decomposition:}
    \Statex \quad Inner subdomain $\Omega_{\mathrm{in}}$: $\eta\in[0,\eta_{\mathrm{top}}]$, stretched coordinates $(\xi,\eta)$
    \Statex \quad Outer subdomain $\Omega_{\mathrm{out}}$: $y\ge\delta\eta_{\mathrm{interface}}$, physical coordinates $(x,y)$
    \Statex \quad Overlap $\Omega_{\mathrm{overlap}}=\Omega_{\mathrm{in}}\cap\Omega_{\mathrm{out}}$: $\eta_{\mathrm{interface}}\le\eta\le\eta_{\mathrm{top}}$, i.e.\ $\delta\eta_{\mathrm{interface}}\le y\le\delta\eta_{\mathrm{top}}$
    \State \textbf{Solver Construction:}
    \Statex \quad Discretize and factorize each subdomain operator once; reuse across all iterations
    \Statex \quad Precompute the transfer maps between subdomains via $x=L\xi$, $y=\delta\eta$
    \State \textbf{Initialization:}
    \Statex \quad Overlap field set to the far-field concentration: $C_{\mathrm{overlap}}^{(0)}$
    \State \textbf{Schwarz Iteration:}
    \For{$k = 1$ to $K$}
        \State \textbf{Step A -- Probe inner data:}
        \Statex \qquad Evaluate $C_{\mathrm{overlap}}^{(k-1)}$ at $\eta=\eta_{\mathrm{top}}$: $\;C_{\mathrm{overlap}}^{\mathrm{top}}$
        \State \textbf{Step B -- Inner solve:}
        \Statex \qquad Solve Eq.~\ref{eq:inner_pde} (slip wall) or Eq.~\ref{eq:inner_pde_noslip} (no-slip wall) on $\Omega_{\mathrm{in}}$
        \Statex \qquad with $\partial C/\partial\eta=\delta\,g(L\xi)$ at $\eta=0$ and $C=C_{\mathrm{overlap}}^{\mathrm{top}}$ at $\eta=\eta_{\mathrm{top}}$
        \Statex \qquad Restrict the inner solution to the overlap: $\;\widehat{C}_{\mathrm{overlap}}$
        \Statex \qquad Under-relax (Eq.~\ref{eq:relax}): $C_{\mathrm{overlap}} \gets \omega\,\widehat{C}_{\mathrm{overlap}} + (1-\omega)\,C_{\mathrm{overlap}}^{(k-1)}$
        \State \textbf{Step C -- Probe outer data:}
        \Statex \qquad Evaluate the updated $C_{\mathrm{overlap}}$ at $\eta=\eta_{\mathrm{interface}}$: $\;C_{\mathrm{overlap}}^{\mathrm{interface}}$
        \State \textbf{Step D -- Outer solve:}
        \Statex \qquad Solve the SUPG-stabilized Eq.~\ref{eq:adv_diff} on $\Omega_{\mathrm{out}}$ with $C=C_{\mathrm{overlap}}^{\mathrm{interface}}$ at $y=\delta\eta_{\mathrm{interface}}$
        \Statex \qquad Restrict the outer solution to the overlap: $\;\widehat{C}_{\mathrm{overlap}}$
        \Statex \qquad Under-relax (Eq.~\ref{eq:relax}): $C_{\mathrm{overlap}}^{(k)} \gets \omega\,\widehat{C}_{\mathrm{overlap}} + (1-\omega)\,C_{\mathrm{overlap}}$
        \State \textbf{Convergence Check:}
        \Statex \qquad $r \gets \big\|C_{\mathrm{overlap}}^{(k)}-C_{\mathrm{overlap}}^{(k-1)}\big\|_{\Omega_{\mathrm{overlap}}}$
        \Statex \qquad \textbf{if} $k > 1$ \textbf{and} $r < \varepsilon$ \textbf{then break}
    \EndFor
    \State \textbf{Output:} converged inner and outer solutions on $\Omega_{\mathrm{in}}$ and $\Omega_{\mathrm{out}}$, in agreement on $\Omega_{\mathrm{overlap}}$ to within $\varepsilon$
    \Statex \quad \textit{One-way method:} $K=1$, so that the inner solution is transferred to the outer subdomain, but the outer solution is never returned to the inner subdomain
\end{algorithmic}
\end{algorithm}

\subsection{Inverse solver}
Algorithm~\ref{alg:inverse} summarizes the adjoint-based optimization procedure built on the converged forward solution. 

\begin{algorithm}[H]
\caption{\textbf{Schwarz-Coupled Adjoint Inverse Recovery}}
\label{alg:inverse}
\begin{algorithmic}
    \State \textbf{Input:}
    \Statex \quad Measurements $C_{\mathrm{obs}}$ on the observation region $\Omega_{\mathrm{obs}}$
    \Statex \quad Unknown parameter $\theta$, initial guess $\theta^{0}$; \; update operator $\mathcal{O}$ (Methods 1--3); \; number of epochs $N$
    \State \textbf{Setup:}
    \Statex \quad Assemble and factorize the inner and outer operators
    \Statex \quad Register $\theta$ as the control and $J$ (Eq.~\ref{eq:misfit}) as the reduced functional
    \State \textbf{Optimization Loop:}
    \For{$n = 0$ to $N-1$}
        \State \textbf{Step 1 -- Forward evaluation:}
        \Statex \qquad Advance the Schwarz iteration (Algorithm~\ref{alg:forward}) at $\theta^{n}$ to convergence without recording
        \Statex \qquad Freeze the converged interface trace feeding the subdomain containing $\theta$
        \Statex \qquad Record a single directed evaluation of the coupled model with this trace held fixed \vspace{2mm}
        \Statex \hspace{\algorithmicindent}%
        \renewcommand{\arraystretch}{1.3}%
        \begin{tabular}{@{}l >{\raggedright\arraybackslash}p{0.36\linewidth} >{\raggedright\arraybackslash}p{0.36\linewidth}@{}}
        \toprule
        & \textbf{Common subdomain} & \textbf{Distinct subdomains} \\
        \midrule
        \textbf{Record}  & the $\theta$-subdomain solve & both solves, joined by $\mathbf{P}$ \\
        \textbf{Misfit}  & $J(\theta^{n})$ on $\Omega_{\mathrm{obs}}$ & $J(\theta^{n})$ on $\Omega_{\mathrm{obs}}$ \\
        \textbf{Adjoint} & closes locally $\Rightarrow \nabla_\theta J(\theta^{n})$ & crosses via $\mathbf{P}^{\!\top} \Rightarrow \nabla_\theta J(\theta^{n})$ \\
        \bottomrule
        \end{tabular}%
        \renewcommand{\arraystretch}{1.0}
        \Statex
        \State \textbf{Steps 2--3 -- Misfit and adjoint gradient:}
        \Statex \qquad Evaluate $J(\theta^{n})$ (Eq.~\ref{eq:misfit}) and $\nabla_\theta J(\theta^{n})$ by reverse-mode differentiation (Eq.~\ref{eq:adjoint})
        \State \textbf{Step 4 -- Parameter update:}
        \Statex \qquad $\theta^{n+1} \gets \mathcal{O}\big(J,\,\nabla_\theta J,\,\theta^{n}\big)$
    \EndFor
    \State \textbf{Output:} recovered parameter $\theta^{N}$
\end{algorithmic}
\end{algorithm}

\section{Supplementary Results}
\label{app:supplement_data}

This section reports the discrete reconstruction of the spatially varying wall flux with overlap data. The relative $L_\infty$ error and the wall time of the methods are presented in Table~\ref{tab:overlap_inverse_agentic}.


\begin{sidewaystable*}
\centering
\small
\setlength{\tabcolsep}{3pt}
\renewcommand{\arraystretch}{1.2}
\begin{tabular}{|*{11}{c|}}
\hline
\multicolumn{11}{|c|}{\textbf{Spatially varying wall flux: Discrete profile reconstruction -- Overlap data}}\\
\hline
\multirow{2}{*}{\textbf{Profile}} &
\multirow{2}{*}{\textbf{Init.}} &
\multicolumn{3}{c|}{\textbf{Single-domain -- Adam}} &
\multicolumn{3}{c|}{\textbf{Schwarz -- Adam}} &
\multicolumn{3}{c|}{\textbf{Schwarz -- Agentic}}\\
\cline{3-11}
& &
$\alpha^*$ & Rel.\ $L_\infty$ (\%) & Wall time (min) &
$\alpha^*$ & Rel.\ $L_\infty$ (\%) & Wall time (min) &
$\alpha_{\rm ref}$ & Rel.\ $L_\infty$ (\%) & Wall time (min)\\
\hline

\multicolumn{11}{|c|}{$\boldsymbol{D=10^{-4}}$}\\
\hline
\multirow{3}{*}{Polynomial}
& Flat & 1e-7 & 2.691 & 48.036 & 1e-8 & 2.292 & 24.494
& 1e-8 & 2.787 [0.556] & 61.680 [0.330] \\
\cline{2-11}
& P10 & 1e-6 & 1.485 & 45.983 & 5e-8 & 2.550 & 21.336
& 5e-8 & 3.097 [1.913] & 63.495 [6.171] \\
\cline{2-11}
& P20 & 5e-7 & 2.337 & 60.165 & 1e-8 & 2.785 & 19.499
& 1e-8 & 3.003 [0.090] & 55.248 [0.590] \\
\hline
\multirow{3}{*}{Fourier}
& Flat & 5e-8 & 16.179 & 49.277 & 1e-8 & 20.905 & 33.165
& 1e-8 & 9.167 [0.751] & 61.340 [0.817] \\
\cline{2-11}
& P10 & 1e-7 & 3.113 & 49.569 & 1e-6 & 2.740 & 27.017
& 1e-6 & 4.024 [0.348] & 55.892 [1.009] \\
\cline{2-11}
& P20 & 1e-7 & 2.887 & 48.818 & 5e-7 & 5.735 & 26.843
& 5e-7 & 8.077 [2.079] & 53.282 [3.707] \\
\hline
\multirow{3}{*}{Hybrid}
& Flat & 1e-7 & 14.471 & 49.663 & 5e-8 & 16.955 & 30.017
& 5e-8 & 13.157 [0.911] & 61.061 [1.021] \\
\cline{2-11}
& P10 & 1e-6 & 1.028 & 89.273 & 5e-7 & 0.704 & 27.436
& 5e-7 & 3.127 [0.054] & 54.512 [1.594] \\
\cline{2-11}
& P20 & 1e-7 & 2.996 & 59.883 & 1e-7 & 2.943 & 25.965
& 1e-7 & 4.086 [1.092] & 51.464 [3.220] \\
\hline

\multicolumn{11}{|c|}{$\boldsymbol{D=10^{-5}}$}\\
\hline
\multirow{3}{*}{Polynomial}
& Flat & 1e-8 & 47.430 & 50.892 & 1e-8 & 44.744 & 33.385
& 1e-8 & 19.294 [5.307] & 57.106 [0.959] \\
\cline{2-11}
& P10 & 1e-7 & 2.542 & 51.159 & 1e-7 & 2.771 & 31.088
& 1e-7 & 3.906 [1.193] & 51.371 [2.009] \\
\cline{2-11}
& P20 & 1e-7 & 3.416 & 50.782 & 1e-7 & 4.880 & 32.857
& 1e-7 & 5.753 [1.931] & 54.347 [1.227] \\
\hline
\multirow{3}{*}{Fourier}
& Flat & 1e-9 & 48.974 & 51.036 & 1e-9 & 50.896 & 26.739
& 1e-9 & 15.486 [2.877] & 56.619 [1.794] \\
\cline{2-11}
& P10 & 1e-7 & 5.381 & 51.280 & 1e-7 & 3.249 & 50.327
& 1e-7 & 4.057 [0.692] & 51.728 [0.337] \\
\cline{2-11}
& P20 & 1e-7 & 9.012 & 51.421 & 1e-7 & 8.812 & 52.911
& 1e-7 & 8.793 [0.920] & 58.256 [4.115] \\
\hline
\multirow{3}{*}{Hybrid}
& Flat & 1e-9 & 31.980 & 59.788 & 1e-9 & 33.687 & 32.962
& 1e-9 & 11.205 [0.800] & 57.626 [3.273] \\
\cline{2-11}
& P10 & 1e-7 & 2.538 & 65.945 & 1e-7 & 1.252 & 48.591
& 1e-7 & 2.227 [0.086] & 49.303 [0.810] \\
\cline{2-11}
& P20 & 1e-7 & 4.642 & 71.361 & 1e-7 & 4.167 & 50.878
& 1e-7 & 5.386 [0.937] & 56.045 [0.616] \\
\hline

\multicolumn{11}{|c|}{$\boldsymbol{D=10^{-6}}$}\\
\hline
\multirow{3}{*}{Polynomial}
& Flat & 1e-9 & 51.845 & 378.316 & 1e-9 & 51.571 & 177.348
& 1e-9 & 49.885 [6.864] & 69.313 [4.802] \\
\cline{2-11}
& P10 & 5e-8 & 9.275 & 380.690 & 5e-8 & 9.484 & 160.746
& 5e-8 & 8.488 [0.557] & 52.114 [2.089] \\
\cline{2-11}
& P20 & 5e-8 & 18.304 & 381.471 & 5e-8 & 18.853 & 171.227
& 5e-8 & 18.141 [0.770] & 50.404 [0.017] \\
\hline
\multirow{3}{*}{Fourier}
& Flat & 1e-9 & 54.168 & 219.523 & 1e-9 & 54.371 & 93.101
& 1e-9 & 50.908 [0.287] & 70.562 [2.256] \\
\cline{2-11}
& P10 & 1e-7 & 9.674 & 223.528 & 1e-7 & 9.659 & 95.743
& 1e-7 & 9.930 [1.880] & 55.268 [2.330] \\
\cline{2-11}
& P20 & 1e-7 & 19.260 & 217.978 & 1e-7 & 19.453 & 100.433
& 1e-7 & 18.897 [0.080] & 55.290 [0.997] \\
\hline
\multirow{3}{*}{Hybrid}
& Flat & 1e-9 & 53.800 & 212.492 & 1e-9 & 54.892 & 84.974
& 1e-9 & 30.392 [0.417] & 68.286 [1.258] \\
\cline{2-11}
& P10 & 1e-7 & 9.563 & 224.159 & 1e-7 & 9.374 & 87.823
& 1e-7 & 6.905 [0.288] & 65.292 [8.157] \\
\cline{2-11}
& P20 & 1e-7 & 18.776 & 223.789 & 1e-7 & 18.845 & 101.212
& 1e-7 & 8.496 [1.163] & 61.472 [4.679] \\
\hline
\end{tabular}%

\caption{Discrete reconstruction of the spatially varying wall flux with overlap data for the polynomial, Fourier, and hybrid profiles, started from the Flat, P10, and P20 initial estimates. For the single-domain and the Schwarz-coupled solver with Adam ($2000$ epochs), $\alpha^*$ is the optimal regularization parameter of the sweep. The active agentic AI optimization ($300$ epochs, five trials) uses $\alpha_{\rm ref}=\alpha^*$ of the Schwarz-coupled solver with Adam and is reported as median [interquartile range]. The relative $L_\infty$ error is the peak error relative to the maximum magnitude of the true profile, and the wall time covers all epochs, including the inference of the language model for the agent.}
\label{tab:overlap_inverse_agentic}
\end{sidewaystable*}

\section{Validation of the language-model-based optimizations}
\label{app:llm-validation}

\begin{figure}[htbp]
\centering
\includegraphics[width=\textwidth]{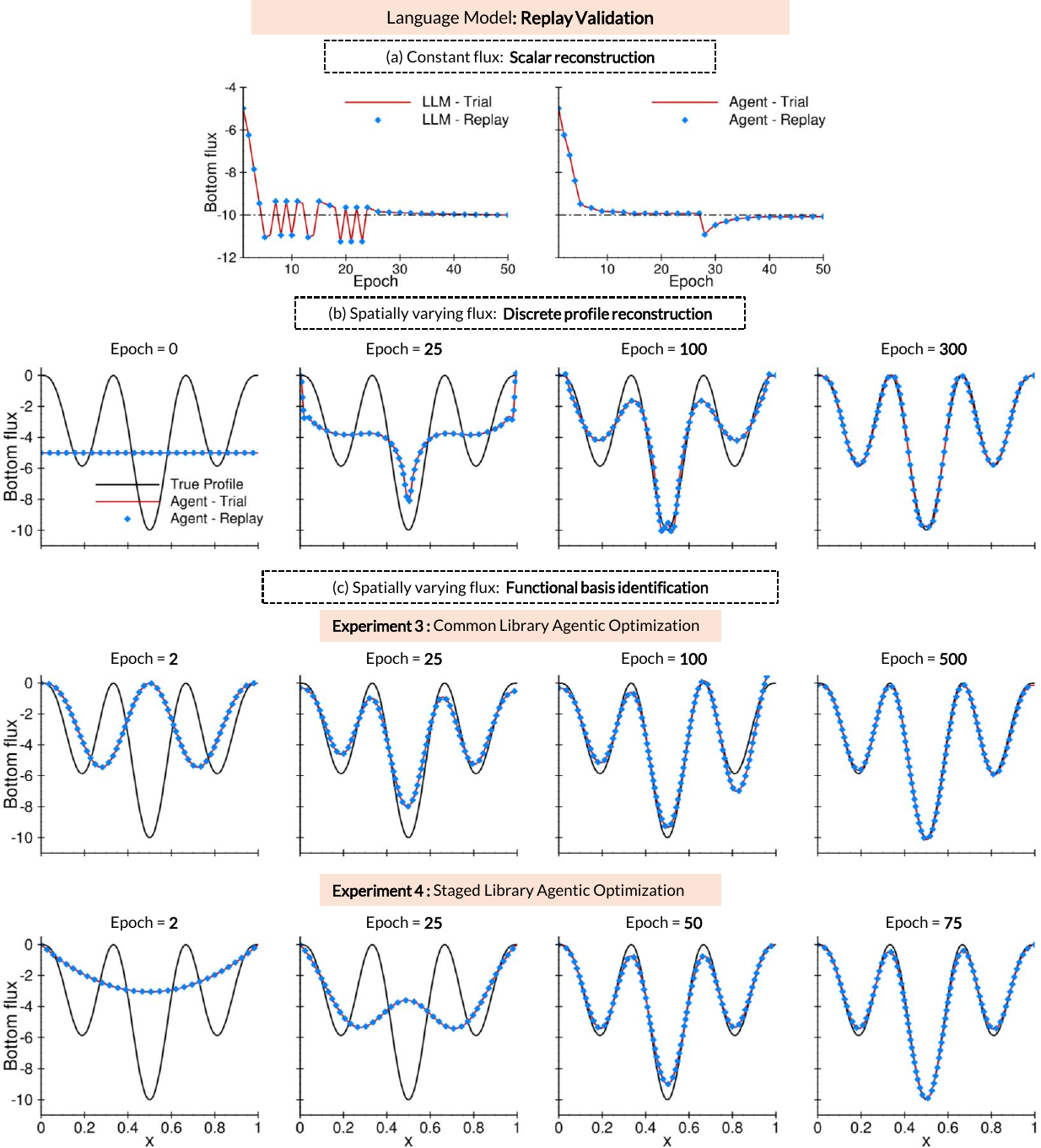}
\caption{Replay validation of the selected successful trials. (a)~Constant wall flux: recovered wall flux against the epoch for the passive LLM optimization (left; overlap data, $D=10^{-5}$, M1) and the active agentic optimization (right; overlap data, $D=10^{-6}$, M2), with the true value $g=-10$ (dash-dotted); the first $50$ of $150$ epochs are shown. (b)~Discrete reconstruction of the hybrid profile from the Flat initialization (inner data, $D=10^{-5}$), at epochs $0$, $25$, $100$, and $300$. (c)~Functional basis identification of the hybrid profile with inner data, with the common library (Experiment~3, $D=10^{-4}$), at epochs $2$, $25$, $100$, and $500$, and with the library grown by the agent (Experiment~4, $D=10^{-6}$), at epochs $2$, $25$, $50$, and $75$. Red curves show the logged trial, blue markers the replay without querying the language model, and black curves the true profile.}
\label{fig:app-validation}
\end{figure}

The optimizations based on language models are probabilistic. Their validation therefore addresses a single question: does every update of the wall flux result from a method of the optimization policy, executed with the controls returned by the language model, or could the language model have supplied a value of the wall flux directly? This appendix addresses this question for different wall flux recoveries. In each case, one successful trial is selected randomly and replayed. At every epoch of the replay, the update is recomputed from the logged method and controls without querying the language model, and the forward and adjoint problems are solved. The replayed trajectory is then compared with that of the trial, as shown for all three recoveries in Fig.~\ref{fig:app-validation}. In all tables of this appendix, the relative error of the recovered wall flux is evaluated offline, after the trial, from the logged controls and the true wall flux. For all selected trials, the replayed trajectory follows that of the trial closely at every epoch shown (Fig.~\ref{fig:app-validation}), including the oscillations, transient departures, and library revisions, which confirms that every update results from the logged method and controls rather than from values supplied by the language model.

\subsection{Constant flux: scalar reconstruction}
\label{app:llm-const}


The constant wall flux $g$ is recovered with the true value $-10$, the initial estimate $-5$, and the bounds $[-20,20]$, over a budget of $150$ epochs, on M1 at $D=10^{-4}$ and $10^{-5}$ and on M2 at $D=10^{-6}$, with inner data and with outer data under the overlap strategy. In both configurations, the language model is coupled to the Schwarz-coupled solver through a controller that contains the optimization policy and executes the update.  

The language model returns a structured response containing one available method, the damping, the step fraction, the history window, and a reasoning of at most $30$ words. The damping and the step fraction are limited to $[0.02,1]$, and the history window to between $2$ and $10$ epochs. The response contains no field for the value of the wall flux, and both prompts instruct the language model explicitly not to propose one. The language model can therefore influence the wall flux only through the selected method and its controls. From these, the controller computes the proposal of the selected method, with the step scaled by the product of the damping and the step fraction. It then restricts the proposal to the trust region and to the bounds. The trust radius is half the current magnitude of the wall flux, limited to between $0.25$ and $2.5$, so that the change of the wall flux in a single update is proportional to its current magnitude. The updated wall flux is returned to the solver for the next forward and adjoint solve.

Each epoch evaluates one value of the wall flux, starting from the initial estimate at epoch~1. The update at epoch~1 is a deterministic step executed by the controller, which yields $g=-6.25$. It is not a decision of the language model, and it provides the second pair of wall flux and gradient required by the methods that use several entries of the epoch history. The language model therefore selects the updates at epochs $2$ to $149$, and epoch~$150$ evaluates the final wall flux. 

\subsubsection*{Optimization policy}
\label{app:llm-const-policy}

The optimization policy consists of nine gradient-based methods, shared by the passive LLM optimization and the active agentic optimization, which fall into two groups. The gradient-step methods move the wall flux downhill by a step scaled with the damping and step fraction returned by the language model. They comprise (a)~the normalized gradient method, (b)~Adam, and (c)~momentum, whose moments and step directions are maintained by the controller, and (d)~the damped Newton method. The target-point methods estimate a stationary point of the misfit from the epoch history and move the wall flux toward it. They comprise (e)~the secant method, which takes the root of the line through the gradients of the two most recent epochs, (f)~the linear regression of the gradient, which takes the root of a least-squares line through the gradients over the selected history window, and (g)~the quadratic fit of the misfit, which takes the vertex of the parabola through the misfits of the three most recent epochs. These target points are defined only if the epoch history contains at least two different values of the wall flux for the secant method and the linear regression, and three for the quadratic fit, and if the estimated slope of the gradient or curvature of the misfit is nonzero. The quadratic fit additionally requires a positive curvature, so that its vertex is a minimum rather than a maximum of the misfit. They also comprise (h)~the bracketed root finding of the gradient and (i)~the bracketed golden-section refinement of the misfit, which require an observed sign change of the gradient or an observed bracket of the minimum of the misfit, respectively.  The Hessian of the damped Newton method is computed by the controller only after this method is selected, safeguarded against non-positive and vanishing values, and not passed to the language model. A reference to a Newton step in the logged reasoning therefore does not imply that the language model observed a second derivative.

\subsubsection*{Passive and active configurations}
\label{app:llm-const-configs}

Within this common framework, the two configurations were kept as similar as possible, so that their results are directly comparable. They differ only in the aspects listed in Table~\ref{tab:app-const-configs}. The synthetic observations are generated separately by the single-domain solver on M4, and the agent can access only the resulting data file, which is required for the misfit computation. 

\begin{table}[!ht]
\centering
\small
\begin{tabular}{@{}lp{0.36\textwidth}p{0.36\textwidth}@{}}
\toprule
 & Passive LLM optimization & Active agentic optimization \\
\midrule
Language model     & Qwen3-8B, served locally through Ollama & GPT~5.6 Terra, through Codex \\
Sampling           & Temperature $0.5$, fixed seed per trial & No seed \\
Trials per case    & $30$ & $5$, each in a fresh project \\
Access to solver   & None & Read access to the inverse and adjoint solvers; write access only to the definition of $\theta$, not exercised in this recovery \\
System instruction & None & Role as an auditable controller, confined to the optimization step (Table~\ref{tab:app-const-prompts}) \\
Task instruction   & Selection of a method and its controls & As for the passive configuration, restricted to the observed epoch history \\
Epoch history      & Wall flux, misfit, and adjoint gradient of the $10$ most recent epochs & Wall flux, misfit, adjoint gradient, and the previously selected method, of all preceding epochs \\
\bottomrule
\end{tabular}
\caption{Differences between the passive LLM optimization and the active agentic AI optimization for the constant wall flux recovery. All other aspects are common to both.}
\label{tab:app-const-configs}
\end{table}

\subsubsection*{Prompts}

The prompts of the passive LLM optimization and the active agentic optimization are compared block by block in Table~\ref{tab:app-const-prompts}.  The prompts differ in three respects. Since the agent has access to the inverse and adjoint solvers, it alone receives a system instruction (block 1), which confines it to the optimization step: it must not modify the solvers or the inverse workflow, and it acts only after the forward and adjoint computations of each epoch are complete. Its task instruction (block 2) restricts its decision to the observed epoch history, and its epoch history (block 3) contains all preceding epochs rather than the $10$ most recent ones, together with the column \texttt{prior{\char95}action}, which reports the method selected at each preceding epoch. Both configurations therefore select their methods within the same optimization policy and from the same type of epoch history, the agent with access to a longer record.

\begin{table}[!ht]
\centering
\scriptsize
\renewcommand{\arraystretch}{1.25}
\begin{tabular}{@{}>{\raggedright\arraybackslash}p{0.12\textwidth}p{0.41\textwidth}p{0.41\textwidth}@{}}
\toprule
\textbf{Block} & \textbf{Passive LLM optimization} & \textbf{Active agentic optimization} \\
\midrule
1. System instruction &
\itshape None. &
\prompttt You are an auditable controller for a one-dimensional numerical inverse problem. Do not modify the forward solver, the adjoint solver, or the inverse workflow. Act only after the forward and adjoint computations of each epoch are complete, and only on the optimization step. Select exactly one available numerical action and bounded controls from the observed history only. The true parameter is unavailable. Do not invent a parameter value; deterministic code will compute it. Return only the requested JSON object. \\
\midrule
2. Task instruction &
\prompttt Select the next controller action for a scalar inverse problem. The supplied objective and gradient are raw physical quantities; do not assume a particular numerical scale. Do not invent a parameter value. Choose one available action and bounded controls that are likely to produce stable objective reduction and a stationary parameter. &
\prompttt Control one scalar inverse parameter using only the observed history below. The supplied objective and gradient are raw physical quantities; do not assume a particular numerical scale. Choose one available action and bounded controls that are likely to produce stable objective reduction and a stationary parameter. \\
\midrule
3. Epoch history &
\prompttt Observed history, oldest to newest (latest 10 epochs):\newline epoch | G | J{\char95}raw | dJ{\char95}dG\newline {[\ldots]} &
\prompttt Observed history, oldest to newest (all epochs):\newline epoch | G | J{\char95}raw | dJ{\char95}dG | prior{\char95}action\newline {[\ldots]} \\
\midrule
4. Bounds and trust radius &
\multicolumn{2}{p{0.84\textwidth}}{{\normalfont\itshape Common to both:}\newline \prompttt Current hard bounds: [-20, 20].\newline Current trust radius: [r{\char95}n].} \\
\midrule
5. Optimization policy &
\multicolumn{2}{p{0.84\textwidth}}{{\normalfont\itshape Common to both:}\newline \prompttt Available actions:\newline
- normalized{\char95}gradient: normalized downhill step of the trust radius\newline
- adam: controller Adam moment update\newline
- momentum{\char95}gradient: normalized downhill step with persistent momentum\newline
- damped{\char95}newton: controller-computed reduced-Hessian step\newline
- secant{\char95}gradient{\char95}root: two-point root estimate of the gradient\newline
- gradient{\char95}linear{\char95}regression: least-squares fit of gradient vs.\ G\newline
- objective{\char95}quadratic: three-point quadratic fit of the objective\newline
- bracketed{\char95}gradient{\char95}root: root inside an observed gradient-sign bracket\newline
- objective{\char95}bracket{\char95}refine: golden-ratio refinement of an objective bracket} \\
\midrule
6. Output schema &
\multicolumn{2}{p{0.84\textwidth}}{{\normalfont\itshape Common to both:}\newline \prompttt Return exactly one JSON object. action must be available. damping and step{\char95}fraction must be in [0.02, 1.0]. history{\char95}window must be 2 to 10. reasoning must be a concise justification based on the observed history, at most 30 words.} \\
\bottomrule
\end{tabular}
\caption{Prompts of the passive LLM optimization and the active agentic optimization for the constant wall flux recovery, compared block by block. Blocks 1--3 differ between the two configurations, and blocks 4--6, spanning both columns, are common to both. The block headings are added for comparison and are not part of the prompts. In the prompts, \texttt{G}, \texttt{J{\char95}raw}, and \texttt{dJ{\char95}dG} denote the wall flux, the misfit, and its adjoint gradient, and ``one-dimensional'' refers to the single unknown parameter. Fields in brackets are filled by the controller at every epoch. All nine methods of the optimization policy (Section~\ref{app:llm-const-policy}) are shown; at each epoch, the controller lists only those that are available.}
\label{tab:app-const-prompts}
\end{table}

\subsubsection*{Selected trials}

For the passive LLM optimization, one successful trial was drawn at random, namely the trial with overlap data at $D=10^{-5}$ on M1. For the active agentic optimization, the most demanding case was chosen, namely overlap data at $D=10^{-6}$ on M2, and one successful trial was drawn at random from its five trials. Table~\ref{tab:app-const-trials} lists selected epochs at which each trajectory changes character, namely the first decision, selected switches of method, and the approach to the final interval within the $1\%$ band. For the passive trial, it also lists the last decision. For the active trial, it lists the decision at epoch~$55$, after which the wall flux remains essentially stationary until the end of the budget.

\begin{table}[!ht]
\centering
\scriptsize
\setlength{\tabcolsep}{3pt}
\renewcommand{\arraystretch}{1.2}
\begin{tabular}{@{}c >{\centering\arraybackslash}p{0.15\textwidth} >{\centering\arraybackslash}p{0.17\textwidth} c p{0.40\textwidth} c@{}}
\toprule
Epoch & Method & Update & Error & \multicolumn{1}{c}{Logged reasoning} & Ratio \\
$n$ & & $g_n\rightarrow g_{n+1}$ & $\varepsilon_{n+1}$ (\%) & & $J_{n+1}/J_n$ \\
\midrule
\multicolumn{6}{c}{\textbf{(a) Passive LLM optimization} (overlap data, $D=10^{-5}$, M1)} \\
\midrule
2   & Normalized gradient & $-6.2500\rightarrow-7.8500$ & 21.5000 & Gradient descent with damping is stable given the recent trend in objective values and gradient magnitude. & 0.327 \\
6   & Normalized gradient & $-10.9500\rightarrow-9.3500$ & 6.5000 & Gradient descent with damping follows the negative gradient direction to move downhill. & 0.434 \\
26  & Damped Newton & $-9.8500\rightarrow-9.8616$ & 1.3840 & Recent evaluations show oscillation around a local minimum. Damped Newton with small damping can help stabilize convergence. & 0.948 \\
30  & Damped Newton & $-9.8926\rightarrow-9.9017$ & 0.9826 & Gradient and objective are improving; Newton step with damping ensures stability. & 0.968 \\
149 & Damped Newton & $-10.0507\rightarrow-10.0507$ & 0.5069 & Gradient and objective values show consistent decrease, suggesting Newton step is viable with damping for stability. & 1.000 \\
\midrule
\multicolumn{6}{c}{\textbf{(b) Active agentic optimization} (overlap data, $D=10^{-6}$, M2)} \\
\midrule
2  & Secant & $-6.2500\rightarrow-7.1885$ & 28.1150 & Objective fell and the positive gradient decreased across both downhill evaluations; a damped two-point secant root step should continue reduction while limiting extrapolation within the trust radius. & 0.557 \\
4  & Damped Newton & $-8.3857\rightarrow-9.4771$ & 5.2290 & Objective has decreased monotonically while positive gradients shrink; recent gradient slopes imply locally positive curvature. A damped, trust-limited Newton step should reduce $J$ without overshooting the stationary region. & 0.125 \\
13 & Linear regression of the gradient & $-9.8762\rightarrow-9.9334$ & 0.6660 & Recent gradients decrease smoothly and remain positive as $g$ decreases, while objective reduction is flattening. A short-window gradient regression gives a stable local root trend; fractional damping avoids overshoot. & 0.989 \\
42 & Linear regression of the gradient & $-10.0853\rightarrow-10.0852$ & 0.8520 & Recent gradients remain negative but shrink smoothly toward zero as $J$ decreases; a short-window gradient fit gives a conservative root-directed update while limiting extrapolation near stationarity. & 1.000 \\
55 & Linear regression of the gradient & $-10.0811\rightarrow-10.0811$ & 0.8105 & Recent objective values decrease monotonically while the negative gradient shrinks smoothly toward zero; a short local regression estimates the stationary point while minimum trust clipping limits overshoot. & 1.000 \\
\bottomrule
\end{tabular}
\caption{Selected epochs of the selected successful trials of (a)~the passive LLM optimization and (b)~the active agentic optimization for the constant wall flux recovery. The update is the transition computed by the controller from the selected method of the optimization policy and its logged controls, $\varepsilon_{n+1}$ is the relative error of the wall flux, computed after the trial and not available to the language model, and $J_{n+1}/J_n$ is the observed misfit ratio.}
\label{tab:app-const-trials}
\end{table}

\textit{Passive LLM optimization:} The language model selects the normalized gradient method at epochs $2$ to $25$. This method reduces the relative error from $37.5\%$ after the deterministic first update to $6.5\%$ after epoch~6, but it overshoots the true value, so that the wall flux oscillates about the true value up to epoch~25 (Fig.~\ref{fig:app-validation}a, left). Once the epoch history contains repeated sign changes of the gradient, the language model switches to the damped Newton method at epoch~26, with a small damping, and retains it until the last decision. The update at epoch~26 reduces the error to $1.384\%$, and the update at epoch~30 yields $g=-9.90174$, the first value of the final interval within the $1\%$ band, with an error of $0.9826\%$. The wall flux remains within this band until the final epoch, ending at $g=-10.05069$ with a relative error of $0.5069\%$. The last update changes the wall flux by only $8\times10^{-6}$, so that the misfit ratio of $1.000$ reflects a stationary estimate. Of the $148$ decisions, one at each epoch from $2$ through $149$, the language model thus selects the normalized gradient method in $24$ and the damped Newton method in $124$, consistent with the selection mix of Table~\ref{tab:qwen-classical-codex-work-comparison}.

\textit{Active agentic optimization:} In contrast to the passive LLM optimization, the agent relies mainly on methods that combine several entries of the epoch history, namely the secant method and the linear regression of the gradient, consistent with the selection mix of Table~\ref{tab:qwen-classical-codex-work-comparison}. The secant and damped Newton steps of epochs $2$ to $4$ reduce the relative error from $37.5\%$ to $5.229\%$, and the regression of the gradient brings the error below $1\%$ at epoch~13. The wall flux thus approaches the true value without oscillation (Fig.~\ref{fig:app-validation}a, right). This crossing is transient, however. The update at epoch~27 moves the wall flux to $g=-10.9180$ and raises the error to $9.18\%$, which is visible as the jump at epoch~28 in the figure. The agent then corrects its trajectory by selecting bracketed root finding and the regression of the gradient from the optimization policy, and by epoch~42 the wall flux lies within the final interval of the $1\%$ band, with an error of $0.852\%$. The decision at epoch~$55$ changes the wall flux by only $5\times10^{-6}$ and yields $g=-10.08105$ with a relative error of $0.8105\%$, after which the wall flux remains essentially stationary until epoch~$150$. 


\subsection{Spatially varying flux: discrete profile reconstruction}
\label{app:llm-discrete}


The discrete profile reconstruction recovers the wall flux at every wall node. At each epoch, the agent receives the prompt (Table~\ref{tab:app-discrete-prompt}), the available methods of the optimization policy, the current nodal profile of the wall flux and its adjoint gradient, the history of the misfit, and the previously selected methods. It returns one available method, its controls, and a brief reasoning. The controller then computes the update of the entire profile, with the regularization parameter fixed at $\alpha_{\mathrm{ref}}$. As before, the agent has access neither to the true profile nor to the error of the recovered profile, and it cannot modify the inverse workflow.

\subsubsection*{Prompt}

The prompt is shown in Table~\ref{tab:app-discrete-prompt} and follows the structure of the active prompt in Table~\ref{tab:app-const-prompts}. The task instruction (block~2) refers to the entire discrete profile and specifies the fixed regularization weight, the epoch history (block~3) contains the current nodal profile and its adjoint gradient, together with the history of the misfit, and the optimization policy (block~5) comprises methods that update the entire profile simultaneously.

\begin{table}[!ht]
\centering
\scriptsize
\renewcommand{\arraystretch}{1}
\begin{tabular}{@{}>{\raggedright\arraybackslash}p{0.2\textwidth}p{0.75\textwidth}@{}}
\toprule
\textbf{Block} & \textbf{Active agentic optimization: discrete profile reconstruction} \\
\midrule
1. System instruction &
\prompttt You are an auditable controller for a discretized numerical inverse problem. Do not modify the forward solver, the adjoint solver, or the inverse workflow. Act only after the forward and adjoint computations of each epoch are complete, and only on the optimization step. Select exactly one available numerical action and bounded controls from the observed history only. The true profile is unavailable. Do not invent nodal values of the profile; deterministic code will compute them. Return only the requested JSON object. \\
\midrule
2. Task instruction &
\prompttt Control the discrete wall flux profile, consisting of [N] nodal values, using only the observed history below. The supplied objective and gradient are raw physical quantities; do not assume a particular numerical scale. The objective includes a curvature regularization with fixed weight alpha{\char95}ref = [alpha{\char95}ref]. Choose one available action and bounded controls that are likely to produce stable objective reduction and a stationary profile. \\
\midrule
3. Epoch history &
\prompttt Current profile: g = [nodal values]\newline Current gradient: dJ{\char95}dg = [nodal values]\newline Observed history, oldest to newest (all epochs):\newline epoch | J{\char95}raw | prior{\char95}action\newline {[\ldots]} \\
\midrule
4. Bounds and trust radius &
\prompttt Current hard bounds: [bounds on the nodal values].\newline Current trust radius: [r{\char95}n]. \\
\midrule
5. Optimization policy &
\prompttt Available actions:\newline
- adam: controller Adam moment update of the profile\newline
- normalized{\char95}gradient: normalized downhill step of the profile within the trust radius\newline
- momentum{\char95}gradient: normalized downhill step with persistent momentum\newline
- barzilai{\char95}borwein: step with the Barzilai--Borwein step length from the two most recent epochs\newline
- conjugate{\char95}gradient: step along a conjugate direction of the recent gradients\newline
- projected{\char95}gradient: gradient step projected onto the hard bounds \\
\midrule
6. Output schema &
\prompttt Return exactly one JSON object. action must be available. damping and step{\char95}fraction must be in [0.02, 1.0]. reasoning must be a concise justification based on the observed history, at most 30 words. \\
\bottomrule
\end{tabular}
\caption{Prompt of the active agentic optimization for the discrete reconstruction of the spatially varying wall flux. The block headings are added for clarity and are not part of the prompt. Fields in brackets are filled by the controller at every epoch and only available methods are listed.}
\label{tab:app-discrete-prompt}
\end{table}

\subsubsection*{Selected trial}

The trial was drawn at random from the five trials for the hybrid profile initialized from the Flat estimate, with inner data at $D=10^{-5}$ on M1, for which the regularization parameter is $\alpha_{\mathrm{ref}}=10^{-9}$ (Table~\ref{tab:inner_inverse_agentic}). This case is among the most demanding of the discrete reconstruction, since the flat initial estimate differs entirely from the true profile and Adam alone does not recover it. Table~\ref{tab:app-discrete-trial} lists the first decision, the changes of method, and selected epochs at which the course of the recovery changes, including the epochs shown in Fig.~\ref{fig:app-validation}b.

\begin{table}[!ht]
\centering
\scriptsize
\setlength{\tabcolsep}{7.5pt}
\renewcommand{\arraystretch}{1.3}
\begin{tabular}{@{}c >{\centering\arraybackslash}p{0.15\textwidth} c c p{0.44\textwidth} c@{}}
\toprule
Epoch & Method & Update & Error & \multicolumn{1}{c}{Logged Reasoning} & Ratio \\
$n$ & & $\mathrm{RMS}(\Delta g_n)$ & $\varepsilon_{n+1}$ (\%) & & $J_{n+1}/J_n$ \\
\midrule
\multicolumn{6}{c}{\textbf{Discrete reconstruction} (hybrid profile, Flat initialization, inner data, $D=10^{-5}$, M1)} \\
\midrule
1   & Normalized gradient & $1.245\times10^{-1}$ & 55.192 & Normalized-gradient probe from the flat profile that establishes the history of iterates and gradients required by the subsequent methods. & 0.968 \\
5   & Barzilai--Borwein & $1.915\times10^{-2}$ & 52.003 & Pairs of successive iterates and gradients are available, and the objective is decreasing; a damped Barzilai--Borwein step uses the newest pair without destabilizing the profile. & 0.987 \\
10  & Barzilai--Borwein & $2.076\times10^{-2}$ & 47.947 & Barzilai--Borwein has reduced the objective at every epoch, with declining gradient norms; the step-length estimate remains positive, so a conservative step fraction preserves the stable descent. & 0.983 \\
25  & Momentum gradient & $1.987\times10^{-1}$ & 36.024 & The objective decreases slowly under small steps; a bounded momentum-gradient step averages the recent descent directions and reduces dependence on the step-length estimate while maintaining directional continuity. & 0.749 \\
50  & Barzilai--Borwein & $5.653\times10^{-3}$ & 29.024 & The objective and gradient norms decline consistently under Barzilai--Borwein; use the minimum step fraction to limit disruption of the profile as recent improvements diminish. & 0.992 \\
75  & Normalized gradient & $7.014\times10^{-3}$ & 26.923 & Recent normalized-gradient steps consistently lowered the objective; still-elevated gradients favor a damped step within a small trust radius to preserve stability. & 0.999 \\
105 & Adam & $3.244\times10^{-2}$ & 17.026 & Gradients are elevated and spatially nonuniform; a small Adam step provides a node-wise scaled update, strictly bounded to avoid amplifying variations of the profile. & 0.943 \\
200 & Barzilai--Borwein & $2.048\times10^{-2}$ & 7.287 & Barzilai--Borwein has produced a consistent reduction of the objective, with falling gradient norms; a damped step of moderate size preserves this stable trend. & 0.945 \\
300 & Barzilai--Borwein & $1.852\times10^{-3}$ & 4.295 & Barzilai--Borwein has delivered a monotonic reduction of the objective with falling gradient norms; use a conservative damped step fraction as gradients and improvements diminish. & -- \\
\bottomrule
\end{tabular}
\caption{Selected epochs of the selected successful trial of the discrete reconstruction. The update is computed by the controller from the selected method of the optimization policy and its logged controls, and its size is the RMS of the nodal update over all $257$ nodal values. $\varepsilon_{n+1}$ is the relative $L_\infty$ error of the recovered profile, computed after the trial (not available to the agent), and $J_{n+1}/J_n$ is the observed misfit ratio.}
\label{tab:app-discrete-trial}
\end{table}

Among the listed epochs, the agent selects four of the six methods of the optimization policy, with the Barzilai--Borwein method selected most often. Following the initial normalized-gradient step, the Barzilai--Borwein method reduces the misfit steadily, but with small updates, so that the error decreases only from $55.2\%$ to $47.9\%$ by epoch~10. At epoch~25, the agent selects the momentum gradient method, which yields the largest update among the listed epochs, reduces the misfit by $25\%$ (misfit ratio $0.749$), and initiates the formation of the central minimum of the profile (Fig.~\ref{fig:app-validation}b, epoch~25). Between epochs~50 and~75, the reduction of the misfit becomes marginal, with misfit ratios of $0.992$ and $0.999$, and the error decreases only from $29.0\%$ to $26.9\%$. By epoch~100, the central minimum is recovered, whereas the outer minima remain too shallow (Fig.~\ref{fig:app-validation}b, epoch~100). At epoch~105, the agent selects Adam, whose node-wise scaling of the step resumes the reduction of the misfit, with a misfit ratio of $0.943$, and decreases the error to $17.0\%$. The agent subsequently returns to the Barzilai--Borwein method, under which the outer minima approach those of the true profile, and the trial concludes with an error of $4.295\%$ (Fig.~\ref{fig:app-validation}b, epoch~300). This value is consistent with the median of $4.203\%$ over the five trials (Table~\ref{tab:inner_inverse_agentic}), whereas the Schwarz-coupled solver with Adam alone yields $32.9\%$ for the same case. The trial, therefore, indicates that changing the method when the reduction of the misfit becomes marginal enables the recovery of the profile from a distant initial estimate.


\subsection{Spatially varying flux: functional-basis identification}
\label{app:llm-basis}

\begin{table}[!ht]
\centering
\scriptsize
\renewcommand{\arraystretch}{1.}
\begin{tabular}{@{}>{\raggedright\arraybackslash}p{0.12\textwidth}p{0.41\textwidth}p{0.41\textwidth}@{}}
\toprule
\textbf{Block} & \textbf{Experiment 3: common library} & \textbf{Experiment 4: library grown by the agent} \\
\midrule
1. System instruction &
\prompttt You are an auditable controller for a numerical inverse problem in which the wall flux is represented by a linear combination of basis functions. Do not modify the forward solver, the adjoint solver, or the inverse workflow. Act only after the forward and adjoint computations of each epoch are complete, and only on the optimization step. Select exactly one available numerical action and bounded controls from the observed history only. The true wall flux is unavailable. Do not invent values of the coefficients or frequencies; deterministic code will compute them. Return only the requested JSON object. &
\prompttt You are an auditable controller for a numerical inverse problem in which the wall flux is represented by a linear combination of basis functions. Do not modify the forward solver, the adjoint solver, or the inverse workflow. Act only after the forward and adjoint computations of each epoch are complete, and only on the optimization step. Select exactly one available numerical action and bounded controls from the observed history only. The true wall flux is unavailable. Except for the initial values of newly admitted basis functions, do not invent values of the coefficients or frequencies; deterministic code will compute them. Return only the requested JSON object. \\
\midrule
2. Task instruction &
\prompttt Control the linear coefficients and frequencies of the fixed basis library below, using only the observed history. The supplied objective and gradient are raw physical quantities; do not assume a particular numerical scale. Choose one available action and bounded controls that are likely to produce stable objective reduction and stationary coefficients and frequencies.\newline Library: [the basis functions of the common library]. &
\prompttt Control the linear coefficients and frequencies of the current basis library below, using only the observed history. The supplied objective and gradient are raw physical quantities; do not assume a particular numerical scale. Choose one available action and bounded controls that are likely to produce stable objective reduction and stationary coefficients and frequencies.\newline Current library: [basis functions admitted so far]. \\
\midrule
3. Epoch history &
\multicolumn{2}{p{0.84\textwidth}}{{\normalfont\itshape Common to both:}\newline \prompttt Observed history, oldest to newest (all epochs):\newline epoch | coefficients | frequencies | J{\char95}raw | dJ{\char95}dtheta | prior{\char95}action\newline {[\ldots]}} \\
\midrule
4. Bounds and trust radius &
\multicolumn{2}{p{0.84\textwidth}}{{\normalfont\itshape Common to both:}\newline \prompttt Current hard bounds: [bounds on coefficients and frequencies].\newline Current trust radius: [r{\char95}n].} \\
\midrule
5. Optimization policy &
\multicolumn{2}{p{0.84\textwidth}}{{\normalfont\itshape Common to both:}\newline \prompttt Available actions:\newline
- adam: controller Adam moment update of the controls\newline
- normalized{\char95}gradient: normalized downhill step of the controls within the trust radius\newline
- momentum{\char95}gradient: normalized downhill step with persistent momentum\newline
- barzilai{\char95}borwein: step with the Barzilai--Borwein step length from the two most recent epochs\newline
- levenberg{\char95}marquardt: damped Gauss--Newton step of the controls\newline
- conjugate{\char95}gradient: step along a conjugate direction of the recent gradients\newline
- projected{\char95}gradient: gradient step projected onto the hard bounds} \\
\midrule
6. Basis admission policy &
\itshape None. &
\prompttt The library starts from [1, x, x\^{}2] and may be revised at most once every [Delta n] epochs. If the objective has reached a plateau over the last [Delta n] epochs, admit the next family of basis functions in the order: polynomials of order 3 to 5; Fourier functions; products of Fourier functions and polynomials. Admitted functions are retained. For each admitted function, decide the initial values of its coefficient and frequency, and write the admitted functions and their initial values into the definition of theta. \\
\midrule
7. Output schema &
\prompttt Return exactly one JSON object. action must be available. damping and step{\char95}fraction must be in [0.02, 1.0]. reasoning must be a concise justification based on the observed history, at most 30 words. &
\prompttt Return exactly one JSON object. action must be available. damping and step{\char95}fraction must be in [0.02, 1.0]. reasoning must be a concise justification based on the observed history, at most 30 words. At a revision epoch, also return admit (true or false); if admit is true, return admitted{\char95}functions and initial{\char95}values for their coefficients and frequencies. \\
\bottomrule
\end{tabular}
\caption{Prompts of the active agentic optimization for the functional-basis identification of the spatially varying wall flux in Experiment~3, where the agent updates the controls of the fixed common library, and in Experiment~4, where the agent also grows the library. The controls are the linear coefficients and frequencies of the basis functions, and both experiments share the complete epoch history, the bounds on the controls, and the optimization policy. Experiment~4 adds the basis admission policy, under which the library grows from $\{1,\,x,\,x^2\}$ by admitting the next family of basis functions at a revision epoch if the misfit has reached a plateau. For each admitted function, the agent decides the initial values of its coefficient and frequency, returns them in the admission fields of the output schema, and writes them into the definition of $\theta$. Since $\theta$ parametrizes only the wall flux boundary condition, this edit does not modify the forward solver, the adjoint solver, or the inverse workflow referred to in block~1.}
\label{tab:app-basis-prompts}
\end{table}

The functional-basis identification represents the wall flux by a linear combination of basis functions, whose linear coefficients and frequencies are the controls (Section~\ref{sec:llm-basis}). The previously described Experiment~3 and Experiment~4 are validated. 

At each epoch, the controller passes the prompt (Table~\ref{tab:app-basis-prompts}), the optimization policy methods available at that epoch, and the epoch history to the agent. The epoch history comprises the linear coefficients and frequencies of all preceding epochs, the misfit, and its adjoint gradient, together with the method selected at each preceding epoch. The agent returns a structured response containing one available method, its controls, and a brief reasoning. In Experiment~4, the response additionally contains, at each revision, the admission decision and the initial values of the controls of the admitted basis functions, which the agent decides and writes into the definition of $\theta$.

\subsubsection*{Prompts}

The prompts of Experiments~3 and~4 are compared block by block in Table~\ref{tab:app-basis-prompts}. Both follow the structure of the active prompt in Table~\ref{tab:app-const-prompts} and share the epoch history, the bounds and trust radius, and the optimization policy (blocks~3--5). The prompt of Experiment~4 differs in four respects: its system instruction permits the agent to assign the initial values of newly admitted basis functions (block~1), its task instruction refers to the current rather than a fixed library (block~2), it contains the basis admission policy (block~6), and its output schema includes the admission fields (block~7). The word limit for the reasoning was not enforced by the controller, so that the logged reasoning of Experiment~3 partly exceeds it. In Table~\ref{tab:app-basis-trials-exp3}, the logged reasoning is therefore condensed.

\subsubsection*{Selected trials}

One trial of each experiment was drawn at random from its five trials for the hybrid profile with inner data: for Experiment~3 at $D=10^{-4}$ on M1, and for Experiment~4 at $D=10^{-6}$ on M2, the lowest diffusivity, at which Experiments~1 and~2 recover the profile only with errors of $15$--$25\%$ (Table~\ref{tab:basis_recovery}). Tables~\ref{tab:app-basis-trials-exp3} and~\ref{tab:app-basis-trials-exp4} list, for both trials, the first decision, the changes of method or library, and selected epochs at which the course of the recovery changes, including the epochs shown in Fig.~\ref{fig:app-validation}c. For Experiment~4, the recovered wall flux equation is given explicitly, so that the evolution of the library can be followed.

\begin{table} [!ht]
\centering
\scriptsize
\setlength{\tabcolsep}{3pt}
\renewcommand{\arraystretch}{1.2}
\begin{tabular}{@{}c >{\centering\arraybackslash}p{0.13\textwidth} c p{0.60\textwidth} c@{}}
\toprule
Epoch & Method & Error & \multicolumn{1}{c}{Logged Reasoning} & Ratio \\
$n$ & & $\varepsilon_{n+1}$ (\%) & & $J_{n+1}/J_n$ \\
\midrule
\multicolumn{5}{c}{\textbf{Experiment~3: common library} (hybrid profile, inner data, $D=10^{-4}$, M1)} \\
\midrule
1   & Adam & 99.391 & Only one epoch is available; Adam suits the highly anisotropic scaled gradient, whose grouped cosine-frequency component is much smaller than the coefficient components. Use a conservative first step. & 0.977 \\
5   & Barzilai--Borwein & 72.823 & Adam steps reduced the objective, but the latest produced sign reversals in the dominant gradient components; a safeguarded Barzilai--Borwein step with conservative damping limits overshoot. & 0.821 \\
10  & Barzilai--Borwein & 56.807 & All Barzilai--Borwein steps were accepted, the objective decreased monotonically, and the gradient norm contracted; continue with moderate damping to limit oscillation while preserving progress. & 3.021 \\
25  & Levenberg--Marquardt & 20.926 & Levenberg--Marquardt has reduced the objective at every epoch since epoch 19; the coefficient gradients reversed after the large step at epoch 23, so take a damped half step. & 0.855 \\
50  & Levenberg--Marquardt & 15.242 & Levenberg--Marquardt has consistently reduced the objective; continue it with moderate damping and a cautious step after the recent successful steps. & 0.990 \\
100 & Levenberg--Marquardt & 15.167 & Levenberg--Marquardt has produced the strongest sustained decrease of the objective, while recent full steps oscillate mildly; retain it with modest regularization and a half step. & 0.982 \\
250 & Normalized gradient & 12.683 & The current objective is the lowest observed, and the latest normalized-gradient step reduced it, whereas recent Levenberg--Marquardt steps were volatile; take a conservative gradient step. & 8.913 \\
500 & Normalized gradient & 3.684 & Recent larger Levenberg--Marquardt steps overshot, and the gradient points back toward the earlier low-misfit region; a damped normalized-gradient step recovers it and clears stale curvature memory. & -- \\
\bottomrule
\end{tabular}
\caption{Selected epochs of the selected successful trial of Experiment~3. The update is computed by the controller from the selected method of the optimization policy and its logged controls. $\varepsilon_{n+1}$ is the relative $L_\infty$ error of the recovered profile, evaluated offline after the trial and not available to the agent, and $J_{n+1}/J_n$ is the observed misfit ratio. The library is the common library of Eq.~\eqref{eq:common_library}, fixed throughout the trial; since no sparsity-promoting ($\ell_1$ norm) penalty is applied, all of its basis functions remain active, and the recovered expression is therefore not listed.}
\label{tab:app-basis-trials-exp3}
\end{table}

\begin{table} [!ht]
\centering
\scriptsize
\setlength{\tabcolsep}{3pt}
\renewcommand{\arraystretch}{1.15}
\begin{tabular}{@{}c >{\centering\arraybackslash}p{0.09\textwidth} >{\raggedright\arraybackslash}p{0.34\textwidth} c p{0.29\textwidth} c@{}}
\toprule
Epoch & Method & \multicolumn{1}{c}{Recovered wall flux} & Error & \multicolumn{1}{c}{Logged reasoning} & Ratio \\
$n$ & & \multicolumn{1}{c}{$g_{n+1}(x)$} & $\varepsilon_{n+1}$ (\%) & & $J_{n+1}/J_n$ \\
\midrule
\multicolumn{6}{c}{\textbf{Experiment~4: library grown by the agent} (hybrid profile, inner data, $D=10^{-6}$, M2)} \\
\midrule
1 & Normalized gradient &
$-0.005-12.139x+12.139x^2$ &
69.653 & Initial epoch has nonzero coefficient gradient; use the required full normalized coefficient step. & 0.164 \\
\midrule
5 & Normalized gradient &
$-2.282-10.674x+12.956x^2$ &
56.210 & J has decreased every epoch under normalized coupled gradients; retain the established step scales and damping. & 0.744 \\
\midrule

9 & Normalized gradient &
$-4.390 - 6.927x + 12.584x^{2}$ &
53.014 & J not decreased over recent epochs under normalized coupled gradients; retain the established step scales and damping; admitting the next family. & 0.994 \\
\midrule
15 & Normalized gradient &
$\begin{aligned}[t]
&-1.604 - 9.603x - 2.678x^{2}\\
&+ 9.298x^{3} + 9.001x^{4} + 0.241x^{5}
\end{aligned}$ &
46.598 & J has decreased monotonically through the recent normalized-gradient updates; retain the established coupled step scales. & 0.847 \\
\midrule
19 & Normalized gradient &
$\begin{aligned}[t]
&-1.342 - 10.598x - 2.223x^{2}\\
&+ 10.300x^{3} + 9.270x^{4} - 0.759x^{5}
\end{aligned}$ &
46.461 & J is not decreasing appreciably through the recent normalized-gradient updates; deciding to admit the next family. & 0.988 \\
\midrule

25 & Normalized gradient &
$\begin{aligned}[t]
&-1.276-13.119x+14.395x^2\\
&+0.002x^3+0.001x^4+0.001x^5\\
&+1.276\cos(4.054\pi x)+0.005\sin(4.002\pi x)
\end{aligned}$ &
70.346 & J has increased recently at every observed epoch under normalized-gradient updates; retain the established coupled step. Current frequency gradients are nonzero but small, supporting the reduced recent frequency rate. & 1.171 \\
\midrule

29 & Normalized gradient &
$\begin{aligned}[t]
& -0.068 - 11.829x + 10.642x^2 \\
& - 0.002x^3 + 0.004x^4 - 0.001x^5\\
&- 0.679\cos(3.976\pi x) - 0.006\sin(4.005\pi x)
\end{aligned}$ &
60.005 & J has decreased over recent epoch, but not significantly, under normalized-gradient updates; retain the established coupled step; decide to append the next family. & 1.008 \\
\midrule
35 & Normalized gradient &
$\begin{aligned}[t]
&-0.056-19.388x+19.444x^2+0.001x^4-0.001x^5\\
&+0.060\cos(4.070\pi x)-0.052x\cos(4.058\pi x)\\
&+16.450x\cos(5.994\pi x)-16.640x^2\cos(5.974\pi x)\\
&+0.002\sin(4.010\pi x)-0.001x\sin(3.994\pi x)\\
&+0.001x^2\sin(4.018\pi x)
\end{aligned}$ &
10.822 & J has decreased monotonically through the recent normalized-gradient updates; retain the established coupled step scales. Current active coefficient and frequency gradients remain material. & 0.915 \\
\midrule
50 & Normalized gradient &
$\begin{aligned}[t]
&0.095-20.963x+20.868x^2+0.001x^4-0.001x^5\\
&-0.095\cos(4.090\pi x)+0.095x\cos(4.084\pi x)\\
&+17.480x\cos(5.994\pi x)-17.484x^2\cos(5.984\pi x)\\
&+0.002\sin(4.010\pi x)-0.001x\sin(3.994\pi x)\\
&+0.001x^2\sin(4.018\pi x)
\end{aligned}$ &
6.988 & J has decreased for three consecutive epochs under the current reduced normalized-gradient schedule; retain it without a behavioral reset. & 0.962 \\
\midrule
75 & Normalized gradient &
$\begin{aligned}[t]
&0.258-21.985x+21.727x^2+0.001x^4-0.001x^5\\
&-0.258\cos(4.098\pi x)+0.257x\cos(4.094\pi x)\\
&+19.440x\cos(5.996\pi x)-19.560x^2\cos(5.990\pi x)\\
&+0.002\sin(4.010\pi x)-0.001x\sin(3.994\pi x)\\
&+0.001x^2\sin(4.018\pi x)
\end{aligned}$ &
4.160 & J increased at the current state and both observed gradients are zero; retain the established normalized-gradient schedule with a reduced step fraction. & -- \\
\bottomrule
\end{tabular}
\caption{Selected epochs of the selected successful trial of Experiment~4, including the epochs immediately preceding each revision of the library (epochs~9, 19, and~29). The recovered wall flux $g_{n+1}$ is given with the logged coefficients and frequencies of the active basis functions after the update, rounded to three decimals. $\varepsilon_{n+1}$ is the relative $L_\infty$ error of the recovered profile, evaluated offline after the trial (not available to the agent), and $J_{n+1}/J_n$ is the observed misfit ratio. The logged reasoning is reproduced as recorded and refers to the recent misfit history, which may differ from the ratio of the current epoch.}
\label{tab:app-basis-trials-exp4}
\end{table}

\textit{Experiment~3:} The trial ends with an error of $3.684\%$, which equals the median over the five trials (Table~\ref{tab:basis_recovery}). At epoch~2, the recovered profile oscillates with an incorrect frequency and exhibits two minima instead of three (Fig.~\ref{fig:app-validation}c, Experiment~3), with an error close to $100\%$. The agent first selects Adam and then changes to the Barzilai--Borwein method at epoch~5, which reduces the error to $56.8\%$ by epoch~10. The full Barzilai--Borwein step at epoch~10, however, increases the misfit by a factor of three, after which the agent changes to the Levenberg--Marquardt method. Under this method, the three minima of the true profile form, and the error decreases to $20.9\%$ at epoch~25 and to $15.2\%$ at epoch~100. At this stage, the central minimum is nearly recovered, whereas the right minimum is too deep and the profile does not return to zero at the right end of the wall. The agent subsequently changes to the normalized gradient method, whose step at epoch~250 increases the misfit by a factor of about nine. The agent then continues with damped normalized-gradient steps that return the controls toward the region of low misfit, and the error decreases from $12.7\%$ at epoch~250 to $3.684\%$ at epoch~500, at which the recovered profile agrees closely with the true profile.

\textit{Experiment~4:} The agent selects the normalized gradient method at every listed epoch, so that the course of the recovery is determined by the evolution of the library. With $\Delta n=10$, the library passes through four stages:
\begin{equation*}
\begin{aligned}
  \mathcal{D}^{(0)} &= \{1,\;x,\;x^2\}, && 1\le n<10,\\
  \mathcal{D}^{(1)} &= \mathcal{D}^{(0)}\cup\{x^3,\;x^4,\;x^5\}, && 10\le n<20,\\
  \mathcal{D}^{(2)} &= \mathcal{D}^{(1)}\cup\{\cos(\beta_1\pi x),\;\sin(\beta_2\pi x)\}, && 20\le n<30,\\
  \mathcal{D}^{(3)} &= \mathcal{D}^{(2)}\cup\{x\cos(\beta_3\pi x),\;x\cos(\beta_4\pi x),\;x^2\cos(\beta_5\pi x), && \\
  &\qquad\qquad\;\; x\sin(\beta_6\pi x),\;x^2\sin(\beta_7\pi x)\}, && 30\le n\le75,
\end{aligned}
\end{equation*}

where $\beta_1,\ldots,\beta_7$ are the frequencies of the individual Fourier and polynomial--Fourier basis functions. Within each revision interval, the misfit flattened in the epochs preceding the next revision, with misfit ratios of $0.994$, $0.988$, and $1.008$ at epochs~9, 19, and~29, respectively. The agent therefore identified a plateau at each of these epochs and decided to admit the next family, which entered the library at epochs~10, 20, and~30. After epoch~30, when no family remains, the library is no longer changed. Since no sparsity-promoting ($\ell_1$ norm) penalty is applied, every basis function remains in the expression, and its contribution is set by its linear coefficient and, for the Fourier and polynomial--Fourier functions, also by its frequency. A basis function therefore vanishes if its linear coefficient converges to zero or, for the sine-type functions, if its frequency converges to zero. The recovered wall flux at the selected epochs is listed in Table~\ref{tab:app-basis-trials-exp4}. The objective is a recovered profile that best approximates the true profile, rather than the identification of the basis functions of the true envelope. Each stage of the library is therefore assessed by the shape and the relative $L_\infty$ error of the recovered profile that it can represent.

With $\mathcal{D}^{(0)}$, the recovered profile is a single parabolic trough (Fig.~\ref{fig:app-validation}c, Experiment~4, epoch~2). It reproduces the trough of the true profile with a smaller depth but none of its three minima, with an error of $69.7\%$ after epoch~1, $56.2\%$ after epoch~5, and $53.0\%$ after epoch~9, at which the misfit has stagnated. With $\mathcal{D}^{(1)}$, the higher-order polynomials deepen the trough and reduce the error to $46.6\%$ after epoch~15, after which the misfit stagnates again, and the error remains at $46.5\%$ after epoch~19. A polynomial library of this order can, however, represent only a non-oscillatory profile, and not the three minima of the true profile, which limits the attainable error.

The extension to $\mathcal{D}^{(2)}$ admits the first oscillatory basis functions. After epoch~25, the linear coefficients of $x^3$, $x^4$, and $x^5$ have become negligible, and the recovered wall flux is dominated by the basis function $\cos(\beta_1\pi x)$ with $\beta_1\approx4.05$. This basis function produces two minima and a maximum at the center of the wall, where the true profile has its deepest minimum. At this epoch, the misfit increases, with a misfit ratio of $1.171$, and the error rises to $70.3\%$, since the controls of the newly admitted Fourier functions have not yet adjusted. By epoch~29, the coefficient of $\cos(\beta_1\pi x)$ has changed sign, the error has decreased to $60.0\%$, and the misfit has stagnated again, with a misfit ratio of $1.008$, which indicates that the Fourier functions alone cannot represent the three minima of the true profile.

The extension to $\mathcal{D}^{(3)}$ admits the polynomial--Fourier basis functions, with which the library can represent an oscillation whose amplitude varies along the wall. After epoch~35, the recovered profile reproduces the three minima of the true profile, and the error decreases to $10.8\%$. The recovered wall flux is now dominated by the basis functions $x$, $x^2$, $x\cos(\beta_4\pi x)$, and $x^2\cos(\beta_5\pi x)$, with $\beta_4\approx5.99$ and $\beta_5\approx5.97$, which closely resemble those of the true envelope, whereas the linear coefficient of $\cos(\beta_1\pi x)$ has become small. During the subsequent epochs, the depths of the minima are refined, and the error decreases to $7.0\%$ after epoch~50 and to $4.160\%$ after epoch~75, which is below the median of $4.987\%$ over the five trials (Table~\ref{tab:basis_recovery}). The remaining basis functions are retained with smaller linear coefficients, so that the recovered wall flux contains more basis functions than the true envelope, which is acceptable here, since only the approximation of the true profile is of interest.

\bibliographystyle{elsarticle-num}
\bibliography{reference}

\end{document}